\documentclass[aps,prd,a4paper,twocolumn]{revtex4}

\usepackage{graphicx}
\usepackage{bm}
\usepackage{amsfonts}
\usepackage{amsmath}

\usepackage[russian,ngerman,english]{babel}

\newcommand{\muas}[0]{\hbox{\rm $\mu$as}}

\newcommand{\ve}[1]{\mbox{\boldmath$#1$}}
\newcommand*\Laplace{\mathop{}\!\mathbin\bigtriangleup}

\let\oldbibitem\bibitem
\renewcommand\bibitem[2][]{\oldbibitem{#2}}

\begin{document}

\title{Light propagation in 2PN approximation in the monopole and quadrupole field of a body at rest: The basic transformations} 

\author{Sven Zschocke}

\affiliation{Lohrmann Observatory,
Dresden Technical University, Helmholtzstrasse 10, D-01069 Dresden, Germany}


\begin{abstract}
Todays precision in astrometry has reached a level of a few micro-arcseconds in the 
angular measurements of celestial objects. The next generations of astrometric facilities 
are aiming at the sub-micro-arcsecond scale of accuracy. Sub-micro-arcsecond astrometry 
requires a considerable improvement in the theory of light propagation in the curved space-time 
of the solar system. In particular, it is indispensable to determine light trajectories to the 
second order of the post-Newtonian scheme, where the monopole and quadrupole structure 
of some solar system bodies need to be taken into account. In reality, both the light source 
as well as the observer are located at finite spatial distances from the gravitating body. 
This fact implies for the need to solve the boundary value problem of light propagation, where 
the light trajectory is fully determined by the spatial positions of source and observer  
and its unit direction at past infinity. This problem has been solved in a recent investigation. 
A practical relativistic model of observational data reduction necessitates the determination 
of the unit tangent vector along the light trajectory at the spatial position of the observer. 
In this investigation, the unit tangent vector at the observers position is determined by a sequence 
of several basic transformations. The light trajectory is then fully given by this unit tangent vector 
at the observers position and by the spatial positions of the celestial light source and observer. 
The determination of this unit tangent vector allows one to calculate the impact of the monopole and 
quadrupole structure of solar system bodies on light deflection on the sub-micro-arcsecond level, 
both for stellar light sources as well as for light sources located in the solar system. Numerical 
values for the magnitude of light deflection, caused by the monopole and quadrupole structure of 
the gravitating bodies, are given for grazing light rays at the giant planets of the solar system. 
The General Relativistic Model (GREM) is presently used for data reduction of the ESA astrometry mission Gaia. 
It is shown how the implementation of these basic transformations into GREM would proceed for possible 
future space astrometry missions like Theia or GaiaNIR. 
\end{abstract}

\pacs{95.10.Jk, 95.10.Ce, 95.30.Sf, 04.25.Nx, 04.80.Cc}

\maketitle



\section{Introduction}\label{Introduction}

If the space-time scale of variation of the electromagnetic field is much smaller than that of the curvature of space-time, then 
light propagation can be described by null geodesics, which is the limit of geometrical optics \cite{Einstein2,MTW}. 
The calculation of these curvilinear light trajectories through the solar system is of decisive importance in relativistic astrometry in order to 
determine the actual astrometric positions of celestial light sources. A milestone in astrometric precision has been achieved by the space-mission Hipparcos 
of European Space Agency (ESA), which has reached accuracies on the milli-arcsecond (${\rm mas}$) level in angular observations of stellar 
objects \cite{Hipparchos1,Hipparchos2}. The state-of-the-art accuracy in angular measurements has reached the micro-arcsecond level ($\muas$), attained by the ESA 
astrometry mission Gaia \cite{Gaia1,Gaia2}. Also optical telescopes on Earth are aiming at similar levels of accuracies in near future \cite{ELT}. 

The next aim of astrometric science consists of advancing into areas on the sub-micro-arcsecond (sub-\muas) scale in positional measurements  
\cite{Proceeding_Sub_Microarcsecond_Astrometry,article_sub_micro_4}. In fact, there are several space-missions proposed to ESA, aiming at such 
ultra-highly precise astrometric measurements, like NEAT \cite{NEAT1,NEAT2,NEAT3}, Theia \cite{Theia}, Gaia-NIR \cite{Gaia_NIR},  
ASTROD \cite{Astrod1,Astrod2}, LATOR \cite{Lator1,Lator2}, ODYSSEY \cite{Odyssey}, SAGAS \cite{Sagas}, and TIPO \cite{TIPO}. 
Another direction aiming at the sub-\muas{} scale of accuracy is embarked by the proposal MoonLITE \cite{Moonlite}, an optical Michelson 
interferometer on the Moon with a baseline of $100\,{\rm m}$. Also feasibility studies of ground-based telescopes are under consideration aiming 
at accuracies on the sub-\muas{} level \cite{nas_telescope}. Most of these proposals are triggered by the search for Earth-like exoplanets 
in the vicinity of the Sun, detection of gravitational waves by astrometric measurements, measurements of dark matter distributions, and new 
highly precise tests of general relativity in the gravitational fields of the solar system. 

In these proposed missions it is necessary to determine the light trajectories in the first order of the post-Newtonian scheme (1PN) by taking into account the 
first few mass-multipoles of the multipole expansion of the metric tensor of solar system bodies \cite{Poncin_Lafitte_Teyssandier,Zschocke_Total_Light_Deflection}. 
In addition, the light trajectories in the second post-Newtonian scheme (2PN) need also to be determined, where at least the monopole and quadrupole structure of 
solar system bodies is taken into account. 

The problem of light propagation in the second order of the post-Newtonian scheme in the gravitational field of monopoles at rest has first been solved 
in \cite{Brumberg1987,Brumberg1991}. Subsequent investigations have recovered that it is necessary to implement these 2PN solutions in the relativistic model 
of light propagation for astrometry on the \muas{} level of accuracy, at least in the monopole field of the giant planets. In particular, it has been found that 
the 2PN monopole light deflection amounts up to $16.1$ \muas${}$ and $4.4$ \muas${}$ for grazing rays at the giant planets Jupiter and Saturn, respectively. The reason for this 
surprising result is the occurrence of so-called {\it enhanced monopole terms} in the 2PN light trajectory, which have been recovered independently by several 
investigations in \cite{Klioner_Zschocke,AshbyBertotti2010,Teyssandier}. The problem of 2PN light propagation in the monopole field has been reconsidered under 
several aspects and by using different approaches in a set of subsequent investigations
\cite{KlionerKopeikin1992,Deng_Xie,2PN_Light_PropagationA,Deng_2015,Xie_Huang,Minazzoli2,Hees_Bertone_Poncin_Lafitte_2014b}.
In order to account for the motion of the solar system bodies, the positions of the massive bodies are taken at their retarded instant of time in the 
relativistic model of data reduction for the Gaia mission. In order to investigate this issue further, the problem of light propagation in the gravitational fields 
of slowly moving monopoles in 2PN approximation has been solved in \cite{Zschocke3,Zschocke4,Zschocke5}.

The impact of the quadrupole structure of solar system bodies on light trajectories is the most significant effect beyond the monopole. Such a rigorous 2PN solution 
for light trajectories has been obtained only recently in our investigation \cite{Zschocke_Quadrupole_1}, where the initial value problem of 2PN light propagation 
in the quadrupole field of a body at rest has been solved. It has been found that the 2PN quadrupole light deflection amounts up to $0.95$ \muas${}$ 
and $0.29$ \muas${}$ for grazing rays at the giant planets Jupiter and Saturn, respectively, caused by {\it enhanced quadrupole terms}. 

In reality, both the celestial light source and the observer are located at finite distances from the gravitating body. Therefore, the solution of the boundary value 
problem of light propagation is required. This solution has been determined recently in our investigation in \cite{Zschocke_Quadrupole_2}. In particular, the 
2PN light trajectory has been given in terms of the spatial position $\ve{x}_0$ of the light source and of the spatial position $\ve{x}_1$ of the observer. For real 
astrometric measurements it is necessary to determine the following three fundamental transformations of the boundary value problem: 
$\ve{k} \rightarrow \ve{\sigma}$, $\ve{\sigma} \rightarrow \ve{n}$, $\ve{k} \rightarrow \ve{n}$, 
where $\ve{k}$ is the unit direction from the source toward the observer, $\ve{\sigma}$ is the unit tangent vector of light trajectory at minus infinity, 
and $\ve{n}$ is the unit tangent vector of the light ray at the spatial position of the observer. These transformations represent the basis of the Gaia relativistic 
model (GREM) of light propagation \cite{Klioner2003a}, which has later been refined by our investigations in \cite{Klioner_Zschocke,Zschocke_Klioner}. 
The implementation of the quadrupole structure of solar system bodies into these transformations would allow for highly precise measurements of light deflection 
on the sub-\muas${}$ level in the solar system. 
As mentioned, an important step toward the solution of the boundary value problem has been achieved in our recent investigation \cite{Zschocke_Quadrupole_2}, 
where the spatial coordinates of source and observer have been implemented in the solution of the light trajectory. In this present investigation the 
solution of the boundary value problem will be completed, as it has been announced in the summary section in \cite{Zschocke_Quadrupole_2}. 

The manuscript is organized as follows: In Section~\ref{Section1} the metric tensor of a body with monopole and quadrupole structure 
in 2PN approximation is briefly reconsidered and the 2PN solution of the geodesic equation for light rays is presented in the form as 
deduced in our article \cite{Zschocke_Quadrupole_2}. Section~\ref{GREM} gives a brief introduction into the relativistic model and data reduction  
to determine the spatial position of the celestial light sources. In Sections~\ref{Transformation1} - \ref{Transformation3} the approach 
of how to obtain the transformations of the boundary value problem is described and these basic transformations are represented. 
Upper limits of the quadrupole light deflection for celestial light sources in the solar system and numerical values 
are given in Section~\ref{Section3}. The results of this investigation are summarized in Section~\ref{Summary}. The notations, 
some details of the calculations and tensorial coefficients and scalar functions are shifted into a set of several Appendixes.

\section{The metric tensor and geodesic equation and light trajectory}\label{Section1}  

The curved space-time of an isolated body at rest is assumed to be covered by harmonic coordinates $\left(x^0,x^1,x^2,x^3\right)$, where the origin of the spatial coordinates 
is located at the center of mass of the body, which is assumed to be at rest with respect to the global harmonic coordinate system. 
Because the gravitational fields of the solar system and the light trajectories through the solar system cannot be solved
in their exact form one has to resort on approximation schemes. The post-Newtonian scheme is used in case of weak gravitational fields and 
slow motions of matter, where the metric tensor is series expanded in inverse powers of the speed of light. 
Furthermore, the solar system bodies are not simple monopoles, but can be of complicated structure. These physical features of the gravitating bodies are described 
by a set of six source-multipoles: $\{I_L, J_L, W_L, X_L, Y_L, Z_L\}$ \cite{Thorne,Blanchet_Damour1,Multipole_Damour_2,2PN_Metric1}. 
In the canonical harmonic gauge, the metric tensor depends finally only on a set of two multipoles: mass-multipoles, $M_L = I_L + {\cal O}(c^{-5})$ \cite{Blanchet_Faye_Iyer_Sinha}, 
which account for shape and inner structure of the body, and spin-multipoles, $S_L = J_L + {\cal O}(c^{-5})$ \cite{Blanchet_Faye_Iyer_Sinha}, which account for rotational motions 
and inner currents of the body \cite{Thorne,Blanchet_Damour1,Multipole_Damour_2,2PN_Metric1}.

\subsection{The metric tensor} 

In our investigations in \cite{Zschocke_Quadrupole_1,Zschocke_Quadrupole_2}, we have assumed that the solar system body has no rotational motions 
and no inner currents of matter: $S_L = 0$. Furthermore, we have taken into account only the mass-monopole and mass-quadrupole terms of the multipole decomposition, 
while higher multipoles were neglected. Then, the post-Newtonian expansion of the metric tensor reads, up to terms of the order ${\cal O}\left(c^{-6}\right)$,  
\begin{eqnarray}
        g_{\alpha\beta} &=& \eta_{\alpha\beta} + h_{\alpha\beta}^{\left(2\right)}\left(M,M_{ab}\right) 
        + h_{\alpha\beta}^{\left(4\right)}\left(M,M_{ab}\right), 
\label{Metric}
\end{eqnarray}

\noindent 
where $\eta_{\alpha\beta} = {\rm diag}\left(-1,+1,+1,+1\right)$ is the metric of flat space-time, and where the metric perturbations are of the orders  
$h_{\alpha\beta}^{\left(2\right)} = {\cal O}\left(c^{-2}\right)$ and $h_{\alpha\beta}^{\left(4\right)} = {\cal O}\left(c^{-4}\right)$. 
In the stationary case, the mass-monopole and mass-quadrupole are given by 
\begin{eqnarray}
M &=& \int d^3 x \,\frac{T^{00} + T^{kk}}{c^2}\,,
\label{Mass}
\\
M_{ab} &=& \int d^3 x \,\underset{ab}{\rm STF} \,{x}_{a b}\,\frac{T^{00} + T^{kk}}{c^2}\,,
\label{Quadrupole}
\end{eqnarray}

\noindent
where $T^{\alpha\beta}$ is the energy-momentum tensor of the body. 
The integrals in (\ref{Mass}) and (\ref{Quadrupole}) run over the three-dimensional volume of the body, 
and $\underset{ab}{\rm STF} \,{x}_{a b} = x_a x_b - \frac{1}{3}\, r^2 \delta_{ab}$ is the symmetric and trace-free (STF) part of $x_{ab} = x_a x_b$ with $r = |\ve{x}|$. 
The mass-dipole is zero, $M_i = 0$, because the origin of the spatial axes is located the center of mass of the source.
The metric perturbations in (\ref{Metric}) can be deduced from the metric density achieved in the basic investigations 
in \cite{Thorne,Blanchet_Damour1,Multipole_Damour_2,2PN_Metric1} and were explicitly given in \cite{Zschocke_2PM_Metric,Frutos_Alfaro_Soffel}.

\subsection{The geodesic equation for light rays}

The light signals propagate through the curved space-time of the solar system along null-geodesics. The trajectories of these light signals are determined 
by the geodesic equation, which in terms of coordinate time reads \cite{MTW,Brumberg1991,Kopeikin_Efroimsky_Kaplan}: 
\begin{eqnarray}
	\frac{\ddot{x}^i\left(t\right)}{c^2}   
	+ \Gamma^{i}_{\mu\nu} \frac{\dot{x}^{\mu}\left(t\right)}{c} 
        \frac{\dot{x}^{\nu}\left(t\right)}{c}   
	&=& \Gamma^{0}_{\mu\nu} \frac{\dot{x}^{\mu}\left(t\right)}{c} \frac{\dot{x}^{\nu}\left(t\right)}{c} \frac{\dot{x}^{i}\left(t\right)}{c}\,, 
\label{Geodetic_Equation}
\end{eqnarray}

\noindent
where a dot means total derivative with respect to the coordinate time and $\Gamma^{\alpha}_{\mu\nu}$ are the Christoffel symbols, which are functions of 
the metric tensor $g_{\alpha\beta}$. 

The geodesic equation is a differential equation of second order. In order to find an unique solution of the geodesic equation 
in (\ref{Geodetic_Equation}), so-called mixed initial-boundary conditions must be imposed, which have extensively been used in the literature, e.g. 
\cite{Brumberg1991,Kopeikin_Efroimsky_Kaplan,KlionerKopeikin1992,Klioner_Zschocke,Kopeikin1997,KopeikinSchaefer1999_Gwinn_Eubanks,Brumberg1987,KopeikinKorobkovPolnarev2006},
\begin{eqnarray}
        \ve{\sigma} &=& \frac{\dot{\ve{x}}\left(t\right)}{c}\bigg|_{t = - \infty}\,,
\label{Boundary_Condition}
\\
        \ve{x}_0 &=& \ve{x}\left(t\right)\bigg|_{t=t_0}\;. 
\label{Initial_Condition}
\end{eqnarray}

\noindent 
The condition (\ref{Boundary_Condition}) defines the unit-direction, $\ve{\sigma}\cdot\ve{\sigma} = 1$, of the light ray at past null infinity. The condition (\ref{Initial_Condition}) 
appears also in the boundary value problem and defines the spatial coordinates of the photon at the moment of emission of the light signal, $t_0$, that is the spatial position of the 
light source. At this stage, the three-vectors in (\ref{Boundary_Condition}) and (\ref{Initial_Condition}) are still free parameters of the model (constants of integration of the geodesic equation). 
Later on, the three-vector $\ve{\sigma}$ is uniquely be determined by the transformation $\ve{\sigma}$ to $\ve{n}$, given in Section \ref{Transformation2}, 
and the determination of the spatial position $\ve{x}_0$ from the observed light direction by an observer is discussed in Section~\ref{GREM}. 

\begin{figure}[!ht]
\begin{center}
\includegraphics[scale=0.15]{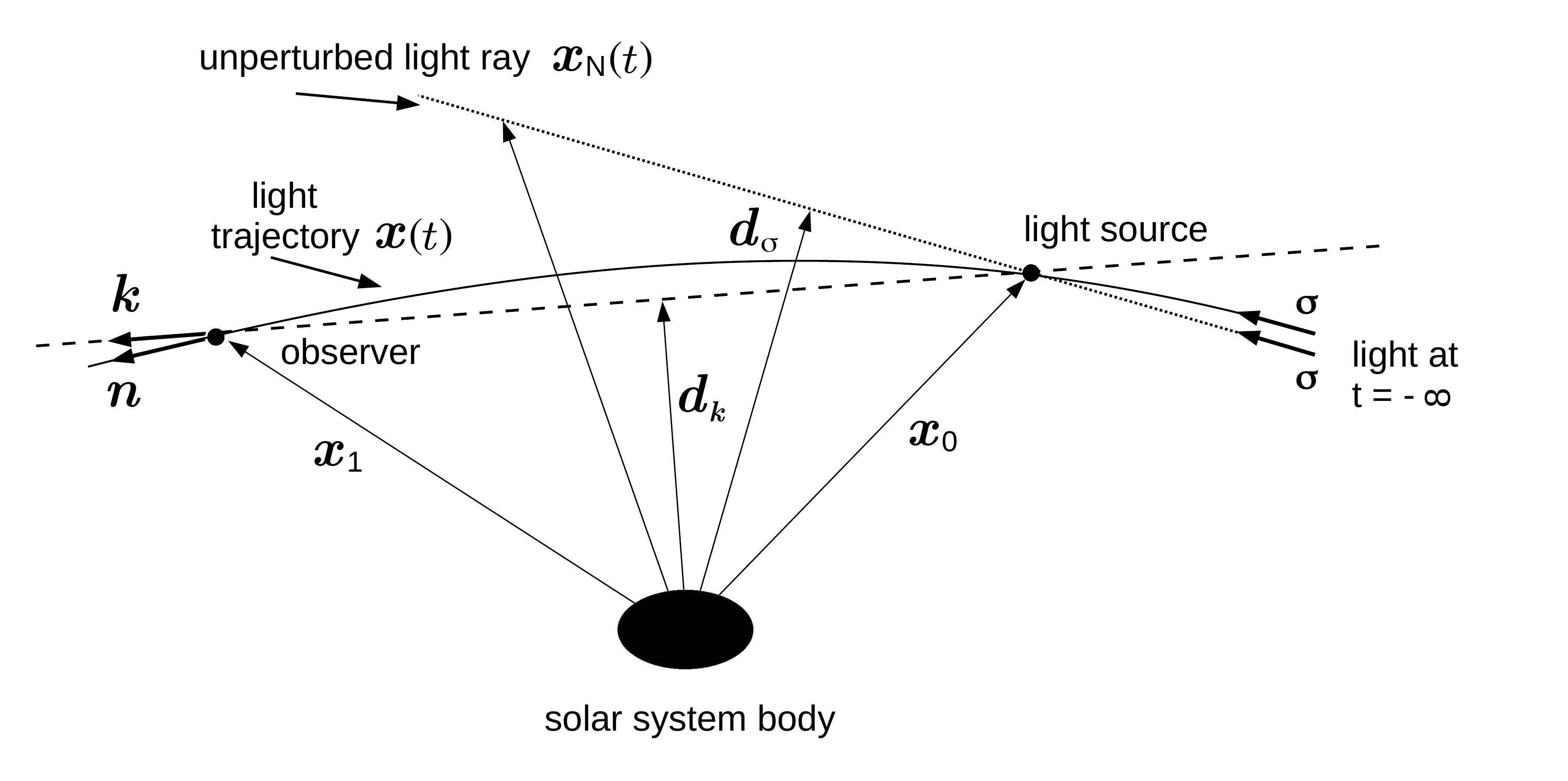}
\end{center}
        \caption{A geometrical representation of the propagation of a light signal through the gravitational field of a solar system body at rest. The origin of the spatial 
	coordinates is located at the barycenter of the body. The spatial position of the light signal $\ve{x}(t)$ is here with respect to the barycenter of the solar system body. 
	The light source is located at $\ve{x}_0$ and the observer is located at spatial position $\ve{x}_1$, both these vectors are with respect to the barycenter of the body. 
	The light signal is emitted by the light source at $\ve{x}_0$ and propagates along the exact light trajectory $\ve{x}\left(t\right)$ 
	(solid line). The unperturbed light ray $\ve{x}_{\rm N}\left(t\right)$ is given by Eq.~(\ref{unperturbed_lightray}) and propagates in the direction 
	of $\ve{\sigma}$ along a straight line through the position of the light source at $\ve{x}_0$ (dotted line). 
	The spatial positions of source and observer are connected by a straight line (fine dashed line). 
        The unit tangent vector $\ve{\sigma}$ along the light trajectory at past null infinity is defined by Eq.~(\ref{Boundary_Condition}). 
	The unit tangent vector $\ve{n}$ along the light trajectory at the observers position is defined by Eq.~(\ref{vector_n}). 
	The unit tangent vector $\ve{k}$ points from the source toward the observer and is defined by Eq.~(\ref{vector_k}). 
        The impact vector $\ve{d}_{\sigma}$ of the unperturbed light ray is given by Eq.~(\ref{impact_vector_unperturbed}). The impact vector $\ve{d}_k$ 
	from the center of mass of the body toward the connecting line between source and observer is given by Eq.~(\ref{impact_vector_k}).} 
\label{Diagram}
\end{figure}

The first integration of geodesic equation (\ref{Geodetic_Equation}) yields the coordinate light velocity, $\dot{\ve{x}}(t)$, and the second integration of geodesic 
equation (\ref{Geodetic_Equation}) yields the light trajectory, $\ve{x}(t)$. In the 2PN approximation they can formally be written in the form 
\begin{eqnarray}
        \frac{\dot{\ve{x}}\left(t\right)}{c} &=& \frac{\dot{\ve{x}}_{\rm 2PN}\left(t\right)}{c} + {\cal O}\left(c^{-6}\right), 
        \label{light_velocity1}
        \\
        \ve{x}\left(t\right) &=& \ve{x}_{\rm 2PN}\left(t\right) + {\cal O}\left(c^{-6}\right),
        \label{light_trajectory1}
\end{eqnarray}

\noindent 
which means in 2PN approximation only terms up to and including order ${\cal O}(c^{-4})$ are taken into account. In what follows, the unperturbed light ray 
\begin{eqnarray}
	\ve{x}_{\rm N} &=& \ve{x}_0 + c \left(t-t_0\right) \ve{\sigma} 
	\label{unperturbed_lightray}
\end{eqnarray}

\noindent
and the impact vector of the unperturbed light ray 
\begin{eqnarray}
	\ve{d}_{\sigma} &=& \ve{\sigma} \times \left( \ve{x}_{\rm N} \times \ve{\sigma} \right)
        \label{impact_vector_unperturbed}
\end{eqnarray}

\noindent
are needed. This impact vector can also be written in the form as given by Eq.~(\ref{impact_vector_0_appendix}). In Fig.~\ref{Diagram} these 
terms in Eqs.~(\ref{Boundary_Condition}) - (\ref{impact_vector_unperturbed}) have been elucidated. 

The coordinate velocity (\ref{light_velocity1}) and the trajectory (\ref{light_trajectory1}) of the light signal in the scheme of the initial value problem 
with the metric tensor (\ref{Metric}) have been determined in the following form  
\begin{eqnarray}
	\frac{\dot{\ve{x}}_{\rm 2PN}\left(t\right)}{c} &=& \ve{\sigma} + \frac{\Delta \dot{\ve{x}}_{\rm 1PN}\left(t\right)}{c} + \frac{\Delta \dot{\ve{x}}_{\rm 2PN}\left(t\right)}{c}\,, 
        \label{light_velocity2}
	\\ 
	\ve{x}_{\rm 2PN}\left(t\right) &=& \ve{x}_0 + c \left(t - t_0\right) \ve{\sigma} 
	\nonumber\\ 
	&& + \Delta\ve{x}_{\rm 1PN}\left(t\right) - \Delta\ve{x}_{\rm 1PN}\left(t_0\right) 
	\nonumber\\
        && + \Delta\ve{x}_{\rm 2PN}\left(t\right) - \Delta\ve{x}_{\rm 2PN}\left(t_0\right), 
        \label{light_trajectory2}
\end{eqnarray}

\noindent 
respectively, where the perturbations in (\ref{light_velocity2}) and (\ref{light_trajectory2}) were presented in \cite{Zschocke_Quadrupole_1}. 

\subsection{The light trajectory of the boundary value problem}

The relativistic model of data reduction requires one to implement the spatial position of the observer, $\ve{x}_1$, given by 
\begin{eqnarray}
        \ve{x}_1 &=& \ve{x}\left(t\right)\,\,\bigg|_{t = t_1}\,. 
        \label{boundary1}
\end{eqnarray}

\noindent 
This equation states, that the spatial position of the light signal at time of reception, $t_1$, is in coincidence with the spatial position of the observer. 
The implementation of the spatial position of the observer, $\ve{x}_1$, has recently been performed in our investigation \cite{Zschocke_Quadrupole_2}. 

The solution of the boundary value problem in 2PN approximation has been presented by Eqs.~(53) - (56) in \cite{Zschocke_Quadrupole_2} and reads 
\begin{eqnarray} 
\frac{\dot{\ve{x}}_{\rm 2PN}\left(t_1\right)}{c} &=& \ve{\sigma} + \frac{\Delta\dot{\ve{x}}_{\rm 1PN}\left(\ve{x}_1\right)}{c} 
+ \frac{\Laplace \dot{\ve{x}}_{\rm 2PN}\left(\ve{x}_1\right)}{c}\;,
\label{First_Integration_2PN_boundary} 
\\
\ve{x}_{\rm 2PN}\left(t_1\right) &=& \ve{x}_0 + c \left(t_1 - t_0\right) \ve{\sigma}
\nonumber\\ 
	&& + \Delta \ve{x}_{\rm 1PN}\left(\ve{x}_1\right) - \Delta \ve{x}_{\rm 1PN}\left(\ve{x}_0\right) 
	\nonumber\\ 
        && + \Laplace\ve{x}_{\rm 2PN}\left(\ve{x}_1\right) - \Laplace\ve{x}_{\rm 2PN}\left(\ve{x}_0\right).  
\label{Second_Integration_2PN_boundary}
\end{eqnarray}

\noindent 
The symbol $\Laplace$ (Laplace) instead of $\Delta$ (Delta) in Eqs.~(\ref{First_Integration_2PN_boundary}) and (\ref{Second_Integration_2PN_boundary}) 
has originally been introduced in our work \cite{Zschocke_Quadrupole_2} in order to distinguish these 2PN terms of the boundary value problem 
from the 2PN terms of the initial value problem; cf. text below Eq.~(39) in \cite{Zschocke_Quadrupole_2}. 
The 1PN perturbation terms in (\ref{First_Integration_2PN_boundary}) and (\ref{Second_Integration_2PN_boundary}) were given by Eqs.~(51) and (52) in \cite{Zschocke_Quadrupole_2}. 
Their spatial components ($i=1,2,3$) read: 
\begin{eqnarray}
        \frac{\Delta\dot{x}^i_{\rm 1PN}\left(\ve{x}_1\right)}{c} &=&
        \frac{G M}{c^2} \sum\limits_{n=1}^{2} U^i_{(n)}\left(\ve{x}_1\right)\,\dot{F}_{(n)}\left(\ve{x}_1\right)
	\nonumber\\ 
	&& \hspace{-1.5cm} + \frac{G M_{ab}}{c^2} \sum\limits_{n=1}^{8} V^{ab\,i}_{(n)}\left(\ve{x}_1\right)\,\dot{G}_{(n)}\left(\ve{x}_1\right),
\label{First_Integration_1PN_Final}
\\
        \Delta x^i_{\rm 1PN}\left(\ve{x}_1\right) &=&
        \frac{G M}{c^2} \sum\limits_{n=1}^{2} U^i_{(n)}\left(\ve{x}_1\right)\,F_{(n)}\left(\ve{x}_1\right)
	\nonumber\\ 
	&& \hspace{-1.5cm} + \frac{G M_{ab}}{c^2} \sum\limits_{n=1}^{8} V^{ab\,i}_{(n)}\left(\ve{x}_1\right)\, G_{(n)}\left(\ve{x}_1\right). 
\label{Second_Integration_1PN_Final}
\end{eqnarray}

\noindent
The term $\Delta \ve{x}_{\rm 1PN}\left(\ve{x}_0\right)$ in Eq.~(\ref{Second_Integration_2PN_boundary}) is obtained from (\ref{Second_Integration_1PN_Final}) 
by the replacement $\ve{x}_1$ by $\ve{x}_0$ in the tensorial coefficients as well as in the scalar functions. 
The 2PN perturbation terms in (\ref{First_Integration_2PN_boundary}) and (\ref{Second_Integration_2PN_boundary}) were given by Eqs.~(55) and (56) in \cite{Zschocke_Quadrupole_2}. 
Their spatial components ($i=1,2,3$) read: 
\begin{eqnarray}
        \frac{\Laplace \dot{x}^i_{\rm 2PN}\left(\ve{x}_1\right)}{c} &=& 
        \frac{G^2 M^2}{c^4} \sum\limits_{n=1}^{2} U^i_{(n)}\left(\ve{x}_1\right)\,\dot{X}_{(n)}\left(\ve{x}_1\right)
	\nonumber\\ 
	&& \hspace{-1.5cm} + \frac{G^2 M M_{ab}}{c^4} \sum\limits_{n=1}^{8} V^{ab\,i}_{(n)}\left(\ve{x}_1\right)\,\dot{Y}_{(n)}\left(\ve{x}_1\right)
        \nonumber\\
        && \hspace{-1.5cm} + \frac{G^2 M_{ab} M_{cd}}{c^4} \sum\limits_{n=1}^{28} W^{abcd\,i}_{(n)}\left(\ve{x}_1\right)\,\dot{Z}_{(n)}\left(\ve{x}_1\right),
\label{First_Integration_2PN_Final}
\\
       \Laplace x^i_{\rm 2PN}\left(\ve{x}_1\right) &=&
       \frac{G^2 M^2}{c^4} \sum\limits_{n=1}^{2} U^i_{(n)}\left(\ve{x}_1\right)\,X_{(n)}\left(\ve{x}_1\right)
       \nonumber\\ 
       && \hspace{-1.5cm} + \frac{G^2 M M_{ab}}{c^4} \sum\limits_{n=1}^{8} V^{ab\,i}_{(n)}\left(\ve{x}_1\right)\,Y_{(n)}\left(\ve{x}_1\right)
       \nonumber\\
        && \hspace{-1.5cm} + \frac{G^2 M_{ab} M_{cd}}{c^4} \sum\limits_{n=1}^{28} W^{abcd\,i}_{(n)}\left(\ve{x}_1\right)\,Z_{(n)}\left(\ve{x}_1\right). 
\label{Second_Integration_2PN_Final}
\end{eqnarray}

\noindent
The term $\Laplace \ve{x}_{\rm 2PN}\left(\ve{x}_0\right)$ in Eq.~(\ref{Second_Integration_2PN_boundary}) is obtained from (\ref{Second_Integration_2PN_Final})
by the replacement $\ve{x}_1$ by $\ve{x}_0$ in the tensorial coefficients as well as in the scalar functions. 

These tensorial coefficients in Eqs.~(\ref{First_Integration_1PN_Final}) - (\ref{Second_Integration_2PN_Final}) were given by Eqs.~(E1) - (E10) and Eqs.~(H1) - (H28) 
in our article \cite{Zschocke_Quadrupole_2}. In general, these tensorial coefficients are products of Kronecker symbols, three-vectors $\ve{\sigma}$, and impact vectors 
\begin{eqnarray}
        \ve{d}^{\,0}_{\sigma} &=& \ve{\sigma} \times \left(\ve{x}_0 \times \ve{\sigma}\right), 
        \label{impact_vector_0}
        \\
        \ve{d}^{\,1}_{\sigma} &=& \ve{\sigma} \times \left(\ve{x}_1 \times \ve{\sigma}\right), 
        \label{impact_vector_1}
\end{eqnarray}

\noindent
where their absolute values, $d^{\,0}_{\sigma} = |\ve{d}^{\,0}_{\sigma}|$ and $d^{\,1}_{\sigma} = |\ve{d}^{\,1}_{\sigma}|$, are determined with the Euclidean metric: 
$|\ve{a}|^2 = a^i a^j \delta_{ij}$. 
Each of these tensorial coefficients has one lower case Latin index $i$, which indicates the spatial index of the unit tangent vector (\ref{Boundary_Condition}),
or the spatial index of the impact vectors (\ref{impact_vector_0}) and (\ref{impact_vector_1}), or is one of the two indices of the Kronecker symbol. 
The lower case Latin indices $a,b$ of the tensorial coefficients $V^{ab\,i}_{(n)}$ are contracted with the corresponding indices of the quadrupole tensor $M_{ab}$. 
Similarly, the lower case Latin indices $a,b,c,d$ of the tensorial coefficients $W^{abcd\,i}_{(n)}$ are contracted with the corresponding indices of the 
product of two quadrupole tensors $M_{ab} \, M_{cd}$, according to Einstein's sum convention.  

The scalar functions in Eqs.~(\ref{First_Integration_1PN_Final}) - (\ref{Second_Integration_2PN_Final}) were given in their explicit form
by Eqs.~(I8) - (I27) and by Eqs.~(J1) - (J76) in Appendixes~I and J, respectively, of our article \cite{Zschocke_Quadrupole_2}.
These scalar functions depend on three-vector $\ve{\sigma}$, on the impact vectors (\ref{impact_vector_0}) and (\ref{impact_vector_1}) 
as well as on the spatial positions of source and observer.

\section{Relativistic models and data reduction}\label{GREM}

The solution for the coordinate velocity (\ref{light_velocity1}) and trajectory (\ref{light_trajectory1}) of the light signal 
represents the main part of relativistic models for high-accuracy astrometry, while the  
final ambition of the model is the determination of the spatial position $\ve{x}_0$ of the celestial light source from 
astrometric measurements. The treatment of this problem is concerned with data reduction of astrometric observations with the aid of such solutions. 
The general principles are described in standard text books \cite{Brumberg1991,Kopeikin_Efroimsky_Kaplan}. Let us consider some basics 
of this treatment as used for the data reduction within the ESA astrometry mission Gaia, which are relevant for our investigations 
aiming at next-generation astrometry missions beyond the Gaia mission. 

The Gaia mission is an all-sky high-precision astrometric satellite, aiming at the largest and most precise three-dimensional map of 
nearly two billion celestial objects, mainly stars of our Galaxy, but also Solar system objects, exo-planets, and quasars.  
In the final star catalog of Gaia, announced for $2030$, the positions and motions of the stellar objects are given with respect to the Barycentric Celestial Reference System (BCRS), 
which is the global standard harmonic coordinate system according to the resolutions of the International Astronomical Union (IAU) \cite{IAU_Resolution1,Kopeikin_Efroimsky_Kaplan}. 
Currently, two independent advanced models for astrometric observations are used for data modeling and reduction pipeline of the astrometry mission Gaia. 
The first model is the General Relativistic Model (GREM), which has been developed in \cite{Klioner2003a,Klioner2004} with several refinements added at a
later stage \cite{Klioner_Zschocke,Zschocke_Klioner}. The second model is the Relativistic Astrometric Model (RAMOD) developed in \cite{RAMOD1,RAMOD2,RAMOD3,RAMOD4}. 
Both these models are designed for relativistic astrometry at micro-arcsecond level of accuracy and allow for an independent check of their results. 
These relativistic models are based on the solution of the geodesic equation for light rays. 
Another independent approach is based on the Time Transfer Function (TTF) and has originally been developed in \cite{TTF1,TTF2}, 
which can also be used to define an astrometric observation using an integral-based method derived from the Synge World Function. 

\begin{figure}[!ht]
\begin{center}
\includegraphics[scale=0.125]{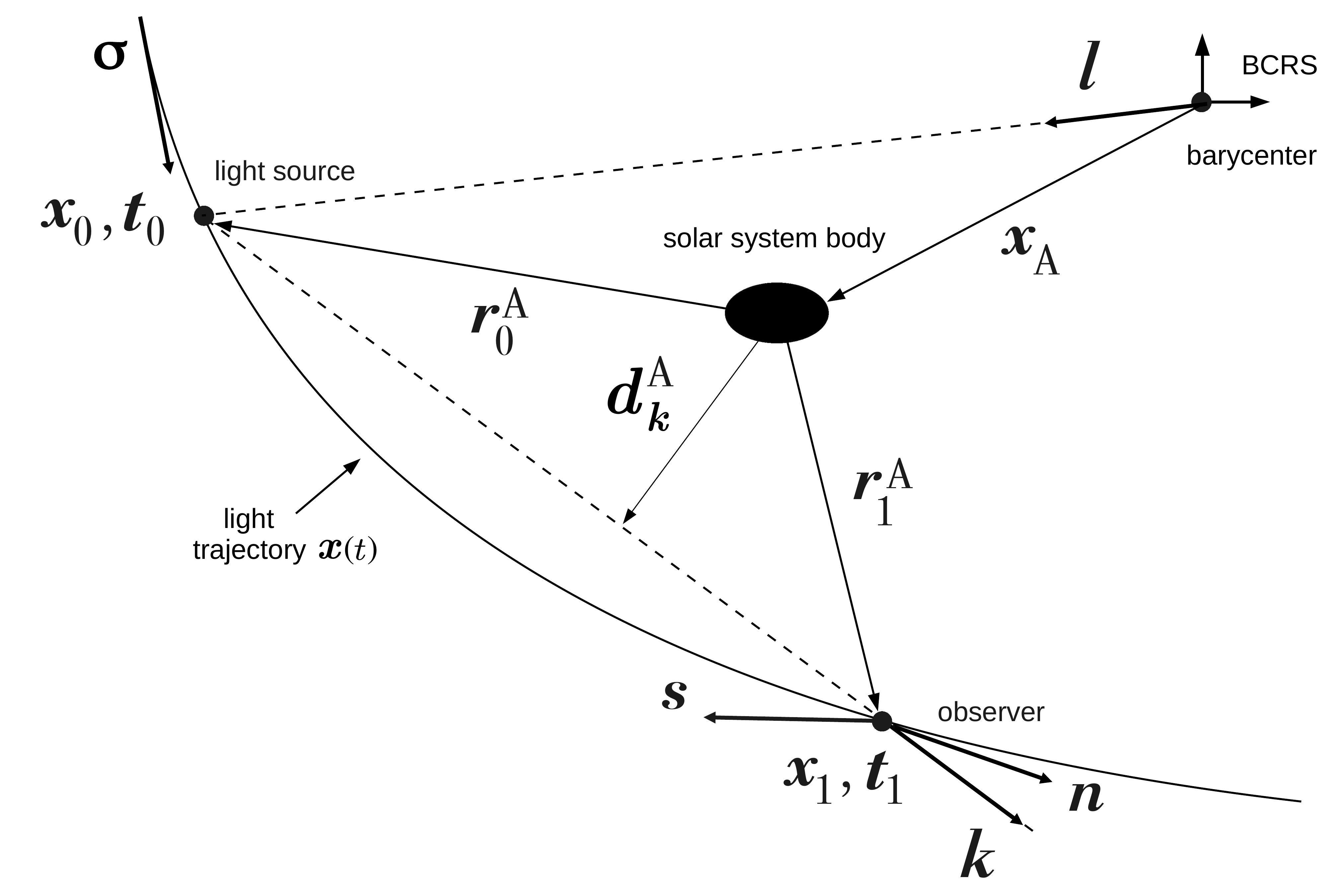}
\end{center}
	\caption{A geometrical representation of the General Relativistic Model (GREM) according to \cite{Klioner2003a,Klioner2004}, which is used for data reduction in the ESA 
	astrometry mission Gaia \cite{Gaia1,Gaia2}. The diagram illustrates how the implementation of the basic transformations into GREM proceeds. The five unit vectors, 
	$\ve{s}$, $\ve{n}$, $\ve{k}$, $\ve{\sigma}$, $\ve{l}$, of the model GREM are explained in the main text.  
	The origin of the spatial coordinates is located at the barycenter of the solar system. The spatial position of the 
	light signal $\ve{x}(t)$ and of the source and observer, $\ve{x}_0$ and $\ve{x}_1$, are with respect to the barycenter of the solar system. 
	The light signal is emitted at coordinate time $t_0$ and received at coordinate time $t_1$. The 
	spatial position of the body in the BCRS is denoted by $\ve{x}_A$. In order to implement the transformations $\ve{k} \rightarrow \ve{\sigma}$, $\ve{\sigma} \rightarrow \ve{n}$, 
	$\ve{k} \rightarrow \ve{n}$, as given by Eqs.~(\ref{transformation_k_to_sigma}), (\ref{transformation_sigma_to_n}), (\ref{transformation_k_to_n}),  
	into GREM, one has to perform a translation of the spatial coordinates from the barycenter of the solar system body into the barycenter of the solar system,  
	which implies a replacement $\ve{x}_0$ by $\ve{r}_0^A = \ve{x}_0 - \ve{x}_A$ and $\ve{x}_1$ by $\ve{r}_1^A = \ve{x}_1 - \ve{x}_A$ in these transformations. 
	These replacements would have to be performed in the scalar functions as well as in the tensorial coefficients. In particular, the impact vector $\ve{d}_k$ 
	in Eq.~(\ref{impact_vector_k}) has to be replaced by $\ve{d}_k^A = \ve{k} \times (\ve{r}_0^A \times \ve{k}) = \ve{k} \times (\ve{r}_1^A \times \ve{k})$.}
\label{Diagram1} 
\end{figure}

In the relativistic model GREM, the determination of the spatial position $\ve{x}_0$ of the light source in the BCRS from the observed direction of the incoming light ray  
at the observer proceeds in a sequence of transformations between a set of altogether five unit vectors. The components of these five unit vectors are with respect to the 
basis vectors of the global coordinate system \cite{Klioner2003a,KlionerKopeikin1992} (see also Fig.~\ref{Diagram1}): the unit vector $\ve{s}$ which denotes the direction 
toward the source as measured by the kinematically non-rotating observer (Gaia satellite), the unit tangent vector $\ve{n}$ of the light ray at the spatial position of the observer, 
the unit direction $\ve{k}$ from the source toward the observer, the unit tangent vector $\ve{\sigma}$ of light trajectory at minus infinity, and the unit vector $\ve{l}$ directed 
from the barycenter of the global coordinate system toward the source. 

\noindent 
In order to determine the spatial position of celestial light sources by astrometric measurements, one needs to account for two primary effects: 
(a) {\it aberration} and (b) {\it parallax}. In what follows, these two effects and their relations to these five unit vectors of GREM will briefly been considered: 

(a) The effect of {\it aberration} is described by the transformation between the unit vectors $\ve{s}$ and $\ve{n}$; cf. Eq.~(7) in \cite{Klioner2003a}. 
To clarify the effect of {\it aberration}, let us consider a fictitious observer, who is located at the same spatial position as the real satellite at the moment of observation. 
This fictitious observer is assumed to be at rest with respect to the BCRS. The direction toward the source as seen by this fictitious observer is denoted by $\ve{s}^{\prime}$.  
Clearly, we have $\ve{s}^{\prime} = - \ve{n}$, which means there is no effect of aberration in this case. The three-vectors $\ve{s}$ and $\ve{s}^{\prime}$ are related to each other 
by a Lorentz boost (Eq.~(10) in \cite{Klioner2003a}), where the three-velocity of the satellite in the BCRS is not the same as in flat space-time, 
but contains a term with the gravitational potential of the solar system at the position of the observer; see Eq.~(12) in \cite{Klioner2003a}. 
This expression follows from the normalization of the four-velocity $u^{\mu}$ of the observer: $g_{\mu\nu}\,u^{\mu} u^{\nu} = - 1$, where $g_{\mu\nu}$ is here the BCRS metric. 
Hence, the three-vector $\ve{s}$ can be obtained from three-vector $\ve{n}$, and vice versa, as soon as the ephemeris of the observers trajectory are known. 
The problem of how to get the components of same three-vector $\ve{s}$ with respect to the tetrads of the observer has been described in detail in \cite{Klioner2004}. 

(b) The effect of {\it parallax} is described by the transformation between the unit vectors $\ve{l}$ and $\ve{k}$; cf. Eq.~(81) in \cite{Klioner2003a}. 
The parallax is defined by $\pi = 1\,{\it A.U.}/|\ve{x}(t_0)|$, where $\ve{x}(t_0)$ is the BCRS coordinate of the celestial light source
at the moment of emission and ${\it A.U.}$ is the astronomical unit. Fixing the parallax requires a set of several observations at different times made by a moving observer 
orbiting the barycenter of the solar system. The knowledge of parallax allows one to calculate the spatial distance of the celestial light source. 

Finally, the spatial position $\ve{x}_0$ of the light source in the BCRS is uniquely given by the knowledge of three-vector $\ve{l}$, which points in the direction of the celestial 
light source as seen from the barycenter of the solar system, and the parallax $\pi$, which yields the spatial distance between the barycenter of solar system and the light source. 
It is clear that the model GREM can also be applied in possible future space astrometry missions like Theia \cite{Theia} or GaiaNIR \cite{Gaia_NIR}. 

As it has just been pointed out, the relativistic model GREM describes the sequence of the transformations 
$\ve{s} \rightarrow \ve{n} \rightarrow \ve{\sigma} \rightarrow \ve{k} \rightarrow \ve{l}$, where the first transformation $\ve{s} \rightarrow \ve{n}$ accounts for the effect of 
aberration and the last transformation $\ve{k} \rightarrow \ve{l}$ accounts for the effect of parallax. The effect of light deflection in the global coordinate system is described 
by the remaining three transformations $\ve{n} \rightarrow \ve{\sigma}$, $\ve{\sigma} \rightarrow \ve{k}$, $\ve{n} \rightarrow \ve{k}$; see also text above Eq.~(16) 
in \cite{Klioner_Zschocke}. These transformations are independent from the final implementation of a moving observer into the relativistic model, which would allow for 
determining the spatial position, $\ve{x}_0$, of the celestial light source. They represent a further extension of the Gaia relativistic model \cite{Klioner2003a,Klioner2004} 
and could also be implemented in relativistic models of data reduction of future space astrometry missions like GaiaNIR \cite{Gaia_NIR} or Theia \cite{Theia}, aiming at astrometry 
on the sub-\muas{} scale of accuracy. The implementation of these transformations into GREM for possible future Gaia-like astrometry missions would proceed in the same way, as it has 
been described in this section, i.e. shifting the origin of spatial coordinates from the barycenter of the body toward the barycenter of the solar system (cf. caption of Fig.~2). 
In the subsequent sections, these transformations will be derived from the boundary value solution in Eqs.~(\ref{First_Integration_2PN_boundary}) 
and (\ref{Second_Integration_2PN_boundary}).

\section{Transformation $\ve{k}$ to $\ve{\sigma}$}\label{Transformation1}

The first basic transformation of the boundary value problem is from $\ve{k}$ to $\ve{\sigma}$, where the unit-vector 
\begin{eqnarray}
\ve{k} &=& \frac{\ve{x}_1 - \ve{x}_0}{ \left|\ve{x}_1 - \ve{x}_0\right|} 
\label{vector_k}
\end{eqnarray}

\noindent
is pointing from light source toward observer, while $\ve{\sigma}$ has been defined by Eq.~(\ref{Boundary_Condition}) as unit tangent vector of the light trajectory 
at past infinity. The introduction of this unit-vector, $\ve{k} \cdot \ve{k} = 1$, implicates a new impact vector (cf. Eq.~(56) in \cite{Klioner_Zschocke}), 
\begin{eqnarray}
	\ve{d}_k &=& \ve{k} \times \left(\ve{x}_0 \times \ve{k}\right) = \ve{k} \times \left(\ve{x}_1 \times \ve{k}\right), 
\label{impact_vector_k}
\end{eqnarray}

\noindent
which points from the origin of the spatial coordinates (center-of-mass of the body) toward the closest point of the coordinate line between $\ve{x}_0$ and $\ve{x}_1$;  
for a graphical elucidation see Fig.~\ref{Diagram}. The absolute value of this impact vector is the impact parameter: $d_k = \left|\ve{d}_k\right|$. The final 
solution of the boundary value problem has to be given in terms of three-vector $\ve{k}$ in (\ref{vector_k}), thence contains this impact vector 
in Eq.~(\ref{impact_vector_k}) rather than the impact vectors in Eqs.~(\ref{impact_vector_0}) and (\ref{impact_vector_1}). 
By using $R = \ve{k} \cdot (\ve{x}_1 - \ve{x}_0) = |\ve{x}_1 - \ve{x}_0|$ and inserting 
\begin{eqnarray}
	\ve{x}_1 &=& \ve{x}_{\rm 2PN}\left(t_1\right) + {\cal O}(c^{-6})
	\label{Replacement_x1}
\end{eqnarray}

\noindent 
into the left-hand side of Eq.~(\ref{Second_Integration_2PN_boundary}), one obtains by iteration  
the following transformation $\ve{k}$ to $\ve{\sigma}$ in 2PN approximation,  
\begin{eqnarray}
       \ve{\sigma} = \ve{k} && - \frac{1}{R}\left[\ve{k}\times \bigg(\Delta\ve{x}_{\rm 1PN}\left(\ve{x}_1,\ve{x}_0\right)\times\ve{k}\bigg)\right]
       \nonumber\\ 
       && - \frac{1}{R}\left[\ve{k}\times \bigg(\Laplace\ve{x}_{\rm 2PN}\left(\ve{x}_1,\ve{x}_0\right)\times\ve{k}\bigg)\right]
       \nonumber\\
       && + \frac{1}{R^2}\,\bigg[\Delta\ve{x}_{\rm 1PN}\left(\ve{x}_1,\ve{x}_0\right) \times
       \bigg(\ve{k}\times\Delta\ve{x}_{\rm 1PN}\left(\ve{x}_1,\ve{x}_0\right)\bigg)\bigg]
       \nonumber\\
       && - \frac{3}{2}\,\frac{1}{R^2}\,\ve{k}\,\bigg| \ve{k} \times \Delta\ve{x}_{\rm 1PN}\left(\ve{x}_1,\ve{x}_0\right)\bigg|^2\,, 
\label{k_sigma_2PN}
\end{eqnarray}

\noindent 
where $\Delta\ve{x}_{\rm 1PN}\left(\ve{x}_1,\ve{x}_0\right) = \Delta\ve{x}_{\rm 1PN}\left(\ve{x}_1\right) - \Delta\ve{x}_{\rm 1PN}\left(\ve{x}_0\right)$ 
and $\Laplace\ve{x}_{\rm 2PN}\left(\ve{x}_1,\ve{x}_0\right) = \Laplace\ve{x}_{\rm 2PN}\left(\ve{x}_1\right) - \Laplace\ve{x}_{\rm 2PN}\left(\ve{x}_0\right)$, 
with the individual terms as given by Eqs.~(\ref{Second_Integration_1PN_Final}) and (\ref{Second_Integration_2PN_Final}), respectively. 
The criterion to stop the iterative process is the 2PN approximation, that means
all terms are neglected that have powers of $c^{-n}$ with $n \ge 6$. It is also noticed here that (\ref{k_sigma_2PN}) implies $\ve{\sigma} \cdot \ve{k} = 1 + {\cal O}(c^{-4})$.

This expression in Eq.~(\ref{k_sigma_2PN}) has first been derived for the case of light propagation in the gravitational field of a monopole at rest by Eq.~(3.2.50) in \cite{Brumberg1991} 
and later been rederived by Eq.~(68) in \cite{Klioner_Zschocke}. The transformation in Eq.~(\ref{k_sigma_2PN}) agrees with Eq.~(76) in \cite{Zschocke_Time_Delay_2PN} and 
generalizes the corresponding relations in \cite{Brumberg1991,Klioner_Zschocke}, because Eq.~(\ref{k_sigma_2PN}) is valid for light propagation 
in the gravitational field of a body with monopole and quadrupole structure.  

The light ray perturbations $\Delta\ve{x}_{\rm 1PN}$ and $\Laplace\ve{x}_{\rm 2PN}$ in Eq.~(\ref{k_sigma_2PN}) are still implicit, because they are given in terms of 
three-vector $\ve{\sigma}$ instead of three-vector $\ve{k}$. In the term $\Laplace\ve{x}_{\rm 2PN}$ one may replace vector $\ve{\sigma}$ by vector $\ve{k}$, because
this term is of order ${\cal O}\left(c^{-4}\right)$, hence such a replacement would cause an error beyond the 2PN approximation. Similarly, one may also replace vector $\ve{\sigma}$ 
by vector $\ve{k}$ in $\Delta\ve{x}_{\rm 1PN}$ for the last two terms in (\ref{k_sigma_2PN}), because these terms are also of the order ${\cal O}\left(c^{-4}\right)$, 
hence such a replacement would cause an error beyond the 2PN approximation. However, such a replacement is not possible for $\Delta\ve{x}_{\rm 1PN}$ in the second term 
on the right-hand side of Eq.~(\ref{k_sigma_2PN}), because this term is of the order ${\cal O}\left(c^{-2}\right)$ and, therefore, such a replacement would cause an error 
of the order ${\cal O}\left(c^{-4}\right)$. This problem has also been mentioned in the text below Eq.~(77) in our work \cite{Zschocke_Time_Delay_2PN}, and has partially been 
solved in that investigation, but only for the scalar product $\ve{k} \cdot \Delta\ve{x}_{\rm 1PN}$. A complete transformation $\ve{k}$ to $\ve{\sigma}$ requires to rewrite expression
$\Delta\ve{x}_{\rm 1PN}$ fully in terms of vector $\ve{k}$ by means of relation (\ref{k_sigma_2PN}) for the quadrupole. 
By using the relations in Appendix~\ref{Appendix1} one finally obtains the following transformation from $\ve{k}$ to $\ve{\sigma}$ in 
2PN approximation:
\begin{eqnarray}
	\sigma^i &=& k^i + \frac{G M}{c^2} \sum\limits_{n=1}^{2} \widehat{A}_{\left(n\right)}\,X^{i}_{\left(n\right)} 
	 + \frac{G M_{ab}}{c^2} \sum\limits_{n=1}^{8} \widehat{B}_{\left(n\right)}\,Y^{ab\,i}_{\left(n\right)} 
	\nonumber\\ 
	&& \hspace{-0.5cm} + \frac{G M}{c^2}\,\frac{G M}{c^2} \sum\limits_{n=1}^{2} \widehat{C}_{\left(n\right)}\,X^{i}_{\left(n\right)} 
	 + \frac{G M}{c^2}\,\frac{G M_{ab}}{c^2} \sum\limits_{n=1}^{8} \widehat{D}_{\left(n\right)}\,Y^{ab\,i}_{\left(n\right)}
	 \nonumber\\
	&& \hspace{-0.5cm}  + \frac{G M_{ab}}{c^2}\,\frac{G M_{cd}}{c^2} \sum\limits_{n=1}^{28} \widehat{E}_{\left(n\right)}\,Z^{abcd\,i}_{\left(n\right)}\,.
	\label{transformation_k_to_sigma}
\end{eqnarray}

\noindent
The tensorial coefficients are given in Appendix~\ref{Appendix7} and the scalar functions are presented in Appendix~\ref{Appendix8}.  
The monopole and quadrupole 1PN terms are given by the last two terms of the first line in (\ref{transformation_k_to_sigma}). These 1PN terms are in agreement with
the results in \cite{Klioner1991,Klioner2003a} (taking GR value $\gamma = 1$ in that references). 
More explicitly, these 1PN monopole and quadrupole terms in (\ref{transformation_k_to_sigma}) 
are in agreement with Eq.~(66) in \cite{Klioner2003a}, if one replaces three-vector $\ve{\sigma}$ by three-vector $\ve{k}$ in Eqs.~(22) 
and (36) in \cite{Klioner2003a}, keeping in mind that such a replacement of $\ve{\sigma}$ by $\ve{k}$ is in line with the 1PN approximation. 
The monopole-monopole terms in 2PN approximation are given by the first term of the second line in (\ref{transformation_k_to_sigma}). These 2PN terms agree with
Eq.~(74) in \cite{Klioner_Zschocke} if one takes the GR values $\alpha = \beta = \gamma = \epsilon = 1$. 
The 2PN monopole-quadrupole and quadrupole-quadrupole terms are given by the last two terms of the second line in (\ref{transformation_k_to_sigma}), 
which are new results of this investigation.

\section{Transformation $\ve{\sigma}$ to $\ve{n}$}\label{Transformation2} 

The unit tangent vector of light trajectory at the moment of observation is one of the most relevant quantities of the relativistic model of light propagation, 
because it is directly related to observables, when it is transformed into the local coordinate system of the space-based observer. 
That unit tangent vector, $\ve{n} \cdot \ve{n} = 1$, at observers position is defined by  
\begin{eqnarray}
        \ve{n} &=& \frac{\dot{\ve{x}}\left(t_1\right)}{\left|\dot{\ve{x}}\left(t_1\right)\right|}\,.
	\label{vector_n}
\end{eqnarray}

\noindent 
The transformation from $\ve{\sigma}$ to $\ve{n}$ is the second basic transformation of the boundary value problem. 
By iteration of Eq.~(\ref{First_Integration_2PN_boundary}) one obtains the transformation $\ve{\sigma}$ to $\ve{n}$ in 2PN approximation,
\begin{eqnarray}
	\ve{n} = \ve{\sigma} 
	&& + \ve{\sigma} \times \bigg(\frac{\Delta\dot{\ve{x}}_{\rm 1PN}\left(t_1\right)}{c} \times \ve{\sigma} \bigg)
	\nonumber\\ 
	&& + \ve{\sigma} \times \bigg(\frac{\Laplace\dot{\ve{x}}_{\rm 2PN}\left(t_1\right)}{c} \times \ve{\sigma} \bigg) 
	\nonumber\\ 
	&& - \frac{\Delta\dot{\ve{x}}_{\rm 1PN}\left(t_1\right)}{c} \left( \frac{\ve{\sigma} \cdot \Delta\dot{\ve{x}}_{\rm 1PN}\left(t_1\right)}{c} \right) 
	\nonumber\\ 
	&& - \frac{1}{2} \left(\frac{\Delta\dot{\ve{x}}_{\rm 1PN}\left(t_1\right)}{c} \cdot \frac{\Delta\dot{\ve{x}}_{\rm 1PN}\left(t_1\right)}{c}\right) \ve{\sigma}  
	\nonumber\\ 
	&& + \frac{3}{2} \left(\frac{\ve{\sigma} \cdot \Delta\dot{\ve{x}}_{\rm 1PN}\left(t_1\right)}{c}\right)^2 \ve{\sigma}\,, 
        \label{transformation_sigma_n}
\end{eqnarray}

\noindent 
where the individual terms $\Delta\dot{\ve{x}}_{\rm 1PN}$ and $\Delta\dot{\ve{x}}_{\rm 2PN}$ are given by Eqs.~(\ref{First_Integration_1PN_Final}) and (\ref{First_Integration_2PN_Final}), 
respectively. The criterion to stop the iterative process is the 2PN approximation, that means all terms are neglected that have powers of $c^{-n}$ with $n \ge 6$.   

The transformation $\ve{\sigma}$ to $\ve{n}$ in 1PN approximation, that means the second term on the right-hand side, was determined long time ago in \cite{Klioner1991,Klioner2003a}, 
and a simplified 1PN transformation has been obtained in \cite{Zschocke_Klioner}. Furthermore, in case of a monopole at rest, the transformation $\ve{\sigma}$ to $\ve{n}$ has been 
determined in 2PN approximation in \cite{Klioner_Zschocke}. The transformation (\ref{transformation_sigma_n}) in 2PN approximation, both for monopole and quadrupole at rest, has recently 
been considered in our investigation \cite{Zschocke_Quadrupole_1}. This transformation was, however, not given in its complete form; cf. Eqs.~(126) - (135) in \cite{Zschocke_Quadrupole_1}. 
In particular, the perturbations $\Delta\dot{\ve{x}}_{\rm 1PN}$ and $\Laplace\dot{\ve{x}}_{\rm 2PN}$ in Eq.~(\ref{transformation_sigma_n}) are still implicit, because they are given in 
terms of three-vector $\ve{\sigma}$ instead of three-vector $\ve{k}$. For the completion of the boundary value problem one needs to express these relativistic terms 
in \cite{Zschocke_Quadrupole_1} as function of three-vector $\ve{k}$ rather than $\ve{\sigma}$; cf. text above Eq.~(81) in \cite{Klioner_Zschocke}. This procedure is described 
in the following. 

In the term $\Laplace\dot{\ve{x}}_{\rm 2PN}$ one may replace vector $\ve{\sigma}$ by vector $\ve{k}$, because
this term is of order ${\cal O}\left(c^{-4}\right)$, hence such a replacement would cause an error beyond the 2PN approximation. Similarly, one may also replace vector $\ve{\sigma}$
by vector $\ve{k}$ in $\Delta\dot{\ve{x}}_{\rm 1PN}$ for the last three terms in (\ref{transformation_sigma_n}), because these terms are also of the order ${\cal O}\left(c^{-4}\right)$,
hence such a replacement would cause an error beyond the 2PN approximation. However, such a replacement is not possible for $\Delta\dot{\ve{x}}_{\rm 1PN}$ in the second term on the 
right-hand side of Eq.~(\ref{transformation_sigma_n}), because this term is of the order ${\cal O}\left(c^{-2}\right)$ and, therefore, such a replacement would cause an error of the 
order ${\cal O}\left(c^{-4}\right)$. A complete transformation $\ve{\sigma}$ to $\ve{n}$ in 2PN approximation requires to rewrite expression
$\Delta\dot{\ve{x}}_{\rm 1PN}$ fully in terms of vector $\ve{k}$ by means of relation (\ref{k_sigma_2PN}) for the quadrupole. 
For this calculation one needs the relations as given in Appendix~\ref{Appendix1}. 
The final result is the transformation $\ve{\sigma}$ to $\ve{n}$ in (\ref{transformation_sigma_n}) fully in terms of three-vector $\ve{k}$ 
in 2PN approximation: 
\begin{eqnarray}
        n^i &=& \sigma^i + \frac{G M}{c^2} \sum\limits_{n=1}^{2} \widehat{H}_{\left(n\right)}\,X^{i}_{\left(n\right)} 
         + \frac{G M_{ab}}{c^2} \sum\limits_{n=1}^{8} \widehat{I}_{\left(n\right)}\,Y^{ab\,i}_{\left(n\right)} 
        \nonumber\\ 
        && \hspace{-0.5cm}  + \frac{G M}{c^2}\,\frac{G M}{c^2} \sum\limits_{n=1}^{2} \widehat{J}_{\left(n\right)}\,X^{i}_{\left(n\right)} 
         + \frac{G M}{c^2}\,\frac{G M_{ab}}{c^2} \sum\limits_{n=1}^{8} \widehat{K}_{\left(n\right)}\,Y^{ab\,i}_{\left(n\right)}
	 \nonumber\\
	 && \hspace{-0.5cm} + \frac{G M_{ab}}{c^2}\,\frac{G M_{cd}}{c^2} \sum\limits_{n=1}^{28} \widehat{L}_{\left(n\right)}\,Z^{abcd\,i}_{\left(n\right)}\,.
        \label{transformation_sigma_to_n}
\end{eqnarray}

\noindent
The tensorial coefficients are given in Appendix~\ref{Appendix7}. The scalar functions are presented in Appendix~\ref{Appendix9}. The monopole and quadrupole terms 
in 1PN are given by the last two terms of the first line in (\ref{transformation_sigma_to_n}). These 1PN terms are in agreement with the results in \cite{Klioner2003a,Zschocke_Klioner} 
if one takes the GR value of the parametrized post-Newtonian (PPN) parameter in these references: $\gamma = 1$. 
In order to confirm this agreement one has to realize that in 1PN approximation the  
three-vector $\ve{\sigma}$ can be replaced by three-vector $\ve{k}$ in Eqs.~(61) and (62) in \cite{Klioner2003a} or in Eqs.~(4) and (7) in \cite{Zschocke_Klioner}. 
The monopole-monopole terms in 2PN approximation are given by the first term of the second line in (\ref{transformation_sigma_to_n}). 
These 2PN terms agree with Eq.~(81) in \cite{Klioner_Zschocke}, if one takes the GR values of the PPN parameters: $\alpha = \beta = \gamma = \epsilon = 1$. 
The 2PN monopole-quadrupole and quadrupole-quadrupole terms 
are given by the last two terms of the second line in (\ref{transformation_sigma_to_n}), which are new results of this investigation. 

The transformation $\ve{\sigma}$ to $\ve{n}$ in Eq.~(\ref{transformation_sigma_to_n}) has already been discussed in our previous work \cite{Zschocke_Quadrupole_1}. 
However, that transformation, given by Eq.~(126) in \cite{Zschocke_Quadrupole_1}, was still incomplete in two respects: (a) only the largest terms of this 
transformation were presented, and (b) the individual expressions of this transformation were still given in terms of vector $\ve{\sigma}$ instead of 
vector $\ve{k}$. The transformation presented here is complete in the sense, that Eq.~(\ref{transformation_sigma_to_n}) contains all 2PN terms 
and the individual expressions are given in terms of vector $\ve{k}$. 

The reason of presenting an incomplete transformation in \cite{Zschocke_Quadrupole_1} was, to demonstrate the occurrence of so-called {\it enhanced terms} in the 2PN quadrupole 
light deflection. The occurrence of {\it enhanced terms} in 2PN calculations of light propagation in the monopole field has been discovered independently by several working groups 
\cite{Klioner_Zschocke,Teyssandier,AshbyBertotti2010}, while the occurrence of {\it enhanced terms} in the quadrupole field has been recovered in our work \cite{Zschocke_Quadrupole_1}. 
The issue of {\it enhanced terms} has been discussed for the transformation $\ve{\sigma}$ to $\ve{n}$ in some detail in Section VIII in our work \cite{Zschocke_Quadrupole_1}. 
More explicitly, the upper limits of {\it enhanced terms} contain a large factor $x_1/d^{\,1}_{\sigma}$ 
(cf. Eqs.~(144) - (146) in \cite{Zschocke_Quadrupole_1}) and, therefore, they are much larger than the usual 2PN terms, which do not contain 
such a large factor (cf. Eqs.~(148) - (150) in \cite{Zschocke_Quadrupole_1}). It should be noticed, that this large factor does not imply that 
these {\it enhanced terms} are divergent, because the spatial positions $x_1$ of an observer are restricted to the case of realistic configurations, 
where the observer is located inside the solar system \cite{Zschocke_Quadrupole_1,Zschocke_Quadrupole_2}.

\section{Transformation $\ve{k}$ to $\ve{n}$}\label{Transformation3}

The third basic transformation of the boundary value problem is the transformation $\ve{k}$ to $\ve{n}$, which is obtained by combining
relations (\ref{transformation_k_to_sigma}) and (\ref{transformation_sigma_to_n}): 
\begin{eqnarray}
        n^i &=& k^i + \frac{G M}{c^2} \sum\limits_{n=1}^{2} \widehat{M}_{\left(n\right)}\,X^{i}_{\left(n\right)}
         + \frac{G M_{ab}}{c^2} \sum\limits_{n=1}^{8} \widehat{N}_{\left(n\right)}\,Y^{ab\,i}_{\left(n\right)}
        \nonumber\\
        && \hspace{-0.5cm} + \frac{G M}{c^2}\,\frac{G M}{c^2} \sum\limits_{n=1}^{2} \widehat{U}_{\left(n\right)}\,X^{i}_{\left(n\right)}
         + \frac{G M}{c^2}\,\frac{G M_{ab}}{c^2} \sum\limits_{n=1}^{8} \widehat{V}_{\left(n\right)}\,Y^{ab\,i}_{\left(n\right)}
	 \nonumber\\ 
	&& \hspace{-0.5cm}  + \frac{G M_{ab}}{c^2}\,\frac{G M_{cd}}{c^2} \sum\limits_{n=1}^{28} \widehat{W}_{\left(n\right)}\,Z^{abcd\,i}_{\left(n\right)}\,.
        \label{transformation_k_to_n}
\end{eqnarray}

\noindent 
The tensorial coefficients are given in Appendix~\ref{Appendix7}. The scalar functions are presented in Appendix~\ref{Appendix10}. The monopole and quadrupole terms in 1PN approximation 
are given by the last two terms of the first line in (\ref{transformation_sigma_to_n}). These 1PN terms are in agreement with the results in \cite{Klioner2003a,Zschocke_Klioner}, if one 
takes the GR value of the PPN parameter in these references: $\gamma = 1$. The 2PN monopole-monopole terms are given by the first term of the second line 
in (\ref{transformation_sigma_to_n}) and they are in agreement with the 2PN monopole-monopole terms as given by Eq.~(87) in \cite{Klioner_Zschocke}, if one takes the GR values of the 
PPN parameters in that reference: $\alpha = \beta = \gamma = \epsilon = 1$. The 2PN 
monopole-quadrupole and quadrupole-quadrupole terms are given by the second and third terms of the second line in (\ref{transformation_sigma_to_n}). These expressions are some of the 
primary new results of this investigation. The transformation (\ref{transformation_k_to_n}) is of fundamental importance for the effects of time delay and light deflection: the terms 
parallel to $\ve{k}$ yield the time delay and the terms perpendicular to $\ve{k}$ yield the light deflection. While the effect of time delay has already been investigated  
in our work \cite{Zschocke_Time_Delay_2PN}, the effect of light deflection will numerically be considered in the next section.

\section{Numerical results for massive solar system bodies}\label{Section3} 

In order to determine the magnitude of light deflection one needs to concretize the quadrupole tensor of the solar system body. To perform these calculations, 
the assumption is adopted that the planets of the solar system are described as axisymmetric ellipsoids, that means they are having 
the following geometrical shape 
\begin{eqnarray}
	\frac{\left(x^1\right)^2}{A^2} + \frac{\left(x^2\right)^2}{B^2} + \frac{\left(x^3\right)^2}{C^2} &=& 1\,,
        \label{Axi_Symmetry}
\end{eqnarray}

\noindent
where for the principal semi-axes of the body we have $A = B \neq C$. The densities of these axisymmetric bodies can still be radial dependent. 
Furthermore, it is assumed that the unit-vector ${\ve e}_3$ is perpendicular to the $A$-$B$-plane of the semi-axes and this unit-vector is parallel 
to the $x^3$-axis of the global harmonic coordinate system. That chosen specific orientation is not a loss of generality; see also the comments on  
Section IV in \cite{Zschocke_Total_Light_Deflection}. Then, the STF quadrupole moment in (\ref{Quadrupole}) and the product of two quadrupole moments 
take the following form \cite{Klioner2003a,Zschocke_Klioner,Zschocke_Time_Delay_2PN} (e.g. Eqs.~(B35) and (B36) in \cite{Zschocke_Time_Delay_2PN}),  
\begin{eqnarray} 
	M_{ab} &=& - M\,J_2 \left(P\right)^2\;\underset{ab}{\rm STF}\;\delta^3_a\,\delta^3_b\,,
        \label{Quadrupole_Tensor_1} 
	\\
	M_{ab}\,M_{cd} &=& + \left(M\right)^2 \left(J_2\right)^2 \left(P\right)^4\;\underset{ab}{\rm STF}\;\delta^3_a\,\delta^3_b\;\;\underset{cd}{\rm STF}\;\delta^3_c\,\delta^3_d\,,
	\nonumber\\ 
        \label{Quadrupole_Tensor_2} 
\end{eqnarray}

\noindent
where $M$ is the mass of the body, $P$ is the equatorial radius of the body, $J_2$ is the actual second zonal harmonic coefficient, and 
the symmetric and tracefree part is given by: $\underset{ab}{\rm STF}\;\delta^3_a\,\delta^3_b = \delta^3_a \delta^3_b - 1/3\,\delta_{ab}$. 
These physical parameters are given in Table~\ref{Table1} for the giant planets of the Solar System. 
\begin{table}[h!]
\caption{\label{Table1} The numerical parameters in Eq.~(\ref{Quadrupole_Tensor_1}): Schwarzschild radius $m = GM/c^2$ (in meter), equatorial radius $P$ (in meter), 
	and the actual second zonal harmonic coefficient $J_2$ of giant planets of the Solar System \cite{JPL}, which is a dimensionless quantity.} 
\footnotesize
\begin{tabular}{@{}|c|c|c|c|}
\hline
	Object & $m$ & $P\,\left[10^6\right]$ & $J_2\,\left[10^{-3}\right]$ \\
&&&\\[-10pt]
\hline
Jupiter & $1.410$ & $ 71.49$ & $14.697$ \\
Saturn & $ 0.422 $ & $ 60.27$ & $16.331$ \\
Uranus & $ 0.064 $ & $ 25.56$ & $3.516$ \\
Neptune & $ 0.076 $ & $ 24.76$ & $3.538$ \\
\hline
\end{tabular}\\
\end{table}
\normalsize

\noindent 
By inserting these expressions (\ref{Quadrupole_Tensor_1}) and (\ref{Quadrupole_Tensor_2}) into Eq.~(\ref{transformation_k_to_n}), one obtains the transformation 
from $\ve{k}$ to $\ve{n}$ for the case of axisymmetric solar system bodies, 
\begin{eqnarray}
	\hspace{-0.25cm} \ve{n} = \ve{k} + \ve{n}^{\rm M}_{\rm 1PN} + \ve{n}^{\rm Q}_{\rm 1PN} 
	+ \ve{n}^{{\rm M} \times {\rm M}}_{\rm 2PN} + \ve{n}^{{\rm M} \times {\rm Q}}_{\rm 2PN} + \ve{n}^{{\rm Q} \times {\rm Q}}_{\rm 2PN},  
        \label{transformation_k_to_n_axisymmetric}
\end{eqnarray}

\noindent
where the individual terms in (\ref{transformation_k_to_n_axisymmetric}) are 
\begin{eqnarray}
	\ve{n}_{\rm 1PN}^{\rm M} &=& \frac{G M}{c^2} \sum\limits_{n=1}^{2} \widehat{M}_{\left(n\right)}\,\ve{X}_{\left(n\right)}\,,
	\label{1PN_M_term}
	\\
	\ve{n}_{\rm 1PN}^{\rm Q} &=& \frac{G M}{c^2}\,J_2 \left(P\right)^2 \sum\limits_{n=1}^{8} \widehat{N}_{\left(n\right)}\,\ve{Y}_{\left(n\right)}\,,
	\label{1PN_Q_term}
        \\
	\ve{n}^{{\rm M} \times {\rm M}}_{\rm 2PN} &=& \frac{G^2 M^2}{c^4} \sum\limits_{n=1}^{2} \widehat{U}_{\left(n\right)}\,\ve{X}_{\left(n\right)}\,,
	\label{2PN_MM_term}
        \\
        \ve{n}^{{\rm M} \times {\rm Q}}_{\rm 2PN} &=& \frac{G^2 M^2}{c^4}\,J_2 \left(P\right)^2 \sum\limits_{n=1}^{8} \widehat{V}_{\left(n\right)}\,\ve{Y}_{\left(n\right)}\,,
	\label{2PN_MQ_term}
	\\
	\ve{n}^{{\rm Q} \times {\rm Q}}_{\rm 2PN} &=& \frac{G^2 M^2}{c^4} \left(J_2\right)^2 \left(P\right)^4 \sum\limits_{n=1}^{28} \widehat{W}_{\left(n\right)}\,\ve{Z}_{\left(n\right)}\,. 
	\label{2PN_QQ_term}
\end{eqnarray}

\noindent
The spatial components of the three vectors in Eqs.~(\ref{1PN_M_term}) and (\ref{2PN_MM_term}) were given by Eqs.~(\ref{coefficient_X1_k}) and (\ref{coefficient_X2_k}), and 
the spatial components of the three vectors in Eqs.~(\ref{1PN_Q_term}), (\ref{2PN_MQ_term}) and (\ref{2PN_QQ_term}) are defined by 
\begin{eqnarray}
	Y^i_{\left(n\right)} &=& - \underset{ab}{\rm STF}\,\delta^3_a\,\delta^3_b\;\; Y^{ab\,i}_{\left(n\right)}\,,
	\label{vector_Y_n}
	\\
	Z^i_{\left(n\right)} &=& + \underset{ab}{\rm STF}\,\delta^3_a\,\delta^3_b\;\; \underset{cd}{\rm STF}\,\delta^3_c\,\delta^3_d \;\; Z^{abcd\,i}_{\left(n\right)}\,,
	\label{vector_Z_n}
\end{eqnarray}

\noindent
with the tensorial coefficients given by Eqs.~(\ref{coefficient_Y1_k}) - (\ref{coefficient_Y8_k}) and Eqs.~(\ref{coefficient_Z1_k}) - (\ref{coefficient_Z28_k}). 
These vectorial coefficients in (\ref{vector_Y_n}) and (\ref{vector_Z_n}) are linear combinations of only three vectors: $\ve{k}$, $\ve{d}_k$, $\ve{e}_3$.
The three-vector $\ve{k}$ points from the source toward the observer and has been defined by
Eq.~(\ref{vector_k}), the impact vector $\ve{d}_k$ points from the center of mass of the body toward the straight
line between source and observer and was defined by Eq.~(\ref{impact_vector_k}), and the three-vector $\ve{e}_3$
was introduced in the text below Eq.~(\ref{Axi_Symmetry}). It is noticed here, that vector $\ve{k}$
and impact vector $\ve{d}_k$ are perpendicular to each other, but in general the three-vector $\ve{e}_3$ can point
toward any spatial direction relative to the three-vectors $\ve{k}$ and $\ve{d}_k$. 

The light deflection in the global harmonic coordinate system is defined by the angle $\varphi$ between the three-vectors $\ve{k}$ and $\ve{n}$, which means 
\begin{eqnarray}
	\varphi &=& \arcsin \left| \ve{k} \times \ve{n}\right| \,,
	\label{light_deflection}
\end{eqnarray}

\noindent
where the unit vector $\ve{n}$ in 2PN approximation is given by the expression in Eq.~(\ref{transformation_k_to_n}). For the numerical evaluation, however, we will implement 
the unit vector $\ve{n}$ as given by Eq.~(\ref{transformation_k_to_n_axisymmetric}) for the case of an axisymmetric solar system body. 
Clearly, terms proportional to $\ve{k}$ in (\ref{transformation_k_to_n_axisymmetric}) do not contribute to the light deflection angle in (\ref{light_deflection}). 
The upper limits of the 1PN terms are given by 
\begin{eqnarray}
	\varphi_{\rm 1PN}^{{\rm M}} &=& \left| \ve{k} \times \ve{n}^{{\rm M}}_{\rm 1PN}\right| \le 4\,\frac{m}{d_{k}}\,,
\label{upper_limit_M}
\\
        \varphi_{\rm 1PN}^{{\rm Q}} &=& \left| \ve{k} \times \ve{n}^{Q}_{\rm 1PN}\right| \le
	4\,\frac{m}{d_k}\,\left|J_2\right| \left(\frac{P}{d_k}\right)^2 ,
\label{upper_limit_Q}
\end{eqnarray}

\noindent
where $m = G M /c^2$ is the Schwarzschild radius of the solar system body. 
Equation~(\ref{upper_limit_M}) represents the famous result of general theory of gravity \cite{Einstein2}. 
These limits in (\ref{upper_limit_M}) and (\ref{upper_limit_Q}) are in agreement with Eqs.~(139) and (140) in \cite{Zschocke_Quadrupole_1}. 

In the past, some remarkable work has been done regarding the 1PN quadrupole light deflection. In view of this longer history, we will be focused 
here on very few articles, where this effect has been investigated, mainly with respect to its detectability by the Gaia mission or future space astrometry missions. 
The 1PN light trajectory as function of coordinate time in the quadrupole field of a body at rest has first been determined in \cite{Klioner1991} 
and later again in \cite{KlionerKopeikin1992,Klioner2003a}, where numerical values for light deflection of grazing rays at the Sun and planets of the Solar System were presented. 
A rigorous solution for 1PN (mass-multipoles) and 1.5PN (spin-multipoles) light trajectories in the field of bodies at rest with full multipole structure has been worked out 
in \cite{Kopeikin1997}, explicitly mentioning an agreement with the previous results regarding the 1PN quadrupole in \cite{Klioner1991}. New insights about the detectability 
of the quadrupole light deflection 
at Jupiter within the Gaia mission have been investigated in \cite{Quadrupole1}. The impact of the first few higher multipoles of solar system bodies on light deflection has carefully 
been investigated in \cite{Poncin_Lafitte_Teyssandier}. Later it has been discovered that the total light deflection is given in terms of Chebyshev polynomials, 
which allows for strict upper limits of light deflection to any order of the multipoles \cite{Zschocke_Total_Light_Deflection}.  
The Erez-Rosen metric, which is an exact solution of the field equations, and the role of the quadrupole on light propagation has been investigated in \cite{Quadrupole2}. 
Finally, the recent investigation in \cite{Quadrupole3} is mentioned, which represents the first measurement of 
total light deflection at Jupiter by means of Gaia data, which lends strong reliability to the numerical values reported in Table~II there and 
provides also a significant observational context for the future. 
For a detailed representation of the impact of the 1PN quadrupole field on light deflection we refer to \cite{Klioner1991,Zschocke_Klioner}. 

The upper limits of the 2PN terms are given by 
\begin{eqnarray}
	&& \hspace{-0.55cm} \varphi_{\rm 2PN}^{{\rm M} \times {\rm M}} = \left| \ve{k} \times \ve{n}^{{\rm M} \times {\rm M}}_{\rm 2PN}\right| \le
	16 \left(\frac{m}{d_k}\right)^2 \frac{x_1}{d_k} ,
\label{upper_limit_MM}
\\
	&& \hspace{-0.5cm} \varphi_{\rm 2PN}^{{\rm M} \times {\rm Q}} = \left| \ve{k} \times \ve{n}^{{\rm M} \times {\rm Q}}_{\rm 2PN}\right| \le
	64 \left(\frac{m}{d_k}\right)^2 \left|J_2\right| \left(\frac{P}{d_k}\right)^2 \frac{x_1}{d_k} ,
\label{upper_limit_MQ}
\\
	&& \hspace{-0.5cm} \varphi_{\rm 2PN}^{{\rm Q} \times {\rm Q}} = \left| \ve{k} \times \ve{n}^{{\rm Q} \times {\rm Q}}_{\rm 2PN}\right| \le
	48 \left(\frac{m}{d_k}\right)^2 \left|J_2\right|^2 \left(\frac{P}{d_k}\right)^4 \frac{x_1}{d_k} . 
\label{upper_limit_QQ}
\end{eqnarray}

\noindent 
In the estimation of the upper limits in (\ref{upper_limit_MM}) - (\ref{upper_limit_QQ}) all those terms have been neglected, which do not contain the large 
factor $x_1/d_k$, which means terms of the order ${\cal O}\left(m^2/d_k^2\right)$ are negligible. 
What important here is, that the upper limits of 2PN terms in (\ref{upper_limit_MM}) - (\ref{upper_limit_QQ}) contain this factor $x_1/d_k$, where $x_1$ is the 
distance between the solar system body and the observer, while $d_k$ is the impact vector of the light ray. This factor becomes numerically large, $x_1/d_k \gg 1$. For instance, 
in case of crazing light rays, the impact vector $d_k$ equals the radius $P$ of the body and $x_1/P \gg 1$. Accordingly, those terms in Eq.~(\ref{transformation_k_to_n}) that contain 
such a large factor, are so-called {\it enhanced terms}. 
As mentioned in the introductory section, the occurrence of {\it enhanced terms} in 2PN calculations has been recovered by different working groups in case of light propagation 
in the gravitational field of a monopole at rest \cite{Klioner_Zschocke,AshbyBertotti2010,Teyssandier}. Later, in our work \cite{Zschocke_Quadrupole_1} it has been found that 
these {\it enhanced terms} do also occur in case of light propagation in the gravitational field of a quadrupole at rest. It is noticed, that these {\it enhanced terms} 
do not diverge, because the observer is assumed to be located in the solar system so that $x_1$ can not be arbitrarily large; see also text below Eq.~(\ref{relation_absolute_d1_dk}) 
in appendix~\ref{Appendix1}. For the numerical evaluation of these terms we have assumed a Gaia-like observer located at the Lagrange point $L_2$.  

In our investigation \cite{Klioner_Zschocke} the remarkable result has been found that in case of 2PN monopole light propagation, these {\it enhanced terms} are related to the choice 
of the coordinate dependent impact parameter; see text below Eq.~(111) {\it ibid}. From this fact it can be conjectured with certainty, that the appearance of these {\it enhanced terms} 
originates in the case of 2PN quadrupole light propagation also from the choice of coordinate dependent impact vectors. As emphasized in \cite{Klioner_Zschocke}, the use of coordinate 
dependent impact vectors, as they appear in the basic transformations given by Eqs.~(\ref{transformation_k_to_sigma}), (\ref{transformation_sigma_to_n}) and (\ref{transformation_k_to_n}), 
are much more convenient for practical astrometric calculations. 

Some further remarks about these upper limits in Eqs.~(\ref{upper_limit_M}) - (\ref{upper_limit_QQ}) should be included here. 
The determination of the upper limits in Eqs.~(\ref{upper_limit_M}) and (\ref{upper_limit_Q}) has been described in detail in \cite{Zschocke_Klioner}. The calculation of the upper 
limits of the 2PN terms in Eqs.~(\ref{upper_limit_MM}) - (\ref{upper_limit_QQ}) proceeds in a similar way as the approach described in \cite{Upper_Limits1,Upper_Limits2}. In particular, 
in order to find the upper limits, one has to identify those 2PN terms in the scalar functions $\widehat{U}_{(n)}$, $\widehat{V}_{(n)}$, $\widehat{W}_{(n)}$ of Eqs.~(\ref{2PN_MM_term}) - (\ref{2PN_QQ_term}), 
whose upper limits contain the large factor $x_1/d_k$. All the other individual terms in these functions have upper limits of the order ${\cal O}\left(m^2/d_k^2\right)$ and 
can be neglected. The upper limit in Eq.~(\ref{upper_limit_MM}) has already been presented in our investigation \cite{Klioner_Zschocke}, while the upper limits in 
Eqs.~(\ref{upper_limit_MQ}) and (\ref{upper_limit_QQ}) are new results of this investigation. The numerical values of the upper limits of the 1PN terms in 
Eqs.~(\ref{upper_limit_M}) - (\ref{upper_limit_Q}) and of the 2PN terms in Eqs.~(\ref{upper_limit_MM}) - (\ref{upper_limit_QQ}) are given in Table~\ref{Table2} and Table~\ref{Table3}, 
respectively. 
\begin{table}[h!]
        \caption{\label{Table2} 
	The upper limits of the of 1PN light deflection given by Eqs.~(\ref{upper_limit_M}) - (\ref{upper_limit_Q}) for grazing light rays
        (impact vector $d_k$ equals the equatorial radius of the planet $P$) at the giant planets of the Solar System. 
	All values for the light deflections are given in micro-arcseconds (\muas).}
\footnotesize
\begin{tabular}{@{}|c|c|c|}
\hline
&&\\[-7pt]
        Object & $\varphi_{\rm 1PN}^{{\rm M}}$ & $\varphi_{\rm 1PN}^{{\rm Q}}$ \\
&&\\[-7pt]
\hline  
        Jupiter & $16.3 \times 10^3$ & $239$ \\
        Saturn  & $5.8 \times 10^3$ &  $94$ \\
        Uranus  & $2.1 \times 10^3$ &  $6.9$ \\
        Neptune & $2.5 \times 10^3$ &  $8.6$ \\
\hline
\end{tabular}\\
\end{table}
\begin{table}[h!]
        \caption{\label{Table3}The upper limits of the {\it enhanced 2PN terms} of light deflection given by
        Eqs.~(\ref{upper_limit_MM}) - (\ref{upper_limit_QQ}) for grazing light rays (impact vector $d_k$ equals the equatorial radius of the planet $P$) 
        at the giant planets of the Solar System. For the distance between planet and observer, a Gaia-like mission operating at the Lagrange point $L_2$ 
	is assumed, where the maximal distances between the observer and the giant planets have been implemented in the upper limits: $x_1 = 6\,{\it A.U.}$, 
	$x_1 = 11\,{\it A.U.}$, $x_1 = 21\,{\it A.U.}$, and $x_1 = 31\,{\it A.U.}$ for Jupiter, Saturn, Uranus, and Neptune ({\it A.U.} is the astronomical unit). All 
	values for the light deflections are given in micro-arcseconds (\muas).}
\footnotesize
\begin{tabular}{@{}|c|c|c|c|}
\hline
&&&\\[-7pt]
        Object & $\varphi_{\rm 2PN}^{{\rm M} \times {\rm M}}$ & $\varphi_{\rm 2PN}^{{\rm M} \times {\rm Q}}$ & $\varphi_{\rm 2PN}^{{\rm Q} \times {\rm Q}}$ \\
&&&\\[-7pt]
\hline
        Jupiter & $16.11$ & $0.95$ & $0.010$ \\
        Saturn  & $4.42$ &  $0.29$ & $0.003$ \\
        Uranus  & $2.58$ &  $0.04$ & $0.001$ \\
        Neptune & $5.83$ &  $0.08$ & $0.002$ \\
\hline
\end{tabular}\\
\end{table}

\noindent 
The upper limits presented by Eqs.~(\ref{upper_limit_M}) - (\ref{upper_limit_Q}) for the 1PN terms and by Eqs.~(\ref{upper_limit_MM}) - (\ref{upper_limit_QQ}) 
for the 2PN terms, turn out to be in agreement with the upper limits of Eqs.~(106) - (107) for the 1PN terms and by Eqs.~(144) - (146) for the 2PN terms in our previous 
work \cite{Zschocke_Quadrupole_1}. Therefore, the numerical results presented by Table~\ref{Table2} for the 1PN terms and by Table~\ref{Table3} for the 2PN terms are 
in agreement with Table~II and Table~IV, respectively, in our previous work \cite{Zschocke_Quadrupole_1}. But one has to keep in mind, that these upper limits 
in \cite{Zschocke_Quadrupole_1} and the upper limits in this investigation belong to two different transformations: in \cite{Zschocke_Quadrupole_1} the upper limits 
of the transformation $\ve{\sigma}$ to $\ve{n}$ have been determined, while in this investigation the upper limits of the transformation $\ve{k}$ to $\ve{n}$ have been determined. 
The transformation $\ve{\sigma}$ to $\ve{n}$ allows one to determine the light deflection for very distant celestial objects like stars and quasars, 
while the transformation $\ve{k}$ to $\ve{n}$ allows one to determine the light deflection for very distant celestial objects like stars and quasars 
as well as for solar system objects, which is the primary focus of this investigation. 

A final comment should be in order here about the possibility to disentangle the quadrupole effect from the monopole effect. In Section~\ref{GREM} it has been described in some
detail how the unit vector $\ve{n}$ is related to the observed direction $\ve{s}$ of the celestial light source. The light deflection vector $\ve{n}$ is given by 
Eq.~(\ref{transformation_k_to_n}), which is an integral effect of both the monopole and quadrupole terms. For grazing rays the numerical values are presented in 
Table~\ref{Table2} for the 1PN terms and in Table~\ref{Table3} for the 2PN terms. For grazing rays at Jupiter, for instance, the 2PN quadrupole effect amounts to 
about $1\,\muas$. If such a scale of accuracy can be achieved then, in principle, it is possible to confirm that the quadrupole structure of the body is responsible 
for this additional effect to the total net effect of light deflection. However, in order to disentangle the quadrupole effect from the impact of the monopole term 
one has to measure the scaling behavior of the light deflection. That means, each individual 1PN multipole or 2PN multipole term has its own characteristic scaling slope 
with increasing angular distance between source and barycenter of the gravitating body. It is clear that such an entanglement of the individual terms will be a very difficult task.

\section{Summary and Outlook}\label{Summary}

The todays state-of-the-art in astrometry has reached the micro-arcsecond (\muas) level in the precision of angular measurements. 
In the near future, astrometry is aiming at astrometric precisions on the sub-\muas{} level. This aim is mainly triggered by completely new and  
inspiring science cases of astronomy, like tests of general relativity in the solar system, more accurate determination of natural constants, 
discovery of Earth-like exoplanets by means of astrometric positional measurements of their host star, detection of gravitational waves by 
pairwise angular observations of stars, measurements of dark matter distributions in our galaxy, and extension and improvements of 
model-independent cosmic ladder scales, in order to mention just a few of these issues. 

Such ultra-high accuracies on the sub-\muas{} level necessitate the determination of the curvilinear light trajectories from the celestial light sources 
through the curved space-time of the solar system toward the observer, not only in the first (1PN) but also in the second (2PN) order of the post-Newtonian 
scheme. In particular, as described in some detail in the introductory Section, it is necessary to determine light trajectories in 2PN approximation 
in the monopole as well as in the quadrupole field of solar system bodies. The solution of the initial value problem for light trajectories in 2PN approximation 
in the monopole and quadrupole field of a solar system body at rest has been obtained in our recent investigation \cite{Zschocke_Quadrupole_1}. 
In reality, both the celestial light source and the observer are located at finite spatial distances. That fact implies the need to solve the 
boundary value problem of light propagation. An important step toward the solution of the boundary value problem has been achieved in our recent investigation 
\cite{Zschocke_Quadrupole_2}, where the spatial positions of source and observer were implemented in the solution 
of the light trajectory. 

In this investigation, and as announced in the summary section of our previous work \cite{Zschocke_Quadrupole_2}, the solution of the boundary value problem has been completed. 
In particular, the three basic transformations of the boundary value problem have been derived: 
$\ve{k} \rightarrow \ve{\sigma}$, $\ve{\sigma} \rightarrow \ve{n}$, $\ve{k} \rightarrow \ve{n}$, where $\ve{k}$ is the unit direction from the source toward the observer, 
$\ve{\sigma}$ is the unit tangent vector of light trajectory at minus infinity, and $\ve{n}$ is the unit tangent vector of the light ray at the spatial position of the observer. 
As discussed in Section~\ref{GREM}, these transformations are the basis of GREM \cite{Klioner2003a}, which is the relativistic model of light propagation used for data reduction 
of the astrometry mission Gaia \cite{Gaia1,Gaia2}. This relativistic model has later been refined by our investigations in \cite{Klioner_Zschocke,Zschocke_Klioner}, where simplified 
formulas for the 1PN quadrupole and the 2PN monopole have been implemented into GREM. Here we have implemented the 2PN quadrupole structure of solar system bodies into these 
transformations, which allows for highly precise measurements of light deflection on the sub-\muas${}$ level in the solar system. These transformations are given by 
Eqs.~(\ref{transformation_k_to_sigma}), (\ref{transformation_sigma_to_n}), and (\ref{transformation_k_to_n}). Finally, upper limits for the 2PN light deflection of solar system bodies 
have been presented by Eqs.~(\ref{upper_limit_MM}) - (\ref{upper_limit_QQ}) and their numerical values were given for grazing rays at the giant planets in Table~\ref{Table2}. These 
transformations and the upper limits represent the primary results of this investigation. 

The development of future astrometry can embark into several directions. Mainly two of them seem to come into real focus in near future: (1) an all-sky survey in the 
optical and near-infrared band, like the Gaia-NIR project \cite{Gaia_NIR} and (2) relative astrometry missions, like the proposed missions NEAT \cite{NEAT1,NEAT2,NEAT3} 
and Theia \cite{Theia}, aiming at ultra highly precise angular measurements on the sub-\muas{} level. It might also well be that some concepts of these missions will be combined 
with each other or will at least support each other. The transformations and the upper limits of this investigation could directly be implemented in relativistic models 
of light propagation within such near-future space astrometry missions. For instance, the Gaia-NIR project \cite{Gaia_NIR} is one of the most promising candidates to get realized 
in the foreseeable future as a medium-sized mission of ESA. This mission has infrared telescopes on-board and is designed to unveil the nature of our galaxy and aims mainly at scanning 
the galactic center and stellar sources near the galactic plane. Besides that the primary intention of Gaia-NIR is not to improve the astrometric accuracies of the Gaia mission, 
it is almost certain that the end-of-mission precision will arrive at the sub-\muas{} scale in positional measurements. Furthermore, in an optimistic scenario the GaiaNIR mission could 
already be launched as soon as $2045$ \cite{Eric_Hoeg}.

\section*{Acknowledgments}

This work was funded by the German Research Foundation (Deutsche Forschungsgemeinschaft DFG) under Grant No. 447922800. Sincere gratitude is expressed to
Professor Sergei A. Klioner for continual support and inspiring discussions about astrometry and general theory of relativity. Professor Michael H. Soffel,
Professor Ralf Sch\"utzhold, Professor William G. Unruh, Priv.-Doz. G\"unter Plunien, Dr. Alexey Butkevich, Dipl-Inf. Robin Geyer, Dr. Jan Meichsner, 
Professor Burkhard K\"ampfer, and Professor Laszlo Csernai, are greatly acknowledged for interesting discussions about general theory of relativity.

\appendix

\section{Notation}\label{Appendix0}

Throughout the investigation the following notation is in use:

\begin{itemize}
\item $G$ is the Newtonian constant of gravitation.
\item $c$ is the vacuum speed of light in Minkowskian space-time.
\item Lower case Latin indices $i$, $j$, \dots take values $1,2,3$.
\item $\delta_{ij} = \delta^{ij} = {\rm diag} \left(+1,+1,+1\right)$ is Kronecker delta.
\item $\varepsilon_{ijk} = \varepsilon^{ijk}$ with $\varepsilon_{123} = + 1$ is the fully anti-symmetric Levi-Civita symbol.
\item Three-vectors are in boldface: e.g. $\ve{a}$, $\ve{b}$, $\ve{\sigma}$, $\ve{x}$.
\item Contravariant components of three-vectors: $a^{i} = \left(a^{\,1},a^2,a^3\right)$.
\item Absolute value of a three-vector: $a = |\ve{a}| = \sqrt{a^{\,1}\,a^{\,1}+a^2\,a^2+a^3\,a^3}$.
\item Scalar product of three-vectors: $\ve{a}\,\cdot\,\ve{b}=\delta_{ij}\,a^i\,b^j$.
\item Vector product of two three-vectors: $\left(\ve{a}\times\ve{b}\right)^i=\varepsilon_{ijk}\,a^j\,b^k$.
\item Lower case Greek indices take values 0,1,2,3.
\item Contravariant components of four-vectors: $a^{\mu} = \left(a^{\,0},a^{\,1},a^2,a^3\right)$.
\item milli-arcsecond (mas): $1\,{\rm mas} \simeq 4.85 \times 10^{-9}\,{\rm rad}$.
\item micro-arcsecond (\muas): $1\,\muas \simeq 4.85 \times 10^{-12}\,{\rm rad}$.
\item ${\it A.U.} = 1.496 \times 10^{11}\,{\rm m}$ is the astronomical unit
\item repeated indices are implicitly summed over ({\it Einstein's} sum convention).
\end{itemize}

\section{Some useful relations}\label{Appendix1}

The impact vectors in Eqs.~(\ref{impact_vector_0}) and (\ref{impact_vector_1}) are defined by 
\begin{eqnarray}
        \ve{d}^{\,0}_{\sigma} &=& \ve{\sigma} \times \left(\ve{x}_0 \times \ve{\sigma}\right),
        \label{impact_vector_0_appendix}
        \\
        \ve{d}^{\,1}_{\sigma} &=& \ve{\sigma} \times \left(\ve{x}_1 \times \ve{\sigma}\right),
        \label{impact_vector_1_appendix}
\end{eqnarray}

\noindent
with their absolute values $d^{\,0}_{\sigma} = |\ve{d}^{\,0}_{\sigma}|$ and $d^{\,1}_{\sigma} = |\ve{d}^{\,1}_{\sigma}|$. 
The impact vector (\ref{impact_vector_0_appendix}) is the same as the impact vector defined by Eq.~(\ref{impact_vector_unperturbed}). 
The impact vectors in Eq.~(\ref{impact_vector_k}) 
\begin{eqnarray}
        \ve{d}_k &=& \ve{k} \times \left(\ve{x}_0 \times \ve{k}\right) = \ve{k} \times \left(\ve{x}_1 \times \ve{k}\right), 
\label{impact_vector_k_appendix}
\end{eqnarray}

\noindent
are equal to each other. From the absolute value $d_k = \left|\ve{d}_k\right|$ one gets the relations 
$(\ve{k} \cdot \ve{x}_1)^2 = (x_1)^2 - (d_k)^2$ and $(\ve{k} \cdot \ve{x}_0)^2 = (x_0)^2 - (d_k)^2$. 
From Eq.~(\ref{k_sigma_2PN}) follows in 1PN approximation 
\begin{eqnarray}
	\ve{\sigma} &=& \ve{k} - \frac{1}{R}\left[\ve{k}\times \bigg(\Delta\ve{x}_{\rm 1PN} \times \ve{k}\bigg)\right]
\label{k_sigma_2PN_appendix}
\end{eqnarray}

\noindent
where $\Delta\ve{x}_{\rm 1PN} = \Delta\ve{x}_{\rm 1PN}\left(\ve{x}_1,\ve{x}_0\right)$. 
By inserting (\ref{k_sigma_2PN_appendix}) into (\ref{impact_vector_0_appendix}) and (\ref{impact_vector_1_appendix}) one obtains 
(cf. Eqs.~(I6) and (I7) in \cite{Zschocke_Time_Delay_2PN})
\begin{eqnarray}
	\ve{d}^{\,0}_{\sigma} &=& \ve{d}_k + \frac{\ve{d}_k \cdot \Delta\ve{x}_{\rm 1PN}}{R}\,\ve{k} 
	+ \left(\ve{k} \cdot \ve{x}_0\right)\,\frac{\ve{k} \times \left(\Delta\ve{x}_{\rm 1PN} \times \ve{k} \right)}{R}\,,
	\nonumber\\ 
        \label{relation_d0_dk}
        \\
        \ve{d}^{\,1}_{\sigma} &=& \ve{d}_k + \frac{\ve{d}_k \cdot \Delta\ve{x}_{\rm 1PN}}{R}\,\ve{k} 
        + \left(\ve{k} \cdot \ve{x}_1\right)\,\frac{\ve{k} \times \left(\Delta\ve{x}_{\rm 1PN} \times \ve{k} \right)}{R}\,, 
	\nonumber\\ 
        \label{relation_d1_dk}
\end{eqnarray}

\noindent
up to terms of the order ${\cal O}(c^{-4})$. From these equations one obtains the following relations for the absolute values 
(cf. Eqs.~(I11) and (I12) in \cite{Zschocke_Time_Delay_2PN}),  
\begin{eqnarray}
	\frac{1}{\left(d^{\,0}_{\sigma}\right)^n} &=& \frac{1}{\left(d_k\right)^n} - \frac{n}{R}\,
	\frac{\left(\ve{k} \cdot \ve{x}_0\right)\,\left(\ve{d}_k \cdot \Delta \ve{x}_{\rm 1PN}\right) }{\left(d_k\right)^{n+2}} \,,
        \label{relation_absolute_d0_dk}
	\\
	\frac{1}{\left(d^{\,1}_{\sigma}\right)^n} &=& \frac{1}{\left(d_k\right)^n} - \frac{n}{R}\,
        \frac{\left(\ve{k} \cdot \ve{x}_1\right)\,\left(\ve{d}_k \cdot \Delta \ve{x}_{\rm 1PN}\right) }{\left(d_k\right)^{n+2}} \,,
        \label{relation_absolute_d1_dk}
\end{eqnarray}

\noindent 
up to terms of the order ${\cal O}(c^{-4})$ and where $n$ can be any integer. These equations represent the first two terms 
of an infinite series expansion. This series is convergent for observers, which are assumed to be located somewhere in the solar system or in the vicinity of the solar 
system; cf. text below Eq.~(J7) in \cite{Zschocke_Quadrupole_1} as well as text below Eq.~(35) in \cite{Zschocke_Quadrupole_2}. Furthermore, one also needs the relations 
(cf. Eqs.~(I13) and (I14) in \cite{Zschocke_Time_Delay_2PN})
\begin{eqnarray}
	\ve{\sigma} \cdot \ve{x}_0 &=& \ve{k} \cdot \ve{x}_0 - \frac{\ve{d}_k \cdot \Delta \ve{x}_{\rm 1PN}}{R} \,,
        \label{relation_0_appendix}
	\\
	\ve{\sigma} \cdot \ve{x}_1 &=& \ve{k} \cdot \ve{x}_1 - \frac{\ve{d}_k \cdot \Delta \ve{x}_{\rm 1PN}}{R} \,, 
        \label{relation_1_appendix}
\end{eqnarray}

\noindent
up to terms of the order ${\cal O}(c^{-4})$. These relations follow from Eqs.~(\ref{impact_vector_k_appendix}) and (\ref{k_sigma_2PN_appendix}).

\section{The tensorial coefficients in Eq.~(\ref{transformation_k_to_sigma}) and Eqs.~(\ref{transformation_sigma_to_n}) - (\ref{transformation_k_to_n})}\label{Appendix7} 

The tensorial coefficients with one and three spatial indices in Eq.~(\ref{transformation_k_to_sigma}) and Eqs.~(\ref{transformation_sigma_to_n}) - (\ref{transformation_k_to_n}) 
read as follows:
\begin{eqnarray}
        X^{i}_{\left(1\right)} &=& k^i\;,
        \label{coefficient_X1_k}
        \\
        X^{i}_{\left(2\right)} &=& d_{k}^i \;,
        \label{coefficient_X2_k}
        \\
        Y^{ab\,i}_{\left(1\right)} &=& k^a \delta^{bi}\;, 
        \label{coefficient_Y1_k}
        \\
        Y^{ab\,i}_{\left(2\right)} &=& d_{k}^a \delta^{bi}\;, 
        \label{coefficient_Y2_k}
        \\
        Y^{ab\,i}_{\left(3\right)} &=& k^a k^b k^i\;, 
        \label{coefficient_Y3_k}
        \\
        Y^{ab\,i}_{\left(4\right)} &=& k^a d_{k}^b k^i\;, 
        \label{coefficient_Y4_k}
        \\
        Y^{ab\,i}_{\left(5\right)} &=& d_{k}^a \;d_{k}^b \;k^i\;, 
        \label{coefficient_Y5_k}
        \\
        Y^{ab\,i}_{\left(6\right)} &=& d_{k}^a \;d_{k}^b \;d_{k}^i\;, 
        \label{coefficient_Y6_k}
        \\
        Y^{ab\,i}_{\left(7\right)} &=& k^a k^b \;d_{k}^i\;, 
        \label{coefficient_Y7_k}
        \\
        Y^{ab\,i}_{\left(8\right)} &=& k^a\;d_{k}^b\;d_{k}^i\;.  
        \label{coefficient_Y8_k}
\end{eqnarray}

\noindent
The tensorial coefficients with five spatial indices in Eq.~(\ref{transformation_k_to_sigma}) and Eqs.~(\ref{transformation_sigma_to_n}) - (\ref{transformation_k_to_n}) 
read as follows: 
\begin{eqnarray}
        Z^{abcd\,i}_{\left(1\right)} &=& \delta^{ac} k^b \delta^{di} \;,
        \label{coefficient_Z1_k}
        \\
        Z^{abcd\,i}_{\left(2\right)} &=& \delta^{ac} d_{k}^b \delta^{di} \;,
        \label{coefficient_Z2_k}
        \\
        Z^{abcd\,i}_{\left(3\right)} &=& k^a k^b k^c \delta^{di} \;,
        \label{coefficient_Z3_k}
        \\
        Z^{abcd\,i}_{\left(4\right)} &=& k^a k^b d_{k}^c \delta^{di} \;,
        \label{coefficient_Z4_k}
        \\
        Z^{abcd\,i}_{\left(5\right)} &=& k^a d_{k}^b k^c \delta^{di} \;,
        \label{coefficient_Z5_k}
        \\
        Z^{abcd\,i}_{\left(6\right)} &=& k^a d_{k}^b d_{k}^c \delta^{di} \;,
        \label{coefficient_Z6_k}
	\\
        Z^{abcd\,i}_{\left(7\right)} &=& d_{k}^a d_{k}^b k^c \delta^{di} \;,
        \label{coefficient_Z7_k}
        \\
        Z^{abcd\,i}_{\left(8\right)} &=& d_{k}^a d_{k}^b d_{k}^c \delta^{di} \;,
        \label{coefficient_Z8_k}
        \\
        Z^{abcd\,i}_{\left(9\right)} &=& \delta^{ac} \delta^{bd} k^i \;,
        \label{coefficient_Z9_k}
        \\
        Z^{abcd\,i}_{\left(10\right)} &=& \delta^{ac} k^b k^d k^i \;,
        \label{coefficient_Z10_k}
        \\
        Z^{abcd\,i}_{\left(11\right)} &=& \delta^{ac} k^b d_{k}^d k^i \;,
        \label{coefficient_Z11_k}
        \\
        Z^{abcd\,i}_{\left(12\right)} &=& \delta^{ac} d_{k}^b d_{k}^d k^i \;,
        \label{coefficient_Z12_k}
	\\ 
        Z^{abcd\,i}_{\left(13\right)} &=& k^a k^b k^c k^d k^i \;,
        \label{coefficient_Z13_k}
        \\
        Z^{abcd\,i}_{\left(14\right)} &=& k^a k^b k^c d_{k}^d k^i \;,
        \label{coefficient_Z14_k}
        \\
        Z^{abcd\,i}_{\left(15\right)} &=& k^a k^b d_{k}^c d_{k}^d k^i \;,
        \label{coefficient_Z15_k}
        \\
        Z^{abcd\,i}_{\left(16\right)} &=& k^a d_{k}^b k^c d_{k}^d k^i \;,
        \label{coefficient_Z16_k}
	\\ 
        Z^{abcd\,i}_{\left(17\right)} &=& k^a d_{k}^b d_{k}^c d_{k}^d k^i \;,
        \label{coefficient_Z17_k}
        \\
        Z^{abcd\,i}_{\left(18\right)} &=& d_{k}^a d_{k}^b d_{k}^c d_{k}^d k^i \;,
        \label{coefficient_Z18_k}
        \\
        Z^{abcd\,i}_{\left(19\right)} &=& \delta^{ac} \delta^{bd} d_{k}^i \;,
        \label{coefficient_Z19_k}
	\\ 
        Z^{abcd\,i}_{\left(20\right)} &=& \delta^{ac} k^b k^d d_{k}^i \;,
        \label{coefficient_Z20_k}
        \end{eqnarray} 
        
        \begin{eqnarray} 
        Z^{abcd\,i}_{\left(21\right)} &=& \delta^{ac} k^b d_{k}^d d_{k}^i \;,
        \label{coefficient_Z21_k}
        \\
        Z^{abcd\,i}_{\left(22\right)} &=& \delta^{ac} d_{k}^b d_{k}^d d_{k}^i \;,
        \label{coefficient_Z22_k}
        \\
        Z^{abcd\,i}_{\left(23\right)} &=& k^a k^b k^c k^d d_{k}^i \;,
        \label{coefficient_Z23_k}
        \\
        Z^{abcd\,i}_{\left(24\right)} &=& k^a k^b k^c d_{k}^d d_{k}^i \;,
        \label{coefficient_Z24_k}
        \\
        Z^{abcd\,i}_{\left(25\right)} &=& k^a k^b d_{k}^c d_{k}^d d_{k}^i \;,
        \label{coefficient_Z25_k}
        \\
        Z^{abcd\,i}_{\left(26\right)} &=& k^a d_{k}^b k^c d_{k}^d d_{k}^i \;,
        \label{coefficient_Z26_k}
        \\
        Z^{abcd\,i}_{\left(27\right)} &=& k^a d_{k}^b d_{k}^c d_{k}^d d_{k}^i \;,
        \label{coefficient_Z27_k}
        \\
        Z^{abcd\,i}_{\left(28\right)} &=& d_{k}^a d_{k}^b d_{k}^c d_{k}^d d_{k}^i \;.
        \label{coefficient_Z28_k}
\end{eqnarray}

\section{The scalar functions in Eq.~(\ref{transformation_k_to_sigma})}\label{Appendix8}

In order to simplify the notations, the following abbreviations are introduced, which are identical to our 
former notations (cf. Eqs.~(H20) - (H24) in \cite{Zschocke_Time_Delay_2PN}), while an additional function $k_{\left(1\right)}$ has been introduced:
\begin{eqnarray}
        e_{\left(n\right)} &=& \left(x_1 + \ve{k} \cdot \ve{x}_1\right)^n - \left(x_0 + \ve{k} \cdot \ve{x}_0\right)^n \;,
        \label{e_n}
        \\
        f_{\left(n\right)} &=& \frac{1}{\left(x_1\right)^n} - \frac{1}{\left(x_0\right)^n} \;,
        \label{f_n}
        \\
        g_{\left(n\right)} &=& \frac{\ve{k} \cdot \ve{x}_1}{\left(x_1\right)^n} - \frac{\ve{k} \cdot \ve{x}_0}{\left(x_0\right)^n} \;,
        \label{g_n}
        \\
        h_{\left(1\right)} &=& \arctan \frac{\ve{k} \cdot \ve{x}_1}{d_k} - \arctan \frac{\ve{k} \cdot \ve{x}_0}{d_k} \;,
        \label{h_1}
        \\
        h_{\left(2\right)} &=&
        + \frac{\ve{k} \cdot \ve{x}_1}{d_k} \left(\arctan \frac{\ve{k} \cdot \ve{x}_1}{d_k} + \frac{\pi}{2}\right)
	\nonumber\\ 
	&& - \frac{\ve{k} \cdot \ve{x}_0}{d_k} \left(\arctan \frac{\ve{k} \cdot \ve{x}_0}{d_k} + \frac{\pi}{2}\right),
        \label{h_2}
	\\ 
	k_{\left(1\right)} &=& 
	\left(\ve{k} \cdot \ve{x}_1\right) \left(x_1 + \ve{k} \cdot \ve{x}_1 \right) - \left(\ve{k} \cdot \ve{x}_0\right) \left(x_0 + \ve{k} \cdot \ve{x}_0 \right)
        \nonumber\\ 
	&=& x_1 \left(x_1 + \ve{k} \cdot \ve{x}_1 \right) - x_0 \left(x_0 + \ve{k} \cdot \ve{x}_0 \right). 
	\label{k_2}
\end{eqnarray}

\noindent
The following relations allow one to remove the angles in these functions,  
\begin{eqnarray}
	\ve{k} \cdot \ve{x}_1 &=& \frac{\left(x_1\right)^2 - \left(x_0\right)^2 + \left(R\right)^2}{2\,R}\;,
	\label{k_dot_x1}
	\\
	\ve{k} \cdot \ve{x}_0 &=& \frac{\left(x_1\right)^2 - \left(x_0\right)^2 - \left(R\right)^2}{2\,R}\;.
        \label{k_dot_x0}
\end{eqnarray}

\noindent
If one inserts these relations into the scalar functions (\ref{e_n}) - (\ref{k_2}), then one obtains more involved expressions for these scalar functions, 
which, as mentioned, do not contain any angles. Therefore, it might sometimes be convenient to implement the relations (\ref{k_dot_x1}) and (\ref{k_dot_x0}) 
for practical astrometric calculations. 

The scalar functions in Eq.~(\ref{transformation_k_to_sigma}) read as follows:
\begin{eqnarray}
	\widehat{A}_{\left(1\right)} &=& 0\,,  
	\label{scalar_function_A_sigma_k_1}
        \\
	\widehat{A}_{\left(2\right)} &=& + \frac{2}{\left(d_k\right)^2}\,\frac{e_{\left(1\right)}}{R}\,,
	\label{scalar_function_A_sigma_k_2}
        \\
	\widehat{B}_{\left(1\right)} &=& + \frac{2}{\left(d_k\right)^2}\,\frac{g_{\left(1\right)}}{R}\,, 
	\label{scalar_function_B_sigma_k_1}
        \\
	\widehat{B}_{\left(2\right)} &=& - \frac{4}{\left(d_k\right)^4}\,\frac{e_{\left(1\right)}}{R} 
	+ \frac{2}{\left(d_k\right)^2}\,\frac{f_{\left(1\right)}}{R}\,,
	\label{scalar_function_B_sigma_k_2}
        \\
	\widehat{B}_{\left(3\right)} &=& - \frac{2}{\left(d_k\right)^2}\,\frac{g_{\left(1\right)}}{R}\,,
	\label{scalar_function_B_sigma_k_3}
        \\
	\widehat{B}_{\left(4\right)} &=& + \frac{4}{\left(d_k\right)^4}\,\frac{e_{\left(1\right)}}{R} 
        - \frac{2}{\left(d_k\right)^2}\,\frac{f_{\left(1\right)}}{R}\,, 
	\label{scalar_function_B_sigma_k_4}
        \\
	\widehat{B}_{\left(5\right)} &=& 0\,,  
	\label{scalar_function_B_sigma_k_5}
        \\
	\widehat{B}_{\left(6\right)} &=& + \frac{8}{\left(d_k\right)^6}\,\frac{e_{\left(1\right)}}{R} 
	- \frac{4}{\left(d_k\right)^4}\,\frac{f_{\left(1\right)}}{R}
	- \frac{1}{\left(d_k\right)^2}\,\frac{f_{\left(3\right)}}{R}\,, 
	\label{scalar_function_B_sigma_k_6}
        \\
	\widehat{B}_{\left(7\right)} &=& + \frac{2}{\left(d_k\right)^4}\,\frac{e_{\left(1\right)}}{R} 
	- \frac{1}{\left(d_k\right)^2}\,\frac{f_{\left(1\right)}}{R} + \frac{f_{\left(3\right)}}{R}\,, 
	\label{scalar_function_B_sigma_k_7}
        \\
	\widehat{B}_{\left(8\right)} &=& - \frac{4}{\left(d_k\right)^4}\,\frac{g_{\left(1\right)}}{R} 
	- \frac{2}{\left(d_k\right)^2}\,\frac{g_{\left(3\right)}}{R}\,, 
	\label{scalar_function_B_sigma_k_8}
	\\
	\widehat{C}_{\left(1\right)} &=& - \frac{2}{\left(d_k\right)^2}\,\frac{e_{\left(1\right)}\,e_{\left(1\right)}}{\left(R\right)^2}\,, 
	\label{scalar_function_C_sigma_k_1}
	\\
	\widehat{C}_{\left(2\right)} &=& + \frac{4}{\left(d_k\right)^4}\,\frac{e_{\left(1\right)}\,k_{\left(1\right)}}{\left(R\right)^2} 
	- \frac{4}{\left(d_k\right)^4}\,\frac{e_{\left(2\right)}}{R} - \frac{1}{4}\,\frac{f_{\left(2\right)}}{R}
	\nonumber\\
	&& + \frac{15}{4}\,\frac{1}{\left(d_k\right)^2}\,\frac{h_{\left(2\right)}}{R}\,,
	\label{scalar_function_C_sigma_k_2}
        \end{eqnarray} 

        \begin{widetext} 
	\begin{eqnarray}
	\widehat{D}_{\left(1\right)} &=& - \frac{20}{\left(d_k\right)^4}\,\frac{e_{\left(1\right)}}{R} 
	+ \frac{4}{\left(d_k\right)^2}\,\frac{f_{\left(1\right)}}{R} 
	+ \frac{93}{32}\,\frac{1}{\left(d_k\right)^2}\,\frac{g_{\left(2\right)}}{R} 
	+ \frac{7}{16}\,\frac{g_{\left(4\right)}}{R} 
	+ \frac{285}{32}\,\frac{1}{\left(d_k\right)^3}\,\frac{h_{\left(1\right)}}{R} 
	- \frac{4}{\left(d_k\right)^4}\,\frac{g_{\left(1\right)}\,k_{\left(1\right)}}{\left(R\right)^2}
	\nonumber\\
	&& - \frac{4}{\left(d_k\right)^2}\,\frac{e_{\left(1\right)}\,f_{\left(1\right)}}{\left(R\right)^2}
	+ \frac{8}{\left(d_k\right)^4}\,\frac{e_{\left(1\right)}\,e_{\left(1\right)}}{\left(R\right)^2}\,,
	\label{scalar_function_D_sigma_k_1}
	\\
	\widehat{D}_{\left(2\right)} &=& + \frac{16}{\left(d_k\right)^6}\,\frac{e_{\left(2\right)}}{R} 
	+ \frac{91}{32}\,\frac{1}{\left(d_k\right)^2}\,\frac{f_{\left(2\right)}}{R}
	+ \frac{7}{16}\,\frac{f_{\left(4\right)}}{R}
	- \frac{4}{\left(d_k\right)^4}\,\frac{g_{\left(1\right)}}{R}
	- \frac{465}{32}\,\frac{1}{\left(d_k\right)^4}\, \frac{h_{\left(2\right)}}{R} 
	- \frac{16}{\left(d_k\right)^6}\,\frac{e_{\left(1\right)}\,k_{\left(1\right)}}{\left(R\right)^2}
	\nonumber\\ 
	&& + \frac{8}{\left(d_k\right)^4}\,\frac{e_{\left(1\right)}\,g_{\left(1\right)}}{\left(R\right)^2}
	- \frac{4}{\left(d_k\right)^4}\,\frac{f_{\left(1\right)}\,k_{\left(1\right)}}{\left(R\right)^2}\,,
	\label{scalar_function_D_sigma_k_2}
	\\
	\widehat{D}_{\left(3\right)} &=& + \frac{20}{\left(d_k\right)^4}\,\frac{e_{\left(1\right)}}{R} 
	- \frac{4}{\left(d_k\right)^2}\,\frac{f_{\left(1\right)}}{R}
        - \frac{93}{32}\,\frac{1}{\left(d_k\right)^2}\,\frac{g_{\left(2\right)}}{R}
        - \frac{7}{16}\,\frac{g_{\left(4\right)}}{R}
        - \frac{285}{32}\,\frac{1}{\left(d_k\right)^3}\,\frac{h_{\left(1\right)}}{R} 
	+ \frac{4}{\left(d_k\right)^4}\,\frac{g_{\left(1\right)}\,k_{\left(1\right)}}{\left(R\right)^2}
	\nonumber\\
	&& + \frac{6}{\left(d_k\right)^2}\,\frac{e_{\left(1\right)}\,f_{\left(1\right)}}{\left(R\right)^2}  
	- \frac{2}{\left(R\right)^2}\,e_{\left(1\right)}\,f_{\left(3\right)} 
	- \frac{12}{\left(d_k\right)^4}\,\frac{e_{\left(1\right)}\,e_{\left(1\right)}}{\left(R\right)^2}\,, 
	\label{scalar_function_D_sigma_k_3}
	\\
	\widehat{D}_{\left(4\right)} &=& - \frac{16}{\left(d_k\right)^6}\,\frac{e_{\left(2\right)}}{R} 
        - \frac{91}{32}\,\frac{1}{\left(d_k\right)^2}\,\frac{f_{\left(2\right)}}{R}
        - \frac{7}{16}\,\frac{f_{\left(4\right)}}{R}
        + \frac{4}{\left(d_k\right)^4}\,\frac{g_{\left(1\right)}}{R}
        + \frac{465}{32}\,\frac{1}{\left(d_k\right)^4}\, \frac{h_{\left(2\right)}}{R} 
	+ \frac{16}{\left(d_k\right)^6}\,\frac{e_{\left(1\right)}\,k_{\left(1\right)}}{\left(R\right)^2}
        \nonumber\\ 
        && - \frac{4}{\left(d_k\right)^4}\,\frac{e_{\left(1\right)}\,g_{\left(1\right)}}{\left(R\right)^2}
	+ \frac{4}{\left(d_k\right)^2}\,\frac{e_{\left(1\right)}\,g_{\left(3\right)}}{\left(R\right)^2} 
	+ \frac{4}{\left(d_k\right)^4}\,\frac{f_{\left(1\right)}\,k_{\left(1\right)}}{\left(R\right)^2}\,, 
	\label{scalar_function_D_sigma_k_4}
	\\
	\widehat{D}_{\left(5\right)} &=& - \frac{8}{\left(d_k\right)^6}\,\frac{e_{\left(1\right)}\,e_{\left(1\right)}}{\left(R\right)^2} 
	+ \frac{4}{\left(d_k\right)^4}\,\frac{e_{\left(1\right)}\,f_{\left(1\right)}}{\left(R\right)^2} 
	+ \frac{2}{\left(d_k\right)^2}\,\frac{e_{\left(1\right)}\,f_{\left(3\right)}}{\left(R\right)^2}\,,
	\label{scalar_function_D_sigma_k_5}
	\\
	\widehat{D}_{\left(6\right)} &=& - \frac{48}{\left(d_k\right)^8}\,\frac{e_{\left(2\right)}}{R} 
        - \frac{263}{64}\,\frac{1}{\left(d_k\right)^4}\,\frac{f_{\left(2\right)}}{R}
        - \frac{91}{32}\,\frac{1}{\left(d_k\right)^2}\,\frac{f_{\left(4\right)}}{R}
	- \frac{5}{8}\,\frac{f_{\left(6\right)}}{R} 
        + \frac{16}{\left(d_k\right)^6}\,\frac{g_{\left(1\right)}}{R}
	+ \frac{4}{\left(d_k\right)^4}\,\frac{g_{\left(3\right)}}{R} 
        + \frac{2325}{64}\,\frac{1}{\left(d_k\right)^6}\, \frac{h_{\left(2\right)}}{R} 
	\nonumber\\
	&& - \frac{16}{\left(d_k\right)^6}\,\frac{e_{\left(1\right)}\,g_{\left(1\right)}}{\left(R\right)^2}
        - \frac{2}{\left(d_k\right)^4}\,\frac{e_{\left(1\right)}\,g_{\left(3\right)}}{\left(R\right)^2} 
        - \frac{2}{\left(d_k\right)^4}\,\frac{f_{\left(3\right)}\,k_{\left(1\right)}}{\left(R\right)^2}
	+ \frac{48}{\left(d_k\right)^8}\,\frac{e_{\left(1\right)}\,k_{\left(1\right)}}{\left(R\right)^2}\,,
	\label{scalar_function_D_sigma_k_6}
	\\
	\widehat{D}_{\left(7\right)} &=& - \frac{16}{\left(d_k\right)^6}\,\frac{e_{\left(2\right)}}{R} 
        - \frac{285}{64}\,\frac{1}{\left(d_k\right)^2}\,\frac{f_{\left(2\right)}}{R}
        + \frac{71}{32}\,\frac{f_{\left(4\right)}}{R} 
        + \frac{5}{8} \left(d_k\right)^2 \frac{f_{\left(6\right)}}{R} 
        - \frac{4}{\left(d_k\right)^2}\,\frac{g_{\left(3\right)}}{R} 
        + \frac{855}{64}\,\frac{1}{\left(d_k\right)^4}\,\frac{h_{\left(2\right)}}{R}  
	\nonumber\\ 
	&& - \frac{6}{\left(d_k\right)^4}\,\frac{e_{\left(1\right)}\,g_{\left(1\right)}}{\left(R\right)^2} 
	- \frac{2}{\left(d_k\right)^2}\,\frac{e_{\left(1\right)}\,g_{\left(3\right)}}{\left(R\right)^2} 
	+ \frac{16}{\left(d_k\right)^6}\,\frac{e_{\left(1\right)}\,k_{\left(1\right)}}{\left(R\right)^2} 
	- \frac{2}{\left(d_k\right)^4}\,\frac{f_{\left(1\right)}\,k_{\left(1\right)}}{\left(R\right)^2}
	+ \frac{2}{\left(d_k\right)^2}\,\frac{f_{\left(3\right)}\,k_{\left(1\right)}}{\left(R\right)^2}\,,
	\label{scalar_function_D_sigma_k_7}
	\\
	\widehat{D}_{\left(8\right)} &=& + \frac{48}{\left(d_k\right)^6}\,\frac{e_{\left(1\right)}}{R} 
        - \frac{12}{\left(d_k\right)^4}\,\frac{f_{\left(1\right)}}{R}
        - \frac{8}{\left(d_k\right)^2}\,\frac{f_{\left(3\right)}}{R}
        - \frac{81}{32}\,\frac{1}{\left(d_k\right)^4}\,\frac{g_{\left(2\right)}}{R}
        - \frac{91}{16}\,\frac{1}{\left(d_k\right)^2}\,\frac{g_{\left(4\right)}}{R}
        - \frac{5}{4}\,\frac{g_{\left(6\right)}}{R}
        - \frac{465}{32}\,\frac{1}{\left(d_k\right)^5}\,\frac{h_{\left(1\right)}}{R} 
	\nonumber\\ 
	&& - \frac{4}{\left(d_k\right)^4}\,\frac{k_{\left(1\right)}\,g_{\left(3\right)}}{\left(R\right)^2}\,.  
	\label{scalar_function_D_sigma_k_8}
        \\
	\widehat{E}_{\left(1\right)} &=& - \frac{8}{\left(d_k\right)^6}\,\frac{e_{\left(1\right)}}{R}
        + \frac{8}{\left(d_k\right)^4}\,\frac{f_{\left(1\right)}}{R}
        + \frac{327}{128}\,\frac{1}{\left(d_k\right)^4}\,\frac{g_{\left(2\right)}}{R}
        + \frac{7}{192}\,\frac{1}{\left(d_k\right)^2}\,\frac{g_{\left(4\right)}}{R}
        - \frac{13}{48}\,\frac{g_{\left(6\right)}}{R}
        - \frac{185}{128}\,\frac{1}{\left(d_k\right)^5}\,\frac{h_{\left(1\right)}}{R}
	+ \frac{8}{\left(d_k\right)^6}\,\frac{k_{\left(1\right)}\,g_{\left(1\right)}}{\left(R\right)^2}\,,
	\nonumber\\ 
	\label{scalar_function_E_sigma_k_1}
	\\ 
	\widehat{E}_{\left(2\right)} &=& + \frac{16}{\left(d_k\right)^8}\,\frac{e_{\left(2\right)}}{R} 
	+ \frac{985}{384}\,\frac{1}{\left(d_k\right)^4}\,\frac{f_{\left(2\right)}}{R} 
	+ \frac{5}{192}\,\frac{1}{\left(d_k\right)^2}\,\frac{f_{\left(4\right)}}{R} 
	- \frac{13}{48}\,\frac{f_{\left(6\right)}}{R} 
	- \frac{8}{\left(d_k\right)^6}\,\frac{g_{\left(1\right)}}{R} 
	- \frac{985}{128}\,\frac{1}{\left(d_k\right)^6}\,\frac{h_{\left(2\right)}}{R} 
	\nonumber\\ 
	&& - \frac{16}{\left(d_k\right)^8}\,\frac{e_{\left(1\right)}\,k_{\left(1\right)}}{\left(R\right)^2} 
	+ \frac{8}{\left(d_k\right)^6}\,\frac{f_{\left(1\right)}\,k_{\left(1\right)}}{\left(R\right)^2}\,,
	\label{scalar_function_E_sigma_k_2}
	\\
	\widehat{E}_{\left(3\right)} &=& - \frac{12}{\left(d_k\right)^6}\,\frac{e_{\left(1\right)}}{R}
        - \frac{4}{\left(d_k\right)^2}\,\frac{f_{\left(3\right)}}{R}
        + \frac{2103}{512}\,\frac{1}{\left(d_k\right)^4}\,\frac{g_{\left(2\right)}}{R}
        - \frac{451}{256}\,\frac{1}{\left(d_k\right)^2}\,\frac{g_{\left(4\right)}}{R}
        - \frac{23}{64}\,\frac{g_{\left(6\right)}}{R} 
	- \frac{9}{32} \left(d_k\right)^2 \frac{g_{\left(8\right)}}{R}
	\nonumber\\ 
	&& + \frac{5175}{512}\,\frac{1}{\left(d_k\right)^5}\,\frac{h_{\left(1\right)}}{R}
	+ \frac{8}{\left(d_k\right)^6}\,\frac{e_{\left(1\right)}\,e_{\left(1\right)}}{\left(R\right)^2}
	- \frac{8}{\left(d_k\right)^4}\,\frac{e_{\left(1\right)}\,f_{\left(1\right)}}{\left(R\right)^2}
	+ \frac{4}{\left(d_k\right)^2}\,\frac{e_{\left(1\right)}\,f_{\left(3\right)}}{\left(R\right)^2} 
	+ \frac{2}{\left(d_k\right)^2}\,\frac{f_{\left(1\right)}\,f_{\left(1\right)}}{\left(R\right)^2}
	\nonumber\\ 
	&& - 2\,\frac{f_{\left(1\right)}\,f_{\left(3\right)}}{\left(R\right)^2} 
	- \frac{2}{\left(d_k\right)^4}\,\frac{g_{\left(1\right)}\,g_{\left(1\right)}}{\left(R\right)^2}
	- \frac{2}{\left(d_k\right)^2}\,\frac{g_{\left(1\right)}\,g_{\left(3\right)}}{\left(R\right)^2}
	- \frac{8}{\left(d_k\right)^6}\,\frac{g_{\left(1\right)}\,k_{\left(1\right)}}{\left(R\right)^2}
	- \frac{4}{\left(d_k\right)^4}\,\frac{g_{\left(3\right)}\,k_{\left(1\right)}}{\left(R\right)^2}\,,
	\label{scalar_function_E_sigma_k_3} 
        \end{eqnarray}
        \end{widetext} 

	\begin{widetext}
	\begin{eqnarray}
	\widehat{E}_{\left(4\right)} &=& + \frac{16}{\left(d_k\right)^8}\,\frac{e_{\left(2\right)}}{R}
        + \frac{27019}{1536}\,\frac{1}{\left(d_k\right)^4}\,\frac{f_{\left(2\right)}}{R} 
	- \frac{1585}{768}\,\frac{1}{\left(d_k\right)^2}\,\frac{f_{\left(4\right)}}{R} 
	- \frac{55}{192}\,\frac{f_{\left(6\right)}}{R} 
	- \frac{9}{32} \left(d_k\right)^2 \frac{f_{\left(8\right)}}{R} 
	- \frac{20}{\left(d_k\right)^6}\,\frac{g_{\left(1\right)}}{R}
	\nonumber\\ 
	&& + \frac{8}{\left(d_k\right)^4}\,\frac{g_{\left(3\right)}}{R} 
	- \frac{5515}{512}\,\frac{1}{\left(d_k\right)^6}\,\frac{h_{\left(2\right)}}{R} 
	+ \frac{12}{\left(d_k\right)^6}\,\frac{e_{\left(1\right)}\,g_{\left(1\right)}}{\left(R\right)^2} 
	+ \frac{4}{\left(d_k\right)^4}\,\frac{e_{\left(1\right)}\,g_{\left(3\right)}}{\left(R\right)^2} 
	- \frac{6}{\left(d_k\right)^4}\,\frac{f_{\left(1\right)}\,g_{\left(1\right)}}{\left(R\right)^2}
	\nonumber\\ 
	&& - \frac{2}{\left(d_k\right)^2}\,\frac{f_{\left(1\right)}\,g_{\left(3\right)}}{\left(R\right)^2} 
        + \frac{4}{\left(d_k\right)^2}\,\frac{f_{\left(3\right)}\,g_{\left(1\right)}}{\left(R\right)^2} 
	- \frac{16}{\left(d_k\right)^8}\,\frac{e_{\left(1\right)}\,k_{\left(1\right)}}{\left(R\right)^2} 
	+ \frac{8}{\left(d_k\right)^6}\,\frac{f_{\left(1\right)}\,k_{\left(1\right)}}{\left(R\right)^2}
	- \frac{12}{\left(d_k\right)^4}\,\frac{f_{\left(3\right)}\,k_{\left(1\right)}}{\left(R\right)^2}\,, 
	\label{scalar_function_E_sigma_k_4}
	\\
	\widehat{E}_{\left(5\right)} &=& - \frac{16}{\left(d_k\right)^8}\,\frac{e_{\left(2\right)}}{R} 
	+ \frac{3859}{768}\,\frac{1}{\left(d_k\right)^4}\,\frac{f_{\left(2\right)}}{R}
	- \frac{1609}{384}\,\frac{1}{\left(d_k\right)^2}\,\frac{f_{\left(4\right)}}{R}
	- \frac{79}{96}\,\frac{f_{\left(6\right)}}{R} 
	- \frac{9}{16} \left(d_k\right)^2 \frac{f_{\left(8\right)}}{R} 
	+ \frac{8}{\left(d_k\right)^6}\,\frac{g_{\left(1\right)}}{R}
	+ \frac{8}{\left(d_k\right)^4}\,\frac{g_{\left(3\right)}}{R}
	\nonumber\\
        && + \frac{2285}{256}\,\frac{1}{\left(d_k\right)^6}\,\frac{h_{\left(2\right)}}{R} 
	+ \frac{8}{\left(d_k\right)^6}\,\frac{e_{\left(1\right)}\,g_{\left(1\right)}}{\left(R\right)^2}
	- \frac{8}{\left(d_k\right)^4}\,\frac{e_{\left(1\right)}\,g_{\left(3\right)}}{\left(R\right)^2}
        + \frac{16}{\left(d_k\right)^8}\,\frac{e_{\left(1\right)}\,k_{\left(1\right)}}{\left(R\right)^2} 
	- \frac{4}{\left(d_k\right)^4}\,\frac{f_{\left(1\right)}\,g_{\left(1\right)}}{\left(R\right)^2} 
	- \frac{4}{\left(d_k\right)^2}\,\frac{f_{\left(3\right)}\,g_{\left(1\right)}}{\left(R\right)^2} 
	\nonumber\\
	&& + \frac{4}{\left(d_k\right)^2}\,\frac{f_{\left(1\right)}\,g_{\left(3\right)}}{\left(R\right)^2} 
	- \frac{8}{\left(d_k\right)^6}\,\frac{f_{\left(1\right)}\,k_{\left(1\right)}}{\left(R\right)^2}
	- \frac{8}{\left(d_k\right)^4}\,\frac{f_{\left(3\right)}\,k_{\left(1\right)}}{\left(R\right)^2}\,, 
	\label{scalar_function_E_sigma_k_5}
	\\
	\widehat{E}_{\left(6\right)} &=& + \frac{32}{\left(d_k\right)^8}\,\frac{e_{\left(1\right)}}{R} 
        - \frac{32}{\left(d_k\right)^6}\,\frac{f_{\left(1\right)}}{R}
        + \frac{16}{\left(d_k\right)^4}\,\frac{f_{\left(3\right)}}{R}
        - \frac{6381}{256}\,\frac{1}{\left(d_k\right)^6}\,\frac{g_{\left(2\right)}}{R}
        + \frac{2323}{384}\,\frac{1}{\left(d_k\right)^4}\,\frac{g_{\left(4\right)}}{R}
        + \frac{119}{96}\,\frac{1}{\left(d_k\right)^2}\,\frac{g_{\left(6\right)}}{R}
	\nonumber\\ 
	&& + \frac{9}{16}\,\frac{g_{\left(8\right)}}{R}
	- \frac{2285}{256}\, \frac{1}{\left(d_k\right)^7}\,\frac{h_{\left(1\right)}}{R} 
	- \frac{32}{\left(d_k\right)^8}\,\frac{e_{\left(1\right)}\,e_{\left(1\right)}}{\left(R\right)^2} 
	+ \frac{32}{\left(d_k\right)^6}\,\frac{e_{\left(1\right)}\,f_{\left(1\right)}}{\left(R\right)^2} 
	+ \frac{8}{\left(d_k\right)^4}\,\frac{e_{\left(1\right)}\,f_{\left(3\right)}}{\left(R\right)^2} 
	- \frac{8}{\left(d_k\right)^4}\,\frac{f_{\left(1\right)}\,f_{\left(1\right)}}{\left(R\right)^2} 
	\nonumber\\
	&& - \frac{4}{\left(d_k\right)^2}\,\frac{f_{\left(1\right)}\,f_{\left(3\right)}}{\left(R\right)^2} 
	- \frac{8}{\left(d_k\right)^6}\,\frac{g_{\left(1\right)}\,g_{\left(1\right)}}{\left(R\right)^2}
	- \frac{8}{\left(d_k\right)^4}\,\frac{g_{\left(1\right)}\,g_{\left(3\right)}}{\left(R\right)^2} 
	+ \frac{16}{\left(d_k\right)^8}\,\frac{g_{\left(1\right)}\,k_{\left(1\right)}}{\left(R\right)^2}
	+ \frac{24}{\left(d_k\right)^6}\,\frac{g_{\left(3\right)}\,k_{\left(1\right)}}{\left(R\right)^2}\,, 
	\label{scalar_function_E_sigma_k_6}
	\\
	\widehat{E}_{\left(7\right)} &=& - \frac{32}{\left(d_k\right)^8}\,\frac{e_{\left(1\right)}}{R} 
	+ \frac{8}{\left(d_k\right)^6}\,\frac{f_{\left(1\right)}}{R}
        + \frac{4}{\left(d_k\right)^4}\,\frac{f_{\left(3\right)}}{R} 
        + \frac{1419}{512}\,\frac{1}{\left(d_k\right)^6}\,\frac{g_{\left(2\right)}}{R}
        + \frac{2443}{768}\,\frac{1}{\left(d_k\right)^4}\,\frac{g_{\left(4\right)}}{R}
        + \frac{143}{192}\,\frac{1}{\left(d_k\right)^2}\,\frac{g_{\left(6\right)}}{R}
        \nonumber\\
        && + \frac{9}{32}\,\frac{g_{\left(8\right)}}{R}
        + \frac{5515}{512}\, \frac{1}{\left(d_k\right)^7}\,\frac{h_{\left(1\right)}}{R}
	+ \frac{16}{\left(d_k\right)^8}\,\frac{e_{\left(1\right)}\,e_{\left(1\right)}}{\left(R\right)^2} 
	- \frac{16}{\left(d_k\right)^6}\,\frac{e_{\left(1\right)}\,f_{\left(1\right)}}{\left(R\right)^2} 
	- \frac{4}{\left(d_k\right)^4}\,\frac{e_{\left(1\right)}\,f_{\left(3\right)}}{\left(R\right)^2}
	+ \frac{4}{\left(d_k\right)^4}\,\frac{f_{\left(1\right)}\,f_{\left(1\right)}}{\left(R\right)^2} 
	\nonumber\\
	&& + \frac{2}{\left(d_k\right)^2}\,\frac{f_{\left(1\right)}\,f_{\left(3\right)}}{\left(R\right)^2} 
	+ \frac{8}{\left(d_k\right)^6}\,\frac{g_{\left(1\right)}\,g_{\left(1\right)}}{\left(R\right)^2} 
	+ \frac{2}{\left(d_k\right)^4}\,\frac{g_{\left(1\right)}\,g_{\left(3\right)}}{\left(R\right)^2}
	- \frac{8}{\left(d_k\right)^8}\,\frac{k_{\left(1\right)}\,g_{\left(1\right)}}{\left(R\right)^2}
	+ \frac{4}{\left(d_k\right)^6}\,\frac{k_{\left(1\right)}\,g_{\left(3\right)}}{\left(R\right)^2}\,,
	\label{scalar_function_E_sigma_k_7}
	\\
	\widehat{E}_{\left(8\right)} &=& - \frac{4831}{512}\,\frac{1}{\left(d_k\right)^6}\,\frac{f_{\left(2\right)}}{R}
	+ \frac{877}{256}\,\frac{1}{\left(d_k\right)^4}\,\frac{f_{\left(4\right)}}{R}
	+ \frac{43}{64}\,\frac{1}{\left(d_k\right)^2}\,\frac{f_{\left(6\right)}}{R}
        + \frac{9}{32}\,\frac{f_{\left(8\right)}}{R}
        - \frac{8}{\left(d_k\right)^6}\,\frac{g_{\left(3\right)}}{R} 
        + \frac{2205}{512}\,\frac{1}{\left(d_k\right)^8}\,\frac{h_{\left(2\right)}}{R} 
	\nonumber\\
	&& - \frac{4}{\left(d_k\right)^6}\,\frac{e_{\left(1\right)}\,g_{\left(3\right)}}{\left(R\right)^2} 
	- \frac{4}{\left(d_k\right)^4}\,\frac{f_{\left(3\right)}\,g_{\left(1\right)}}{\left(R\right)^2} 
	+ \frac{2}{\left(d_k\right)^4}\,\frac{f_{\left(1\right)}\,g_{\left(3\right)}}{\left(R\right)^2} 
	+ \frac{12}{\left(d_k\right)^6}\,\frac{f_{\left(3\right)}\,k_{\left(1\right)}}{\left(R\right)^2}\,, 
	\label{scalar_function_E_sigma_k_8}
	\\
	\widehat{E}_{\left(9\right)} &=& 0\,,  
	\label{scalar_function_E_sigma_k_9}
	\\
	\widehat{E}_{\left(10\right)} &=& + \frac{8}{\left(d_k\right)^6}\,\frac{e_{\left(1\right)}}{R} 
	- \frac{8}{\left(d_k\right)^4}\,\frac{f_{\left(1\right)}}{R} 
	- \frac{327}{128}\,\frac{1}{\left(d_k\right)^4}\,\frac{g_{\left(2\right)}}{R} 
	- \frac{7}{192}\,\frac{1}{\left(d_k\right)^2}\,\frac{g_{\left(4\right)}}{R} 
	+ \frac{13}{48}\,\frac{g_{\left(6\right)}}{R} 
	+ \frac{185}{128}\,\frac{1}{\left(d_k\right)^5}\,\frac{h_{\left(1\right)}}{R} 
	\nonumber\\ 
	&& - \frac{2}{\left(d_k\right)^4}\,\frac{g_{\left(1\right)}\,g_{\left(1\right)}}{\left(R\right)^2} 
	- \frac{8}{\left(d_k\right)^6}\,\frac{g_{\left(1\right)}\,k_{\left(1\right)}}{\left(R\right)^2}\,,  
	\label{scalar_function_E_sigma_k_10}
	\\
	\widehat{E}_{\left(11\right)} &=& - \frac{16}{\left(d_k\right)^8}\,\frac{e_{\left(2\right)}}{R} 
	- \frac{985}{384}\,\frac{1}{\left(d_k\right)^4}\,\frac{f_{\left(2\right)}}{R} 
	- \frac{5}{192}\,\frac{1}{\left(d_k\right)^2}\,\frac{f_{\left(4\right)}}{R} 
	+ \frac{13}{48}\, \frac{f_{\left(6\right)}}{R}  
	+ \frac{8}{\left(d_k\right)^6}\,\frac{g_{\left(1\right)}}{R} 
	+ \frac{985}{128}\,\frac{1}{\left(d_k\right)^6}\,\frac{h_{\left(2\right)}}{R} 
	\nonumber\\ 
	&& + \frac{8}{\left(d_k\right)^6}\,\frac{e_{\left(1\right)}\,g_{\left(1\right)}}{\left(R\right)^2} 
	+ \frac{16}{\left(d_k\right)^8}\,\frac{e_{\left(1\right)}\,k_{\left(1\right)}}{\left(R\right)^2} 
	- \frac{4}{\left(d_k\right)^4}\,\frac{f_{\left(1\right)}\,g_{\left(1\right)}}{\left(R\right)^2}
	- \frac{8}{\left(d_k\right)^6}\,\frac{f_{\left(1\right)}\,k_{\left(1\right)}}{\left(R\right)^2}\,,
	\label{scalar_function_E_sigma_k_11}
	\\
	\widehat{E}_{\left(12\right)} &=& - \frac{8}{\left(d_k\right)^8}\,\frac{e_{\left(1\right)}\,e_{\left(1\right)}}{\left(R\right)^2}  
	- \frac{2}{\left(d_k\right)^4}\,\frac{f_{\left(1\right)}\,f_{\left(1\right)}}{\left(R\right)^2} 
	+ \frac{8}{\left(d_k\right)^6}\,\frac{e_{\left(1\right)}\,f_{\left(1\right)}}{\left(R\right)^2} \,, 
	\label{scalar_function_E_sigma_k_12}
        \\
	\widehat{E}_{\left(13\right)} &=& + \frac{12}{\left(d_k\right)^6}\,\frac{e_{\left(1\right)}}{R} 
	+ \frac{4}{\left(d_k\right)^2}\,\frac{f_{\left(3\right)}}{R} 
	- \frac{2103}{512}\,\frac{1}{\left(d_k\right)^4}\,\frac{g_{\left(2\right)}}{R} 
	+ \frac{451}{256}\,\frac{1}{\left(d_k\right)^2}\,\frac{g_{\left(4\right)}}{R}
	+ \frac{23}{64}\,\frac{g_{\left(6\right)}}{R} 
	+ \frac{9}{32} \left(d_k\right)^2 \frac{g_{\left(8\right)}}{R}
	\nonumber\\ 
	&& - \frac{5175}{512}\,\frac{1}{\left(d_k\right)^5}\,\frac{h_{\left(1\right)}}{R} 
	- \frac{10}{\left(d_k\right)^6}\,\frac{e_{\left(1\right)}\,e_{\left(1\right)}}{\left(R\right)^2}
	+ \frac{10}{\left(d_k\right)^4}\,\frac{e_{\left(1\right)}\,f_{\left(1\right)}}{\left(R\right)^2} 
	- \frac{6}{\left(d_k\right)^2}\,\frac{e_{\left(1\right)}\,f_{\left(3\right)}}{\left(R\right)^2}
	- \frac{5}{2}\,\frac{1}{\left(d_k\right)^2}\,\frac{f_{\left(1\right)}\,f_{\left(1\right)}}{\left(R\right)^2}
	\nonumber\\ 
	&& + 3\,\frac{f_{\left(1\right)}\,f_{\left(3\right)}}{\left(R\right)^2}
	- \frac{1}{2} \left(d_k\right)^2 \frac{f_{\left(3\right)}\,f_{\left(3\right)}}{\left(R\right)^2} 
	+ \frac{4}{\left(d_k\right)^4}\,\frac{g_{\left(1\right)}\,g_{\left(1\right)}}{\left(R\right)^2} 
	+ \frac{2}{\left(d_k\right)^2}\,\frac{g_{\left(1\right)}\,g_{\left(3\right)}}{\left(R\right)^2}
	+ \frac{8}{\left(d_k\right)^6}\,\frac{g_{\left(1\right)}\,k_{\left(1\right)}}{\left(R\right)^2} 
	+ \frac{4}{\left(d_k\right)^4}\,\frac{g_{\left(3\right)}\,k_{\left(1\right)}}{\left(R\right)^2}\,,
	\label{scalar_function_E_sigma_k_13}
        \end{eqnarray}
        \end{widetext} 
	
	\begin{widetext} 
	\begin{eqnarray}
	\widehat{E}_{\left(14\right)} &=& + \frac{12}{\left(d_k\right)^6}\,\frac{g_{\left(1\right)}}{R} 
	- \frac{16}{\left(d_k\right)^4}\,\frac{g_{\left(3\right)}}{R} 
	- \frac{11579}{512}\,\frac{1}{\left(d_k\right)^4}\,\frac{f_{\left(2\right)}}{R} 
	+ \frac{1601}{256}\,\frac{1}{\left(d_k\right)^2}\,\frac{f_{\left(4\right)}}{R} 
	+ \frac{71}{64}\,\frac{f_{\left(6\right)}}{R} 
	+ \frac{27}{32} \left(d_k\right)^2 \frac{f_{\left(8\right)}}{R}
	\nonumber\\
	&& + \frac{945}{512}\,\frac{1}{\left(d_k\right)^6}\,\frac{h_{\left(2\right)}}{R} 
	- \frac{16}{\left(d_k\right)^6}\,\frac{e_{\left(1\right)}\,g_{\left(1\right)}}{\left(R\right)^2}
	+ \frac{16}{\left(d_k\right)^4}\,\frac{e_{\left(1\right)}\,g_{\left(3\right)}}{\left(R\right)^2} 
	+ \frac{8}{\left(d_k\right)^4}\,\frac{f_{\left(1\right)}\,g_{\left(1\right)}}{\left(R\right)^2}
	- \frac{2}{\left(d_k\right)^2}\,\frac{f_{\left(3\right)}\,g_{\left(1\right)}}{\left(R\right)^2}
	+ \frac{20}{\left(d_k\right)^4}\,\frac{f_{\left(3\right)}\,k_{\left(1\right)}}{\left(R\right)^2}
	\nonumber\\
        && - \frac{8}{\left(d_k\right)^2}\,\frac{f_{\left(1\right)}\,g_{\left(3\right)}}{\left(R\right)^2} 
	- 2\,\frac{f_{\left(3\right)}\,g_{\left(3\right)}}{\left(R\right)^2}\,, 
	\label{scalar_function_E_sigma_k_14}
	\\
	\widehat{E}_{\left(15\right)} &=& + \frac{32}{\left(d_k\right)^8}\,\frac{e_{\left(1\right)}}{R} 
	- \frac{8}{\left(d_k\right)^6}\,\frac{f_{\left(1\right)}}{R}
	- \frac{4}{\left(d_k\right)^4}\,\frac{f_{\left(3\right)}}{R}
	- \frac{1419}{512}\,\frac{1}{\left(d_k\right)^6}\,\frac{g_{\left(2\right)}}{R} 
	- \frac{2443}{768}\,\frac{1}{\left(d_k\right)^4}\,\frac{g_{\left(4\right)}}{R} 
	- \frac{143}{192}\,\frac{1}{\left(d_k\right)^2}\,\frac{g_{\left(6\right)}}{R}
	- \frac{9}{32}\,\frac{g_{\left(8\right)}}{R} 
	\nonumber\\
	&& - \frac{5515}{512}\,\frac{1}{\left(d_k\right)^7}\,\frac{h_{\left(1\right)}}{R} 
	- \frac{24}{\left(d_k\right)^8}\,\frac{e_{\left(1\right)}\,e_{\left(1\right)}}{\left(R\right)^2}
	+ \frac{24}{\left(d_k\right)^6}\,\frac{e_{\left(1\right)}\,f_{\left(1\right)}}{\left(R\right)^2} 
	+ \frac{2}{\left(d_k\right)^4}\,\frac{e_{\left(1\right)}\,f_{\left(3\right)}}{\left(R\right)^2}
	- \frac{6}{\left(d_k\right)^4}\,\frac{f_{\left(1\right)}\,f_{\left(1\right)}}{\left(R\right)^2}
	- \frac{1}{\left(d_k\right)^2}\,\frac{f_{\left(1\right)}\,f_{\left(3\right)}}{\left(R\right)^2}
	\nonumber\\ 
	&& + 1\,\frac{f_{\left(3\right)}\,f_{\left(3\right)}}{\left(R\right)^2}
	- \frac{4}{\left(d_k\right)^6}\,\frac{g_{\left(1\right)}\,g_{\left(1\right)}}{\left(R\right)^2}
        + \frac{8}{\left(d_k\right)^4}\,\frac{g_{\left(1\right)}\,g_{\left(3\right)}}{\left(R\right)^2}
        + \frac{2}{\left(d_k\right)^2}\,\frac{g_{\left(3\right)}\,g_{\left(3\right)}}{\left(R\right)^2}
	+ \frac{8}{\left(d_k\right)^8}\,\frac{g_{\left(1\right)}\,k_{\left(1\right)}}{\left(R\right)^2}
	- \frac{4}{\left(d_k\right)^6}\,\frac{g_{\left(3\right)}\,k_{\left(1\right)}}{\left(R\right)^2}\,, 
	\label{scalar_function_E_sigma_k_15}
	\\
	\widehat{E}_{\left(16\right)} &=& - \frac{32}{\left(d_k\right)^8}\,\frac{e_{\left(1\right)}}{R} 
	+ \frac{32}{\left(d_k\right)^6}\,\frac{f_{\left(1\right)}}{R} 
	- \frac{16}{\left(d_k\right)^4}\,\frac{f_{\left(3\right)}}{R}
	+ \frac{6381}{256}\,\frac{1}{\left(d_k\right)^6}\,\frac{g_{\left(2\right)}}{R}
        - \frac{2323}{384}\,\frac{1}{\left(d_k\right)^4}\,\frac{g_{\left(4\right)}}{R}
        - \frac{119}{96}\,\frac{1}{\left(d_k\right)^2}\,\frac{g_{\left(6\right)}}{R}
        - \frac{9}{16}\,\frac{g_{\left(8\right)}}{R}
	\nonumber\\ 
	&& + \frac{2285}{256}\,\frac{1}{\left(d_k\right)^7}\,\frac{h_{\left(1\right)}}{R} 
	+ \frac{40}{\left(d_k\right)^8}\,\frac{e_{\left(1\right)}\,e_{\left(1\right)}}{\left(R\right)^2}
	- \frac{28}{\left(d_k\right)^6}\,\frac{e_{\left(1\right)}\,f_{\left(1\right)}}{\left(R\right)^2} 
	- \frac{8}{\left(d_k\right)^4}\,\frac{e_{\left(1\right)}\,f_{\left(3\right)}}{\left(R\right)^2} 
	+ \frac{10}{\left(d_k\right)^4}\,\frac{f_{\left(1\right)}\,f_{\left(1\right)}}{\left(R\right)^2} 
	+ \frac{4}{\left(d_k\right)^2}\,\frac{f_{\left(1\right)}\,f_{\left(3\right)}}{\left(R\right)^2} 
	\nonumber\\ 
	&& + \frac{8}{\left(d_k\right)^6}\,\frac{g_{\left(1\right)}\,g_{\left(1\right)}}{\left(R\right)^2} 
	+ \frac{4}{\left(d_k\right)^4}\,\frac{g_{\left(1\right)}\,g_{\left(3\right)}}{\left(R\right)^2} 
	- \frac{2}{\left(d_k\right)^2}\,\frac{g_{\left(3\right)}\,g_{\left(3\right)}}{\left(R\right)^2}
	- \frac{16}{\left(d_k\right)^8}\,\frac{g_{\left(1\right)}\,k_{\left(1\right)}}{\left(R\right)^2}
	- \frac{24}{\left(d_k\right)^6}\,\frac{g_{\left(3\right)}\,k_{\left(1\right)}}{\left(R\right)^2}\,,
	\label{scalar_function_E_sigma_k_16}
	\\
	\widehat{E}_{\left(17\right)} &=& + \frac{4831}{512}\,\frac{1}{\left(d_k\right)^6}\,\frac{f_{\left(2\right)}}{R} 
	- \frac{877}{256}\,\frac{1}{\left(d_k\right)^4}\,\frac{f_{\left(4\right)}}{R} 
	- \frac{43}{64}\,\frac{1}{\left(d_k\right)^2}\,\frac{f_{\left(6\right)}}{R}
	- \frac{9}{32}\,\frac{f_{\left(8\right)}}{R} 
	+ \frac{8}{\left(d_k\right)^6}\,\frac{g_{\left(3\right)}}{R}
	- \frac{2205}{512}\,\frac{1}{\left(d_k\right)^8}\,\frac{h_{\left(2\right)}}{R}
	- \frac{16}{\left(d_k\right)^8}\,\frac{e_{\left(1\right)}\,g_{\left(1\right)}}{\left(R\right)^2} 
	\nonumber\\
	&& + \frac{4}{\left(d_k\right)^6}\,\frac{e_{\left(1\right)}\,g_{\left(3\right)}}{\left(R\right)^2}
	+ \frac{8}{\left(d_k\right)^6}\,\frac{f_{\left(1\right)}\,g_{\left(1\right)}}{\left(R\right)^2}
	- \frac{18}{\left(d_k\right)^4}\,\frac{f_{\left(1\right)}\,g_{\left(3\right)}}{\left(R\right)^2}
	+ \frac{14}{\left(d_k\right)^4}\,\frac{f_{\left(3\right)}\,g_{\left(1\right)}}{\left(R\right)^2}
        + \frac{2}{\left(d_k\right)^2}\,\frac{f_{\left(3\right)}\,g_{\left(3\right)}}{\left(R\right)^2}
        - \frac{12}{\left(d_k\right)^6}\,\frac{f_{\left(3\right)}\,k_{\left(1\right)}}{\left(R\right)^2}\,,
	\label{scalar_function_E_sigma_k_17}
	\\
	\widehat{E}_{\left(18\right)} &=& + \frac{4}{\left(d_k\right)^6}\,\frac{e_{\left(1\right)}\,f_{\left(3\right)}}{\left(R\right)^2} 
	- \frac{2}{\left(d_k\right)^4}\,\frac{f_{\left(1\right)}\,f_{\left(3\right)}}{\left(R\right)^2}
	- \frac{1}{2}\,\frac{1}{\left(d_k\right)^2}\,\frac{f_{\left(3\right)}\,f_{\left(3\right)}}{\left(R\right)^2}\,,
	\label{scalar_function_E_sigma_k_18}
	\\
	\widehat{E}_{\left(19\right)} &=& + \frac{5}{384}\,\frac{1}{\left(d_k\right)^4}\,\frac{f_{\left(2\right)}}{R} 
	+ \frac{1}{192}\,\frac{1}{\left(d_k\right)^2}\,\frac{f_{\left(4\right)}}{R}
	- \frac{5}{48}\,\frac{f_{\left(6\right)}}{R} 
	- \frac{5}{128}\,\frac{1}{\left(d_k\right)^6}\,\frac{h_{\left(2\right)}}{R}\,,
	\label{scalar_function_E_sigma_k_19}
	\\
	\widehat{E}_{\left(20\right)} &=& - \frac{8}{\left(d_k\right)^6}\,\frac{g_{\left(1\right)}}{R} 
	+ \frac{3997}{768}\,\frac{1}{\left(d_k\right)^4}\,\frac{f_{\left(2\right)}}{R}
	+ \frac{569}{384}\,\frac{1}{\left(d_k\right)^2}\,\frac{f_{\left(4\right)}}{R}
	+ \frac{95}{96}\,\frac{f_{\left(6\right)}}{R}  
	- \frac{15}{16} \left(d_k\right)^2 \frac{f_{\left(8\right)}}{R} 
	- \frac{925}{256}\,\frac{1}{\left(d_k\right)^6}\,\frac{h_{\left(2\right)}}{R} 
	\nonumber\\ 
	&& + \frac{16}{\left(d_k\right)^6}\,\frac{e_{\left(1\right)}\,g_{\left(1\right)}}{\left(R\right)^2} 
	- \frac{8}{\left(d_k\right)^4}\,\frac{f_{\left(1\right)}\,g_{\left(1\right)}}{\left(R\right)^2}\,,  
	\label{scalar_function_E_sigma_k_20}
	\\
	\widehat{E}_{\left(21\right)} &=& + \frac{80}{\left(d_k\right)^8}\,\frac{e_{\left(1\right)}}{R} 
	- \frac{56}{\left(d_k\right)^6}\,\frac{f_{\left(1\right)}}{R}
	- \frac{3033}{128}\,\frac{1}{\left(d_k\right)^6}\,\frac{g_{\left(2\right)}}{R}
	- \frac{601}{192}\,\frac{1}{\left(d_k\right)^4}\,\frac{g_{\left(4\right)}}{R}
	- \frac{5}{48}\,\frac{1}{\left(d_k\right)^2}\,\frac{g_{\left(6\right)}}{R}
	+ \frac{15}{8}\,\frac{g_{\left(8\right)}}{R}
	- \frac{985}{128}\,\frac{1}{\left(d_k\right)^7}\,\frac{h_{\left(1\right)}}{R}
	\nonumber\\
	&& - \frac{32}{\left(d_k\right)^8}\,\frac{e_{\left(1\right)}\,e_{\left(1\right)}}{\left(R\right)^2} 
	+ \frac{32}{\left(d_k\right)^6}\,\frac{e_{\left(1\right)}\,f_{\left(1\right)}}{\left(R\right)^2} 
	- \frac{8}{\left(d_k\right)^4}\,\frac{f_{\left(1\right)}\,f_{\left(1\right)}}{\left(R\right)^2}
	+ \frac{8}{\left(d_k\right)^6}\,\frac{g_{\left(1\right)}\,g_{\left(1\right)}}{\left(R\right)^2} 
	- \frac{32}{\left(d_k\right)^8}\,\frac{g_{\left(1\right)}\,k_{\left(1\right)}}{\left(R\right)^2}\,,
	\label{scalar_function_E_sigma_k_21}
	\\
	\widehat{E}_{\left(22\right)} &=&  - \frac{64}{\left(d_k\right)^{10}}\,\frac{e_{\left(2\right)}}{R} 
        - \frac{13039}{786}\,\frac{1}{\left(d_k\right)^6}\,\frac{f_{\left(2\right)}}{R}
        - \frac{611}{384}\,\frac{1}{\left(d_k\right)^4}\,\frac{f_{\left(4\right)}}{R}
	- \frac{5}{96}\,\frac{1}{\left(d_k\right)^2}\,\frac{f_{\left(6\right)}}{R}
        + \frac{15}{16}\,\frac{f_{\left(8\right)}}{R}
        + \frac{48}{\left(d_k\right)^8}\,\frac{g_{\left(1\right)}}{R}
	\nonumber\\ 
	&& + \frac{6895}{256}\,\frac{1}{\left(d_k\right)^8}\,\frac{h_{\left(2\right)}}{R}
	- \frac{16}{\left(d_k\right)^8}\,\frac{e_{\left(1\right)}\,g_{\left(1\right)}}{\left(R\right)^2}
	+ \frac{64}{\left(d_k\right)^{10}}\,\frac{e_{\left(1\right)}\,k_{\left(1\right)}}{\left(R\right)^2}
	+ \frac{8}{\left(d_k\right)^6}\,\frac{f_{\left(1\right)}\,g_{\left(1\right)}}{\left(R\right)^2}
	- \frac{32}{\left(d_k\right)^8}\,\frac{f_{\left(1\right)}\,k_{\left(1\right)}}{\left(R\right)^2}\,, 
	\label{scalar_function_E_sigma_k_22}
	\\ 
	\widehat{E}_{\left(23\right)} &=& - \frac{12}{\left(d_k\right)^8}\,\frac{e_{\left(2\right)}}{R} 
	- \frac{31153}{2048}\,\frac{1}{\left(d_k\right)^4}\,\frac{f_{\left(2\right)}}{R}
	+ \frac{2371}{1024}\,\frac{1}{\left(d_k\right)^2}\,\frac{f_{\left(4\right)}}{R}
	+ \frac{37}{256}\,\frac{f_{\left(6\right)}}{R}
	+ \frac{111}{128} \left(d_k\right)^2 \frac{f_{\left(8\right)}}{R} 
	- \frac{15}{16} \left(d_k\right)^4 \frac{f_{\left(10\right)}}{R}
	+ \frac{8}{\left(d_k\right)^6}\,\frac{g_{\left(1\right)}}{R}
	\nonumber\\ 
	&& - \frac{4}{\left(d_k\right)^4}\,\frac{g_{\left(3\right)}}{R} 
	+ \frac{25875}{2048}\,\frac{1}{\left(d_k\right)^6}\,\frac{h_{\left(2\right)}}{R}
	- \frac{14}{\left(d_k\right)^6}\,\frac{e_{\left(1\right)}\,g_{\left(1\right)}}{\left(R\right)^2}
	- \frac{10}{\left(d_k\right)^4}\,\frac{e_{\left(1\right)}\,g_{\left(3\right)}}{\left(R\right)^2}
	+ \frac{12}{\left(d_k\right)^8}\,\frac{e_{\left(1\right)}\,k_{\left(1\right)}}{\left(R\right)^2}
        + \frac{7}{\left(d_k\right)^4}\,\frac{f_{\left(1\right)}\,g_{\left(1\right)}}{\left(R\right)^2} 
	\nonumber\\
	&& - \frac{3}{\left(d_k\right)^2}\,\frac{f_{\left(3\right)}\,g_{\left(1\right)}}{\left(R\right)^2}  
	+ \frac{5}{\left(d_k\right)^2}\,\frac{f_{\left(1\right)}\,g_{\left(3\right)}}{\left(R\right)^2} 
	- 1\,\frac{f_{\left(3\right)}\,g_{\left(3\right)}}{\left(R\right)^2}\,
	- \frac{6}{\left(d_k\right)^6}\,\frac{f_{\left(1\right)}\,k_{\left(1\right)}}{\left(R\right)^2}
	+ \frac{6}{\left(d_k\right)^4}\,\frac{f_{\left(3\right)}\,k_{\left(1\right)}}{\left(R\right)^2}\,,
	\label{scalar_function_E_sigma_k_23}
        \end{eqnarray}
        \end{widetext}

        \begin{widetext}
	\begin{eqnarray}
	\widehat{E}_{\left(24\right)} &=& - \frac{24}{\left(d_k\right)^8}\,\frac{e_{\left(1\right)}}{R} 
	+ \frac{36}{\left(d_k\right)^6}\,\frac{f_{\left(1\right)}}{R}
        + \frac{11343}{512}\,\frac{1}{\left(d_k\right)^6}\,\frac{g_{\left(2\right)}}{R}
        - \frac{59}{256}\,\frac{1}{\left(d_k\right)^4}\,\frac{g_{\left(4\right)}}{R}
        + \frac{65}{64}\,\frac{1}{\left(d_k\right)^2}\,\frac{g_{\left(6\right)}}{R}
        + \frac{9}{32}\,\frac{g_{\left(8\right)}}{R} 
        + \frac{15}{4} \left(d_k\right)^2 \frac{g_{\left(10\right)}}{R}
	\nonumber\\ 
	&& - \frac{945}{512}\,\frac{1}{\left(d_k\right)^7}\,\frac{h_{\left(1\right)}}{R} 
	+ \frac{32}{\left(d_k\right)^8}\,\frac{e_{\left(1\right)}\,e_{\left(1\right)}}{\left(R\right)^2}
	- \frac{32}{\left(d_k\right)^6}\,\frac{e_{\left(1\right)}\,f_{\left(1\right)}}{\left(R\right)^2} 
	- \frac{16}{\left(d_k\right)^4}\,\frac{e_{\left(1\right)}\,f_{\left(3\right)}}{\left(R\right)^2}
	+ \frac{8}{\left(d_k\right)^4}\,\frac{f_{\left(1\right)}\,f_{\left(1\right)}}{\left(R\right)^2}
	+ \frac{8}{\left(d_k\right)^2}\,\frac{f_{\left(1\right)}\,f_{\left(3\right)}}{\left(R\right)^2}
	\nonumber\\
	&& + \frac{4}{\left(d_k\right)^6}\,\frac{g_{\left(1\right)}\,g_{\left(1\right)}}{\left(R\right)^2}
	+ \frac{6}{\left(d_k\right)^4}\,\frac{g_{\left(1\right)}\,g_{\left(3\right)}}{\left(R\right)^2}
        + \frac{2}{\left(d_k\right)^2}\,\frac{g_{\left(3\right)}\,g_{\left(3\right)}}{\left(R\right)^2}
	+ \frac{8}{\left(d_k\right)^8}\,\frac{g_{\left(1\right)}\,k_{\left(1\right)}}{\left(R\right)^2}
	+ \frac{4}{\left(d_k\right)^6}\,\frac{g_{\left(3\right)}\,k_{\left(1\right)}}{\left(R\right)^2}\,, 
	\label{scalar_function_E_sigma_k_24}
	\\
	\widehat{E}_{\left(25\right)} &=& - \frac{64}{\left(d_k\right)^{10}}\,\frac{e_{\left(2\right)}}{R} 
	- \frac{93133}{3072}\,\frac{1}{\left(d_k\right)^6}\,\frac{f_{\left(2\right)}}{R} 
	+ \frac{11479}{1536}\,\frac{1}{\left(d_k\right)^4}\,\frac{f_{\left(4\right)}}{R} 
	+ \frac{433}{384}\,\frac{1}{\left(d_k\right)^2}\,\frac{f_{\left(6\right)}}{R} 
	+ \frac{9}{64}\,\frac{f_{\left(8\right)}}{R} 
	+ \frac{15}{8} \left(d_k\right)^2 \frac{f_{\left(10\right)}}{R}
	\nonumber\\
	&& + \frac{40}{\left(d_k\right)^8}\,\frac{g_{\left(1\right)}}{R} 
	- \frac{20}{\left(d_k\right)^6}\,\frac{g_{\left(3\right)}}{R}
	+ \frac{19405}{1024}\,\frac{1}{\left(d_k\right)^8}\,\frac{h_{\left(2\right)}}{R} 
	- \frac{16}{\left(d_k\right)^8}\,\frac{e_{\left(1\right)}\,g_{\left(1\right)}}{\left(R\right)^2}
	- \frac{2}{\left(d_k\right)^6}\,\frac{e_{\left(1\right)}\,g_{\left(3\right)}}{\left(R\right)^2}
	+ \frac{64}{\left(d_k\right)^{10}}\,\frac{e_{\left(1\right)}\,k_{\left(1\right)}}{\left(R\right)^2}
	\nonumber\\
	&& + \frac{8}{\left(d_k\right)^6}\,\frac{g_{\left(1\right)}\,f_{\left(1\right)}}{\left(R\right)^2}
	- \frac{5}{\left(d_k\right)^4}\,\frac{g_{\left(1\right)}\,f_{\left(3\right)}}{\left(R\right)^2}
	+ \frac{1}{\left(d_k\right)^4}\,\frac{g_{\left(3\right)}\,f_{\left(1\right)}}{\left(R\right)^2}
	- \frac{32}{\left(d_k\right)^8}\,\frac{f_{\left(1\right)}\,k_{\left(1\right)}}{\left(R\right)^2}
	+ \frac{18}{\left(d_k\right)^6}\,\frac{f_{\left(3\right)}\,k_{\left(1\right)}}{\left(R\right)^2}\,,
	\label{scalar_function_E_sigma_k_25}
	\\
	\widehat{E}_{\left(26\right)} &=& + \frac{64}{\left(d_k\right)^{10}}\,\frac{e_{\left(2\right)}}{R} 
        - \frac{5893}{1536}\,\frac{1}{\left(d_k\right)^6}\,\frac{f_{\left(2\right)}}{R}
        + \frac{5887}{768}\,\frac{1}{\left(d_k\right)^4}\,\frac{f_{\left(4\right)}}{R}
        + \frac{457}{192}\,\frac{1}{\left(d_k\right)^2}\,\frac{f_{\left(6\right)}}{R}
        + \frac{9}{32}\,\frac{f_{\left(8\right)}}{R}
        + \frac{15}{4} \left(d_k\right)^2 \frac{f_{\left(10\right)}}{R}
        \nonumber\\
        && - \frac{48}{\left(d_k\right)^8}\,\frac{g_{\left(1\right)}}{R}
        - \frac{16}{\left(d_k\right)^6}\,\frac{g_{\left(3\right)}}{R}
        - \frac{6395}{512}\,\frac{1}{\left(d_k\right)^8}\,\frac{h_{\left(2\right)}}{R}
	- \frac{16}{\left(d_k\right)^8}\,\frac{e_{\left(1\right)}\,g_{\left(1\right)}}{\left(R\right)^2} 
	- \frac{64}{\left(d_k\right)^{10}}\,\frac{e_{\left(1\right)}\,k_{\left(1\right)}}{\left(R\right)^2}
	+ \frac{8}{\left(d_k\right)^6}\,\frac{g_{\left(1\right)}\,f_{\left(1\right)}}{\left(R\right)^2}
	\nonumber\\
        && - \frac{4}{\left(d_k\right)^4}\,\frac{g_{\left(1\right)}\,f_{\left(3\right)}}{\left(R\right)^2} 
        + \frac{32}{\left(d_k\right)^8}\,\frac{f_{\left(1\right)}\,k_{\left(1\right)}}{\left(R\right)^2}
        + \frac{32}{\left(d_k\right)^6}\,\frac{f_{\left(3\right)}\,k_{\left(1\right)}}{\left(R\right)^2}\,,
	\label{scalar_function_E_sigma_k_26}
	\\
	\widehat{E}_{\left(27\right)} &=& + \frac{48}{\left(d_k\right)^8}\,\frac{f_{\left(1\right)}}{R} 
	- \frac{56}{\left(d_k\right)^6}\,\frac{f_{\left(3\right)}}{R} 
	+ \frac{26781}{512}\,\frac{1}{\left(d_k\right)^8}\,\frac{g_{\left(2\right)}}{R}
	- \frac{7457}{256}\,\frac{1}{\left(d_k\right)^6}\,\frac{g_{\left(4\right)}}{R}
	- \frac{493}{64}\,\frac{1}{\left(d_k\right)^4}\,\frac{g_{\left(6\right)}}{R}
	- \frac{129}{32}\,\frac{1}{\left(d_k\right)^2}\,\frac{g_{\left(8\right)}}{R}
	\nonumber\\ 
	&& - \frac{15}{4}\,\frac{g_{\left(10\right)}}{R}
	+ \frac{2205}{512}\,\frac{1}{\left(d_k\right)^9}\,\frac{h_{\left(1\right)}}{R} 
        + \frac{64}{\left(d_k\right)^{10}}\,\frac{e_{\left(1\right)}\,e_{\left(1\right)}}{\left(R\right)^2}
	- \frac{64}{\left(d_k\right)^8}\,\frac{e_{\left(1\right)}\,f_{\left(1\right)}}{\left(R\right)^2}
        - \frac{8}{\left(d_k\right)^6}\,\frac{e_{\left(1\right)}\,f_{\left(3\right)}}{\left(R\right)^2}
        + \frac{16}{\left(d_k\right)^6}\,\frac{f_{\left(1\right)}\,f_{\left(1\right)}}{\left(R\right)^2}
	\nonumber\\
	&& + \frac{4}{\left(d_k\right)^4}\,\frac{f_{\left(1\right)}\,f_{\left(3\right)}}{\left(R\right)^2} 
	+ \frac{8}{\left(d_k\right)^8}\,\frac{g_{\left(1\right)}\,g_{\left(1\right)}}{\left(R\right)^2}
	+ \frac{20}{\left(d_k\right)^6}\,\frac{g_{\left(1\right)}\,g_{\left(3\right)}}{\left(R\right)^2}
	+ \frac{2}{\left(d_k\right)^4}\,\frac{g_{\left(3\right)}\,g_{\left(3\right)}}{\left(R\right)^2}
	- \frac{32}{\left(d_k\right)^{10}}\,\frac{g_{\left(1\right)}\,k_{\left(1\right)}}{\left(R\right)^2}
	- \frac{64}{\left(d_k\right)^8}\,\frac{g_{\left(3\right)}\,k_{\left(1\right)}}{\left(R\right)^2}\,, 
	\label{scalar_function_E_sigma_k_27}
	\\
	\widehat{E}_{\left(28\right)} &=& + \frac{55767}{2048}\,\frac{1}{\left(d_k\right)^8}\,\frac{f_{\left(2\right)}}{R} 
	- \frac{10965}{1024}\,\frac{1}{\left(d_k\right)^6}\,\frac{f_{\left(4\right)}}{R} 
	- \frac{579}{256}\,\frac{1}{\left(d_k\right)^4}\,\frac{f_{\left(6\right)}}{R}
	- \frac{129}{128}\,\frac{1}{\left(d_k\right)^2}\,\frac{f_{\left(8\right)}}{R}
	- \frac{15}{16}\,\frac{f_{\left(10\right)}}{R} 
	+ \frac{24}{\left(d_k\right)^8}\,\frac{g_{\left(3\right)}}{R}
	\nonumber\\ 
	&& - \frac{19845}{2048}\,\frac{1}{\left(d_k\right)^{10}}\,\frac{h_{\left(2\right)}}{R} 
	+ \frac{8}{\left(d_k\right)^6}\,\frac{g_{\left(1\right)}\,f_{\left(3\right)}}{\left(R\right)^2} 
	+ \frac{1}{\left(d_k\right)^4}\,\frac{g_{\left(3\right)}\,f_{\left(3\right)}}{\left(R\right)^2} 
	- \frac{24}{\left(d_k\right)^8}\,\frac{f_{\left(3\right)}\,k_{\left(1\right)}}{\left(R\right)^2}\,. 
	\label{scalar_function_E_sigma_k_28}
        \end{eqnarray}
        \end{widetext}

\section{The scalar functions in Eq.~(\ref{transformation_sigma_to_n})}\label{Appendix9}

In order to simplify the notations, the following abbreviations are introduced,  
\begin{eqnarray}
	\widehat{e}_{\left(n\right)} &=& \left(x_1 + \ve{k} \cdot \ve{x}_1\right)^n\;,
        \label{hat_e_n}
        \\
	\widehat{f}_{\left(n\right)} &=& \frac{1}{\left(x_1\right)^n} \;,
        \label{hat_f_n}
        \\
	\widehat{g}_{\left(n\right)} &=& \frac{\ve{k} \cdot \ve{x}_1}{\left(x_1\right)^n}\;,
        \label{hat_g_n}
        \\
	\widehat{h}_{\left(1\right)} &=& \arctan \frac{\ve{k} \cdot \ve{x}_1}{d_k}\;, 
        \label{hat_h_1}
        \\
	\widehat{h}_{\left(2\right)} &=& \arctan \frac{\ve{k} \cdot \ve{x}_1}{d_k} + \frac{\pi}{2}\;, 
        \label{hat_h_2}
        \\
	\widehat{h}_{\left(3\right)} &=&
        + \frac{\ve{k} \cdot \ve{x}_1}{d_k} \left(\arctan \frac{\ve{k} \cdot \ve{x}_1}{d_k} + \frac{\pi}{2}\right), 
        \label{hat_h_3}
        \\
	\widehat{k}_{\left(1\right)} &=& \left(\ve{k} \cdot \ve{x}_1\right) \left(x_1 + \ve{k} \cdot \ve{x}_1 \right). 
        \label{hat_k_2}
\end{eqnarray}

\noindent 
Then, the scalar functions in Eq.~(\ref{transformation_sigma_to_n}) read as follows: 
\begin{eqnarray}
        \widehat{H}_{\left(1\right)} &=& 0\,,  
        \label{scalar_function_H_n_sigma_1}
        \\
	\widehat{H}_{\left(2\right)} &=& - \frac{2}{\left(d_k\right)^2} \left(1 + \widehat{g}_{\left(1\right)} \right)\,,  
        \label{scalar_function_H_n_sigma_2}
        \\
        \widehat{I}_{\left(1\right)} &=& - 2\,\widehat{f}_{\left(3\right)}\,,  
        \label{scalar_function_I_n_sigma_1}
        \\
	\widehat{I}_{\left(2\right)} &=& + \frac{4}{\left(d_k\right)^4}\left(1 + \widehat{g}_{\left(1\right)}\right) + \frac{2}{\left(d_k\right)^2}\,\widehat{g}_{\left(3\right)}\,,  
        \label{scalar_function_I_n_sigma_2}
        \\
        \widehat{I}_{\left(3\right)} &=& + 2\,\widehat{f}_{\left(3\right)}\,,  
        \label{scalar_function_I_n_sigma_3}
        \\
        \widehat{I}_{\left(4\right)} &=& - \frac{4}{\left(d_k\right)^4}\left(1 + \widehat{g}_{\left(1\right)}\right) - \frac{2}{\left(d_k\right)^2}\,\widehat{g}_{\left(3\right)}\,,  
        \label{scalar_function_I_n_sigma_4}
        \\
        \widehat{I}_{\left(5\right)} &=& 0 \,,  
        \label{scalar_function_I_n_sigma_5} 
        \\
        \widehat{I}_{\left(6\right)} &=& - \frac{8}{\left(d_k\right)^6}\left(1 + \widehat{g}_{\left(1\right)}\right) - \frac{4}{\left(d_k\right)^4}\,\widehat{g}_{\left(3\right)} 
	- \frac{3}{\left(d_k\right)^2}\,\widehat{g}_{\left(5\right)}\,,
	\nonumber\\ 
        \label{scalar_function_I_n_sigma_6}
	\\
        \widehat{I}_{\left(7\right)} &=& - \frac{2}{\left(d_k\right)^4}\left(1 + \widehat{g}_{\left(1\right)}\right) - \frac{1}{\left(d_k\right)^2}\,\widehat{g}_{\left(3\right)} 
        + 3\,\widehat{g}_{\left(5\right)}\,,
        \label{scalar_function_I_n_sigma_7}
        \\
        \widehat{I}_{\left(8\right)} &=& + 6\,\widehat{f}_{\left(5\right)}\,.
        \label{scalar_function_I_n_sigma_8}
\end{eqnarray}

\noindent
For the scalar functions in 2PN terms, we need the abbreviations defined by Eqs.~(\ref{hat_e_n}) - (\ref{hat_k_2}) in this Appendix~\ref{Appendix9} 
as well as the abbreviations defined above by Eqs.~(\ref{e_n}) - (\ref{k_2}) of Appendix~\ref{Appendix8}. One obtains 
        \begin{widetext} 
        \begin{eqnarray}
	\widehat{J}_{\left(1\right)} &=& - \frac{2}{\left(d_k\right)^2} \left(1 + \widehat{g}_{\left(1\right)}\right) 
	\left(1 + \widehat{g}_{\left(1\right)} - \frac{2}{R}\,e_{\left(1\right)} \right), 
        \label{scalar_function_J_n_sigma_1}
        \\
	\widehat{J}_{\left(2\right)} &=& - \frac{4}{\left(d_k\right)^4}\,\frac{\widehat{e}_{\left(1\right)}\,e_{\left(1\right)}}{R} 
	+ \frac{8}{\left(d_k\right)^4}\,\widehat{e}_{\left(1\right)} 
	+ \frac{1}{4}\,\frac{\widehat{g}_{\left(2\right)}}{\left(d_k\right)^2} 
	- \frac{\widehat{g}_{\left(4\right)}}{2} 
	- \frac{15}{4}\,\frac{\widehat{h}_{(2)}}{\left(d_k\right)^3}\,, 
	\nonumber\\ 
        \label{scalar_function_J_n_sigma_2}
        \\
	\widehat{K}_{\left(1\right)} &=& + \frac{4}{\left(d_k\right)^4}\,\frac{\widehat{e}_{\left(1\right)}\,g_{\left(1\right)}}{R}  
	- \frac{4}{\left(d_k\right)^2}\,\frac{\widehat{f}_{\left(1\right)}\,g_{\left(1\right)}}{R} 
	- \frac{4}{\left(d_k\right)^4}\,\frac{e_{\left(1\right)}}{R} \left(1 + \widehat{g}_{\left(1\right)}\right) 
	- \frac{4}{\left(d_k\right)^2}\,\frac{\widehat{g}_{\left(3\right)}\,e_{\left(1\right)}}{R} 
	+ \frac{4}{\left(d_k\right)^4}\,\widehat{g}_{\left(1\right)} 
	+ \frac{4}{\left(d_k\right)^2}\,\widehat{g}_{\left(3\right)}
	\nonumber\\ 
	&& + \frac{2}{\left(d_k\right)^2}\,\widehat{f}_{\left(2\right)}
	+ \frac{7}{2}\,\widehat{f}_{\left(4\right)} 
	- \frac{7}{4} \left(d_k\right)^2 \widehat{f}_{\left(6\right)}\,,
        \label{scalar_function_K_n_sigma_1}
        \\
        \widehat{K}_{\left(2\right)} &=& + \frac{16}{\left(d_k\right)^6}\,\frac{\widehat{e}_{\left(1\right)}\,e_{\left(1\right)}}{R} 
	- \frac{4}{\left(d_k\right)^4}\,\frac{\widehat{f}_{\left(1\right)}\,e_{\left(1\right)}}{R} 
	+ \frac{4}{\left(d_k\right)^4}\,\frac{\widehat{e}_{\left(1\right)}\,f_{\left(1\right)}}{R} 
	- \frac{4}{\left(d_k\right)^2}\,\frac{\widehat{f}_{\left(3\right)}\,e_{\left(1\right)}}{R} 
	- \frac{4}{\left(d_k\right)^2}\,\frac{\widehat{f}_{\left(1\right)}\,f_{\left(1\right)}}{R}
	- \frac{32}{\left(d_k\right)^6}\,\widehat{e}_{\left(1\right)} 
	\nonumber\\ 
	&& + \frac{4}{\left(d_k\right)^2}\,\widehat{f}_{\left(3\right)} 
	- \frac{47}{32}\,\frac{\widehat{g}_{\left(2\right)}}{\left(d_k\right)^4} 
	+ \frac{27}{16}\,\frac{\widehat{g}_{\left(4\right)}}{\left(d_k\right)^2} 
	+ \frac{7}{4}\,\widehat{g}_{\left(6\right)} 
	+ \frac{465}{32}\,\frac{\widehat{h}_{(2)}}{\left(d_k\right)^5}\,,
        \label{scalar_function_K_n_sigma_2}
        \\
        \widehat{K}_{\left(3\right)} &=& - \frac{4}{\left(d_k\right)^4}\,\frac{\widehat{e}_{\left(1\right)}\,g_{\left(1\right)}}{R} 
	+ \frac{4}{\left(d_k\right)^2}\,\frac{\widehat{f}_{\left(1\right)}\,g_{\left(1\right)}}{R} 
	+ \frac{16}{\left(d_k\right)^4} \left(1 + \widehat{g}_{\left(1\right)}\right) \frac{e_{\left(1\right)}}{R} 
	- \frac{2}{\left(d_k\right)^2} \left(1 + \widehat{g}_{\left(1\right)}\right) \frac{f_{\left(1\right)}}{R} 
	+ 2 \left(1 + \widehat{g}_{\left(1\right)}\right) \frac{f_{\left(3\right)}}{R} 
	\nonumber\\ 
	&& + \frac{6}{\left(d_k\right)^2}\,\frac{\widehat{g}_{\left(3\right)}\,e_{\left(1\right)}}{R}
	- 6\,\frac{\widehat{g}_{\left(5\right)}\,e_{\left(1\right)}}{R} 
	- \frac{12}{\left(d_k\right)^4} \left(1 + \widehat{g}_{\left(1\right)}\right) 
	- \frac{6}{\left(d_k\right)^2}\,\widehat{g}_{\left(3\right)}
	+ 6\,\widehat{g}_{\left(5\right)} 
	+ \frac{17}{2}\,\widehat{f}_{\left(4\right)} 
	- \frac{17}{4} \left(d_k\right)^2 \widehat{f}_{\left(6\right)}\,, 
        \label{scalar_function_K_n_sigma_3}
        \\
	\widehat{K}_{\left(4\right)} &=& - \frac{4}{\left(d_k\right)^4} \left(1 + \widehat{g}_{\left(1\right)}\right) \frac{g_{\left(1\right)}}{R}
	- \frac{4}{\left(d_k\right)^2} \left(1 + \widehat{g}_{\left(1\right)}\right) \frac{g_{\left(3\right)}}{R} 
	- \frac{16}{\left(d_k\right)^6}\,\frac{\widehat{e}_{\left(1\right)}\,e_{\left(1\right)}}{R} 
	-  \frac{4}{\left(d_k\right)^4}\,\frac{\widehat{e}_{\left(1\right)}\,f_{\left(1\right)}}{R} 
	+ \frac{4}{\left(d_k\right)^4}\,\frac{\widehat{f}_{\left(1\right)}\,e_{\left(1\right)}}{R} 
	\nonumber\\
	&& + \frac{4}{\left(d_k\right)^2}\,\frac{\widehat{f}_{\left(1\right)}\,f_{\left(1\right)}}{R} 
	+ \frac{8}{\left(d_k\right)^2}\,\frac{\widehat{f}_{\left(3\right)}\,e_{\left(1\right)}}{R} 
	- 12\,\frac{\widehat{f}_{\left(5\right)}\,e_{\left(1\right)}}{R} 
	+ \frac{32}{\left(d_k\right)^6}\,\widehat{e}_{\left(1\right)} 
	- \frac{8}{\left(d_k\right)^2}\,\widehat{f}_{\left(3\right)} 
	+ 12\,\widehat{f}_{\left(5\right)} 
	+ \frac{47}{32}\,\frac{\widehat{g}_{\left(2\right)}}{\left(d_k\right)^4} 
	\nonumber\\ 
	&& - \frac{91}{16}\,\frac{\widehat{g}_{\left(4\right)}}{\left(d_k\right)^2} 
	+ \frac{41}{4}\,\widehat{g}_{\left(6\right)} 
	- \frac{465}{32}\,\frac{\widehat{h}_{(2)}}{\left(d_k\right)^5}\,, 
        \label{scalar_function_K_n_sigma_4}
        \\
        \widehat{K}_{\left(5\right)} &=& + \frac{8}{\left(d_k\right)^6} \left(1 + \widehat{g}_{\left(1\right)}\right) \frac{e_{\left(1\right)}}{R} 
	- \frac{4}{\left(d_k\right)^4} \left(1 + \widehat{g}_{\left(1\right)}\right) \frac{f_{\left(1\right)}}{R}
	- \frac{2}{\left(d_k\right)^2} \left(1 + \widehat{g}_{\left(1\right)}\right) \frac{f_{\left(3\right)}}{R} 
	+ \frac{4}{\left(d_k\right)^4}\,\frac{\widehat{g}_{\left(3\right)}\,e_{\left(1\right)}}{R}
	\nonumber\\ 
	&& + \frac{6}{\left(d_k\right)^2}\,\frac{\widehat{g}_{\left(5\right)}\,e_{\left(1\right)}}{R} 
	- \frac{16}{\left(d_k\right)^6} \left(1 + \widehat{g}_{\left(1\right)}\right) 
	- \frac{4}{\left(d_k\right)^4}\,\widehat{g}_{\left(3\right)}
	- \frac{6}{\left(d_k\right)^2}\,\widehat{g}_{\left(5\right)}
	+ \frac{4}{\left(d_k\right)^4}\,\widehat{f}_{\left(2\right)}
	- \frac{2}{\left(d_k\right)^2}\,\widehat{f}_{\left(4\right)}
	+ 6\,\widehat{f}_{\left(6\right)}\,, 
        \label{scalar_function_K_n_sigma_5}
        \\
	\widehat{K}_{\left(6\right)} &=& + \frac{16}{\left(d_k\right)^6}\,\frac{\widehat{f}_{\left(1\right)}\,e_{\left(1\right)}}{R} 
	+ \frac{4}{\left(d_k\right)^4}\,\frac{\widehat{f}_{\left(1\right)}\,f_{\left(1\right)}}{R} 
	- \frac{48}{\left(d_k\right)^8}\,\frac{\widehat{e}_{\left(1\right)}\,e_{\left(1\right)}}{R}
	+ \frac{2}{\left(d_k\right)^4}\,\frac{\widehat{e}_{\left(1\right)}\,f_{\left(3\right)}}{R}
	+ \frac{6}{\left(d_k\right)^4}\,\frac{\widehat{f}_{\left(3\right)}\,e_{\left(1\right)}}{R} 
	+ \frac{96}{\left(d_k\right)^8}\,\widehat{e}_{\left(1\right)}
	\nonumber\\
	&& - \frac{16}{\left(d_k\right)^6}\,\widehat{f}_{\left(1\right)} 
	- \frac{8}{\left(d_k\right)^4}\,\widehat{f}_{\left(3\right)}
	- \frac{277}{16}\,\frac{1}{\left(d_k\right)^6}\,\widehat{g}_{\left(2\right)}
	- \frac{135}{32}\,\frac{1}{\left(d_k\right)^4}\,\widehat{g}_{\left(4\right)}
	+ \frac{5}{8}\,\frac{1}{\left(d_k\right)^2}\,\widehat{g}_{\left(6\right)}
	- \frac{15}{4}\,\widehat{g}_{\left(8\right)} 
	- \frac{2325}{64}\,\frac{1}{\left(d_k\right)^7}\,\widehat{h}_{(2)}\,,
        \label{scalar_function_K_n_sigma_6}
        \\
        \widehat{K}_{\left(7\right)} &=& - \frac{16}{\left(d_k\right)^6}\,\frac{\widehat{e}_{\left(1\right)}\,e_{\left(1\right)}}{R} 
	+ \frac{2}{\left(d_k\right)^4}\,\frac{\widehat{e}_{\left(1\right)}\,f_{\left(1\right)}}{R} 
	- \frac{2}{\left(d_k\right)^2}\,\frac{\widehat{e}_{\left(1\right)}\,f_{\left(3\right)}}{R} 
	- \frac{2}{\left(d_k\right)^2}\,\frac{\widehat{f}_{\left(3\right)}\,e_{\left(1\right)}}{R} 
	+ \frac{6}{\left(d_k\right)^4}\,\frac{\widehat{f}_{\left(1\right)}\,e_{\left(1\right)}}{R} 
	+ \frac{32}{\left(d_k\right)^6}\,\widehat{e}_{\left(1\right)}
	\nonumber\\
	&& - \frac{8}{\left(d_k\right)^4}\,\widehat{f}_{\left(1\right)} 
	+ \frac{4}{\left(d_k\right)^2}\,\widehat{f}_{\left(3\right)} 
	- \frac{343}{64}\,\frac{1}{\left(d_k\right)^4}\,\widehat{g}_{\left(2\right)}
	+ \frac{99}{32}\,\frac{1}{\left(d_k\right)^2}\,\widehat{g}_{\left(4\right)} 
	- \frac{25}{8}\,\widehat{g}_{\left(6\right)} 
	+ \frac{15}{4} \left(d_k\right)^2 \widehat{g}_{\left(8\right)}
	- \frac{855}{64}\,\frac{1}{\left(d_k\right)^5}\,\widehat{h}_{(2)}\,,
        \label{scalar_function_K_n_sigma_7}
        \\
        \widehat{K}_{\left(8\right)} &=& + \frac{4}{\left(d_k\right)^4}\,\frac{\widehat{f}_{\left(1\right)}\,g_{\left(1\right)}}{R} 
	+ \frac{4}{\left(d_k\right)^4}\,\frac{\widehat{e}_{\left(1\right)}\,g_{\left(3\right)}}{R} 
	+ \frac{8}{\left(d_k\right)^4}\,\frac{\widehat{g}_{\left(3\right)}\,e_{\left(1\right)}}{R} 
	+ \frac{24}{\left(d_k\right)^6}\left(1 + \widehat{g}_{\left(1\right)}\right) \frac{e_{\left(1\right)}}{R} 
	- \frac{16}{\left(d_k\right)^6}\left(1 + \widehat{g}_{\left(1\right)}\right) 
	- \frac{4}{\left(d_k\right)^4}\,\widehat{f}_{\left(2\right)}
	\nonumber\\ 
	&& + \frac{8}{\left(d_k\right)^2}\,\widehat{f}_{\left(4\right)} 
        - \frac{15}{2}\,\widehat{f}_{\left(6\right)} 
        + \frac{15}{2} \left(d_k\right)^2 \widehat{f}_{\left(8\right)} 
	- \frac{12}{\left(d_k\right)^4}\,\widehat{g}_{\left(3\right)}\,. 
        \label{scalar_function_K_n_sigma_8}
       \end{eqnarray}
       \end{widetext}

        \begin{widetext}
        \begin{eqnarray}
	\widehat{L}_{\left(1\right)} &=& - \frac{8}{\left(d_k\right)^6}\,\frac{\widehat{e}_{\left(1\right)}\,g_{\left(1\right)}}{R} 
	+ \frac{4}{\left(d_k\right)^4}\,\frac{\widehat{f}_{\left(1\right)}\,g_{\left(1\right)}}{R} 
	+ \frac{8}{\left(d_k\right)^6} \left(1 + \widehat{g}_{(1)}\right) 
        - \frac{4}{\left(d_k\right)^4}\,\widehat{f}_{(2)} - \frac{\widehat{f}_{(4)}}{\left(d_k\right)^2} 
	- \frac{3}{2}\,\widehat{f}_{(6)} + \frac{13}{8}  \left(d_k\right)^2 \widehat{f}_{(8)}\,, 
        \label{scalar_function_L_n_sigma_1}
        \\
	\widehat{L}_{\left(2\right)} &=& + \frac{16}{\left(d_k\right)^8}\,\frac{\widehat{e}_{\left(1\right)}\,e_{\left(1\right)}}{R} 
	- \frac{8}{\left(d_k\right)^6}\,\frac{\widehat{e}_{\left(1\right)}\,f_{\left(1\right)}}{R} 
	- \frac{8}{\left(d_k\right)^6}\,\frac{\widehat{f}_{\left(1\right)}\,e_{\left(1\right)}}{R}
	+ \frac{4}{\left(d_k\right)^4}\,\frac{\widehat{f}_{\left(1\right)}\,f_{\left(1\right)}}{R}
	- \frac{32}{\left(d_k\right)^8}\,\widehat{e}_{\left(1\right)} + \frac{16}{\left(d_k\right)^6}\,\widehat{f}_{\left(1\right)}
	\nonumber\\
	&& + \frac{985}{128}\,\frac{\widehat{g}_{\left(2\right)}}{\left(d_k\right)^6}  + \frac{217}{192}\,\frac{\widehat{g}_{\left(4\right)}}{\left(d_k\right)^4} 
        + \frac{5}{48}\,\frac{\widehat{g}_{\left(6\right)}}{\left(d_k\right)^2} - \frac{13}{8}\,\widehat{g}_{\left(8\right)} 
	+ \frac{985}{128}\,\frac{\widehat{h}_{(2)}}{\left(d_k\right)^7}\,,  
        \label{scalar_function_L_n_sigma_2}
        \\
	\widehat{L}_{\left(3\right)} &=& + \frac{12}{\left(d_k\right)^6}\,\frac{\widehat{e}_{\left(1\right)}\,g_{\left(1\right)}}{R}  
	- \frac{6}{\left(d_k\right)^4}\,\frac{\widehat{f}_{\left(1\right)}\,g_{\left(1\right)}}{R}
	- \frac{4}{\left(d_k\right)^2}\,\frac{\widehat{f}_{\left(3\right)}\,g_{\left(1\right)}}{R}
	+ 6\,\frac{\widehat{f}_{\left(5\right)}\,g_{\left(1\right)}}{R} 
	- \frac{8}{\left(d_k\right)^6} \left(1 + \widehat{g}_{(1)}\right)\,\frac{e_{\left(1\right)}}{R}
	\nonumber\\ 
	&& - \frac{4}{\left(d_k\right)^4} \frac{\widehat{g}_{\left(3\right)}\,e_{\left(1\right)}}{R}
	+ \frac{4}{\left(d_k\right)^4} \left(1 + \widehat{g}_{(1)}\right)\,\frac{f_{\left(1\right)}}{R} 
	- \frac{4}{\left(d_k\right)^2} \left(1 + \widehat{g}_{(1)}\right)\,\frac{f_{\left(3\right)}}{R} 
        + \frac{2}{\left(d_k\right)^2} \frac{\widehat{g}_{\left(3\right)}\,f_{\left(1\right)}}{R} 
        - 2\,\frac{\widehat{g}_{\left(3\right)}\,f_{\left(3\right)}}{R}
	\nonumber\\
	&& - \frac{4}{\left(d_k\right)^6} \left(1 + \widehat{g}_{(1)}\right) 
        + \frac{6}{\left(d_k\right)^4}\,\widehat{f}_{(2)} - \frac{3}{2}\,\frac{\widehat{f}_{(4)}}{\left(d_k\right)^2} 
        + \frac{13}{4}\,\widehat{f}_{(6)} - \frac{93}{16} \left(d_k\right)^2 \widehat{f}_{(8)}
	+ \frac{9}{4} \left(d_k\right)^4 \widehat{f}_{(10)} + \frac{4}{\left(d_k\right)^4} \,\widehat{g}_{(3)}\,,  
        \label{scalar_function_L_n_sigma_3}
        \\
        \widehat{L}_{\left(4\right)} &=& + \frac{16}{\left(d_k\right)^8}\,\frac{\widehat{e}_{\left(1\right)}\,e_{\left(1\right)}}{R} 
	- \frac{8}{\left(d_k\right)^6}\,\frac{\widehat{f}_{\left(1\right)}\,e_{\left(1\right)}}{R}
	+ \frac{4}{\left(d_k\right)^4}\,\frac{\widehat{f}_{\left(3\right)}\,e_{\left(1\right)}}{R} 
	- \frac{12}{\left(d_k\right)^2}\,\frac{\widehat{f}_{\left(5\right)}\,e_{\left(1\right)}}{R} 
	- \frac{8}{\left(d_k\right)^6}\,\frac{\widehat{e}_{\left(1\right)}\,f_{\left(1\right)}}{R} 
	\nonumber\\
	&& + \frac{4}{\left(d_k\right)^4}\,\frac{\widehat{f}_{\left(1\right)}\,f_{\left(1\right)}}{R} 
	- \frac{2}{\left(d_k\right)^2}\,\frac{\widehat{f}_{\left(3\right)}\,f_{\left(1\right)}}{R} 
	+ 6\,\frac{\widehat{f}_{\left(5\right)}\,f_{\left(1\right)}}{R} 
	+ \frac{12}{\left(d_k\right)^4}\,\frac{\widehat{e}_{\left(1\right)}\,f_{\left(3\right)}}{R} 
	- \frac{6}{\left(d_k\right)^2}\,\frac{\widehat{f}_{\left(1\right)}\,f_{\left(3\right)}}{R} 
	- 2\,\frac{\widehat{f}_{\left(3\right)}\,f_{\left(3\right)}}{R} 
	\nonumber\\ 
	&& - \frac{32}{\left(d_k\right)^8}\,\widehat{e}_{(1)}
	+ \frac{16}{\left(d_k\right)^6}\,\widehat{f}_{(1)} - \frac{16}{\left(d_k\right)^4}\,\widehat{f}_{(3)}
        + \frac{12}{\left(d_k\right)^2}\,\widehat{f}_{(5)} + \frac{5515}{512}\,\frac{\widehat{g}_{(2)}}{\left(d_k\right)^6}
        - \frac{9845}{768} \frac{\widehat{g}_{(4)}}{\left(d_k\right)^4} + \frac{1103}{192}\,\frac{\widehat{g}_{(6)}}{\left(d_k\right)^2}
        \nonumber\\
        && + \frac{137}{32}\,\widehat{g}_{(8)} - \frac{9}{4} \left(d_k\right)^2 \widehat{g}_{(10)}
        + \frac{5515}{512}\,\frac{\widehat{h}_{(2)}}{\left(d_k\right)^7}\,,
        \label{scalar_function_L_n_sigma_4}
        \\
        \widehat{L}_{\left(5\right)} &=& - \frac{16}{\left(d_k\right)^8}\,\frac{\widehat{e}_{\left(1\right)}\,e_{\left(1\right)}}{R} 
	+ \frac{8}{\left(d_k\right)^6}\,\frac{\widehat{f}_{\left(1\right)}\,e_{\left(1\right)}}{R} 
	+ \frac{8}{\left(d_k\right)^6}\,\frac{\widehat{e}_{\left(1\right)}\,f_{\left(1\right)}}{R} 
	- \frac{4}{\left(d_k\right)^4}\,\frac{\widehat{f}_{\left(1\right)}\,f_{\left(1\right)}}{R} 
	- \frac{12}{\left(d_k\right)^2}\,\frac{\widehat{g}_{\left(5\right)}\,g_{\left(1\right)}}{R} 
        \nonumber\\
	&& + \frac{8}{\left(d_k\right)^4}\left(1 + \widehat{g}_{\left(1\right)}\right) \frac{g_{\left(3\right)}}{R}
	+ \frac{4}{\left(d_k\right)^2}\,\frac{\widehat{g}_{\left(3\right)}\,g_{\left(3\right)}}{R}
	+ \frac{32}{\left(d_k\right)^8}\,\widehat{e}_{(1)} 
        - \frac{16}{\left(d_k\right)^6}\,\widehat{f}_{(1)} - \frac{2285}{256}\,\frac{\widehat{g}_{(2)}}{\left(d_k\right)^6} 
        - \frac{749}{384}\,\frac{\widehat{g}_{(4)}}{\left(d_k\right)^4} 
	\nonumber\\ 
	&& - \frac{73}{96}\,\frac{\widehat{g}_{(6)}}{\left(d_k\right)^2}
        + \frac{113}{16}\,\widehat{g}_{(8)} - \frac{9}{2} \left(d_k\right)^2 \widehat{g}_{(10)}
        - \frac{2285}{256}\,\frac{\widehat{h}_{(2)}}{\left(d_k\right)^7}\,,
        \label{scalar_function_L_n_sigma_5}
        \\
        \widehat{L}_{\left(6\right)} &=& - \frac{16}{\left(d_k\right)^8}\,\frac{\widehat{e}_{\left(1\right)}\,g_{\left(1\right)}}{R} 
	- \frac{24}{\left(d_k\right)^6}\,\frac{\widehat{e}_{\left(1\right)}\,g_{\left(3\right)}}{R} 
	+ \frac{16}{\left(d_k\right)^8} \left(1 + \widehat{g}_{\left(1\right)}\right) \frac{e_{\left(1\right)}}{R} 
	+ \frac{8}{\left(d_k\right)^6}\,\frac{\widehat{g}_{\left(3\right)}\,e_{\left(1\right)}}{R} 
	+ \frac{24}{\left(d_k\right)^4}\,\frac{\widehat{g}_{\left(5\right)}\,e_{\left(1\right)}}{R}
	\nonumber\\
	&& - \frac{8}{\left(d_k\right)^6} \left(1 + \widehat{g}_{\left(1\right)}\right) \frac{f_{\left(1\right)}}{R} 
	- \frac{4}{\left(d_k\right)^4}\,\frac{\widehat{g}_{\left(3\right)}\,f_{\left(1\right)}}{R} 
	- \frac{12}{\left(d_k\right)^2}\,\frac{\widehat{g}_{\left(5\right)}\,f_{\left(1\right)}}{R}
	+ \frac{8}{\left(d_k\right)^6}\,\frac{\widehat{f}_{\left(1\right)}\,g_{\left(1\right)}}{R} 
	+ \frac{12}{\left(d_k\right)^4}\,\frac{\widehat{f}_{\left(1\right)}\,g_{\left(3\right)}}{R} 
	\nonumber\\
	&& + \frac{4}{\left(d_k\right)^4}\,\frac{\widehat{f}_{\left(3\right)}\,g_{\left(1\right)}}{R} 
	+ \frac{4}{\left(d_k\right)^2}\,\frac{\widehat{f}_{\left(3\right)}\,g_{\left(3\right)}}{R} 
	+ \frac{16}{\left(d_k\right)^6}\,\widehat{f}_{(2)} 
        - \frac{32}{\left(d_k\right)^4}\,\widehat{f}_{(4)} 
        + \frac{10}{\left(d_k\right)^2}\,\widehat{f}_{(6)} 
        + \frac{17}{2}\,\widehat{f}_{(8)} - \frac{9}{2} \left(d_k\right)^2 \widehat{f}_{(10)}
	\nonumber\\
	&& + \frac{16}{\left(d_k\right)^6}\,\widehat{g}_{(3)} 
        - \frac{24}{\left(d_k\right)^4}\,\widehat{g}_{(5)}\,,
        \label{scalar_function_L_n_sigma_6}
        \\
        \widehat{L}_{\left(7\right)} &=& - \frac{16}{\left(d_k\right)^8} \left(1 + \widehat{g}_{\left(1\right)}\right) \frac{e_{\left(1\right)}}{R} 
	- \frac{8}{\left(d_k\right)^6}\,\frac{\widehat{g}_{\left(3\right)}\,e_{\left(1\right)}}{R} 
	+ \frac{8}{\left(d_k\right)^6} \left(1 + \widehat{g}_{\left(1\right)}\right) \frac{f_{\left(1\right)}}{R} 
	+ \frac{4}{\left(d_k\right)^4} \left(1 + \widehat{g}_{\left(1\right)}\right) \frac{f_{\left(3\right)}}{R} 
	\nonumber\\
	&& + \frac{4}{\left(d_k\right)^4}\,\frac{\widehat{g}_{\left(3\right)}\,f_{\left(1\right)}}{R} 
	+ \frac{2}{\left(d_k\right)^2}\,\frac{\widehat{g}_{\left(3\right)}\,f_{\left(3\right)}}{R} 
	+ \frac{16}{\left(d_k\right)^8}\,\frac{\widehat{e}_{\left(1\right)}\,g_{\left(1\right)}}{R} 
	- \frac{8}{\left(d_k\right)^6}\,\frac{\widehat{f}_{\left(1\right)}\,g_{\left(1\right)}}{R} 
	- \frac{2}{\left(d_k\right)^4}\,\frac{\widehat{f}_{\left(3\right)}\,g_{\left(1\right)}}{R}
	\nonumber\\ 
	&& - \frac{6}{\left(d_k\right)^2}\,\frac{\widehat{f}_{\left(5\right)}\,g_{\left(1\right)}}{R} 
	+ \frac{8}{\left(d_k\right)^6}\,\widehat{f}_{(2)}
        - \frac{4}{\left(d_k\right)^4}\,\widehat{f}_{(4)}
        - \frac{1}{\left(d_k\right)^2}\,\widehat{f}_{(6)}
        + \frac{7}{2}\,\widehat{f}_{(8)}
        - \frac{9}{4} \left(d_k\right)^2 \widehat{f}_{(10)}
        + \frac{8}{\left(d_k\right)^6}\,\widehat{g}_{(3)}\,,
        \label{scalar_function_L_n_sigma_7}
        \\
	\widehat{L}_{\left(8\right)} &=& - \frac{4}{\left(d_k\right)^6}\,\frac{\widehat{f}_{\left(3\right)}\,e_{\left(1\right)}}{R} 
        + \frac{12}{\left(d_k\right)^4}\,\frac{\widehat{f}_{\left(5\right)}\,e_{\left(1\right)}}{R} 
	+ \frac{2}{\left(d_k\right)^4}\,\frac{\widehat{f}_{\left(3\right)}\,f_{\left(1\right)}}{R}
	- \frac{6}{\left(d_k\right)^2}\,\frac{\widehat{f}_{\left(5\right)}\,f_{\left(1\right)}}{R}
	- \frac{12}{\left(d_k\right)^6}\,\frac{\widehat{e}_{\left(1\right)}\,f_{\left(3\right)}}{R}
	\nonumber\\ 
	&& + \frac{6}{\left(d_k\right)^4}\,\frac{\widehat{f}_{\left(1\right)}\,f_{\left(3\right)}}{R}
	+ \frac{2}{\left(d_k\right)^2}\,\frac{\widehat{f}_{\left(3\right)}\,f_{\left(3\right)}}{R}
	+ \frac{16}{\left(d_k\right)^6}\,\widehat{f}_{(3)} 
        - \frac{12}{\left(d_k\right)^4}\,\widehat{f}_{(5)}
        - \frac{2205}{512}\,\frac{\widehat{g}_{(2)}}{\left(d_k\right)^8} 
        + \frac{3361}{256}\,\frac{\widehat{g}_{(4)}}{\left(d_k\right)^6}
	\nonumber\\ 
	&& - \frac{403}{64}\,\frac{\widehat{g}_{(6)}}{\left(d_k\right)^4} 
        - \frac{63}{32}\,\frac{\widehat{g}_{(8)}}{\left(d_k\right)^2} 
        + \frac{9}{4}\,\widehat{g}_{(10)} 
        - \frac{2205}{512}\,\frac{\widehat{h}_{(2)}}{\left(d_k\right)^9}\,,
        \label{scalar_function_L_n_sigma_8}
       \end{eqnarray}
       \end{widetext}

       \begin{widetext}
        \begin{eqnarray}
        \widehat{L}_{\left(9\right)} &=& 0\,,
        \label{scalar_function_L_n_sigma_9}
        \\
        \widehat{L}_{\left(10\right)} &=& + \frac{8}{\left(d_k\right)^6}\,\frac{\widehat{e}_{\left(1\right)}\,g_{\left(1\right)}}{R}
	- \frac{4}{\left(d_k\right)^4}\,\frac{\widehat{f}_{\left(1\right)}\,g_{\left(1\right)}}{R}
	+ \frac{4}{\left(d_k\right)^2}\,\frac{\widehat{f}_{\left(3\right)}\,g_{\left(1\right)}}{R} 
	+ \frac{8}{\left(d_k\right)^6} \left(1 + \widehat{g}_{(1)}\right)
        - \frac{4}{\left(d_k\right)^4}\,\widehat{f}_{(2)}
        - \frac{\widehat{f}_{(4)}}{\left(d_k\right)^2}
	\nonumber\\ 
	&& - \frac{9}{2}\,\widehat{f}_{(6)}
        + \frac{95}{8} \left(d_k\right)^2 \widehat{f}_{(8)}
        - \frac{15}{2} \left(d_k\right)^4 \widehat{f}_{(10)}\,,
        \label{scalar_function_L_n_sigma_10}
        \\
        \widehat{L}_{\left(11\right)} &=& - \frac{16}{\left(d_k\right)^8}\,\frac{\widehat{e}_{\left(1\right)}\,e_{\left(1\right)}}{R} 
	+ \frac{8}{\left(d_k\right)^6}\,\frac{\widehat{f}_{\left(1\right)}\,e_{\left(1\right)}}{R}
	- \frac{8}{\left(d_k\right)^4}\,\frac{\widehat{f}_{\left(3\right)}\,e_{\left(1\right)}}{R}
	- \frac{8}{\left(d_k\right)^6}\left(1 + \widehat{g}_{\left(1\right)}\right) \frac{g_{\left(1\right)}}{R} 
        - \frac{4}{\left(d_k\right)^4}\,\frac{\widehat{g}_{\left(3\right)}\,g_{\left(1\right)}}{R}
	\nonumber\\ 
	&& + \frac{8}{\left(d_k\right)^6}\,\frac{\widehat{e}_{\left(1\right)}\,f_{\left(1\right)}}{R} 
	- \frac{4}{\left(d_k\right)^4}\,\frac{\widehat{f}_{\left(1\right)}\,f_{\left(1\right)}}{R} 
	+ \frac{4}{\left(d_k\right)^2}\,\frac{\widehat{f}_{\left(3\right)}\,f_{\left(1\right)}}{R} 
	- \frac{32}{\left(d_k\right)^8}\,\widehat{e}_{(1)} 
        + \frac{16}{\left(d_k\right)^6}\,\widehat{f}_{(1)}
        + \frac{16}{\left(d_k\right)^4}\,\widehat{f}_{(3)}
	\nonumber\\
	&& + \frac{985}{128}\,\frac{\widehat{g}_{(2)}}{\left(d_k\right)^6}
        + \frac{217}{192}\,\frac{\widehat{g}_{(4)}}{\left(d_k\right)^4}
        + \frac{5}{48}\,\frac{\widehat{g}_{(6)}}{\left(d_k\right)^2} 
        - \frac{85}{8}\,\widehat{g}_{(8)}
        + 15 \left(d_k\right)^2 \widehat{g}_{(10)} 
        + \frac{985}{128}\,\frac{\widehat{h}_{(2)}}{\left(d_k\right)^7}\,,
        \label{scalar_function_L_n_sigma_11}
        \\
	\widehat{L}_{\left(12\right)} &=& + \frac{16}{\left(d_k\right)^8}\left(1 + \widehat{g}_{\left(1\right)}\right) \frac{e_{\left(1\right)}}{R}  
	+ \frac{8}{\left(d_k\right)^6}\,\frac{\widehat{g}_{\left(3\right)}\,e_{\left(1\right)}}{R} 
	- \frac{8}{\left(d_k\right)^6}\left(1 + \widehat{g}_{\left(1\right)}\right) \frac{f_{\left(1\right)}}{R}
	- \frac{4}{\left(d_k\right)^4}\,\frac{\widehat{g}_{\left(3\right)}\,f_{\left(1\right)}}{R} 
	- \frac{16}{\left(d_k\right)^8}\left(1 + \widehat{g}_{\left(1\right)}\right) 
	\nonumber\\
	&& + \frac{4}{\left(d_k\right)^6}\,\widehat{f}_{\left(2\right)} 
	+ \frac{2}{\left(d_k\right)^4}\,\widehat{f}_{\left(4\right)} 
	+ \frac{2}{\left(d_k\right)^2}\,\widehat{f}_{\left(6\right)} 
	- \frac{4}{\left(d_k\right)^6}\,\widehat{g}_{\left(3\right)}\,,
        \label{scalar_function_L_n_sigma_12}
        \\
        \widehat{L}_{\left(13\right)} &=& + \frac{12}{\left(d_k\right)^6}\left(1 + \widehat{g}_{\left(1\right)}\right) \frac{e_{\left(1\right)}}{R} 
        + \frac{6}{\left(d_k\right)^4}\,\frac{\widehat{g}_{\left(3\right)}\,e_{\left(1\right)}}{R}
        - \frac{6}{\left(d_k\right)^2}\,\frac{\widehat{g}_{\left(5\right)}\,e_{\left(1\right)}}{R}
	- \frac{6}{\left(d_k\right)^4} \left(1 + \widehat{g}_{\left(1\right)}\right) \frac{f_{\left(1\right)}}{R} 
	+ \frac{6}{\left(d_k\right)^2} \left(1 + \widehat{g}_{\left(1\right)}\right) \frac{f_{\left(3\right)}}{R} 
	\nonumber\\ 
	&& - \frac{12}{\left(d_k\right)^6}\,\frac{\widehat{e}_{\left(1\right)}\,g_{\left(1\right)}}{R} 
	+ \frac{6}{\left(d_k\right)^4}\,\frac{\widehat{f}_{\left(1\right)}\,g_{\left(1\right)}}{R} 
	- 6\,\frac{\widehat{f}_{\left(5\right)}\,g_{\left(1\right)}}{R}
	- \frac{3}{\left(d_k\right)^2}\,\frac{\widehat{g}_{\left(3\right)}\,f_{\left(1\right)}}{R}
	+ 3\,\frac{\widehat{g}_{\left(5\right)}\,f_{\left(1\right)}}{R} 
	+ 3\,\frac{\widehat{g}_{\left(3\right)}\,f_{\left(3\right)}}{R}
	- 3 \left(d_k\right)^2 \frac{\widehat{g}_{\left(5\right)}\,f_{\left(3\right)}}{R}
	\nonumber\\ 
	&&  - \frac{6}{\left(d_k\right)^4}\,\widehat{f}_{(2)} 
	+ 9\,\frac{\widehat{f}_{(4)}}{\left(d_k\right)^2} - \frac{15}{4}\,\widehat{f}_{(6)} - \frac{27}{16} \left(d_k\right)^2 \widehat{f}_{(8)} 
	+ \frac{9}{4} \left(d_k\right)^4 \widehat{f}_{(10)} - \frac{6}{\left(d_k\right)^4} \,\widehat{g}_{(3)} 
	+ \frac{6}{\left(d_k\right)^2} \,\widehat{g}_{(5)}\,,
	\label{scalar_function_L_n_sigma_13}
        \\
        \widehat{L}_{\left(14\right)} &=& + \frac{4}{\left(d_k\right)^6}\left(1 + \widehat{g}_{\left(1\right)}\right) \frac{g_{\left(1\right)}}{R} 
	- \frac{12}{\left(d_k\right)^4}\left(1 + \widehat{g}_{\left(1\right)}\right) \frac{g_{\left(3\right)}}{R} 
	+ \frac{6}{\left(d_k\right)^2}\,\frac{\widehat{f}_{\left(1\right)}\,f_{\left(3\right)}}{R}
	+ \frac{8}{\left(d_k\right)^4}\,\frac{\widehat{f}_{\left(3\right)}\,e_{\left(1\right)}}{R} 
        - \frac{4}{\left(d_k\right)^2}\,\frac{\widehat{f}_{\left(3\right)}\,f_{\left(1\right)}}{R}
	+ 4\,\frac{\widehat{f}_{\left(3\right)}\,f_{\left(3\right)}}{R} 
	\nonumber\\ 
	&& - 6 \left(d_k\right)^2 \frac{\widehat{f}_{\left(5\right)}\,f_{\left(3\right)}}{R} 
	- \frac{12}{\left(d_k\right)^4}\,\frac{\widehat{e}_{\left(1\right)}\,f_{\left(3\right)}}{R} 
	+ \frac{2}{\left(d_k\right)^4}\,\frac{\widehat{g}_{\left(3\right)}\,g_{\left(1\right)}}{R} 
	- \frac{6}{\left(d_k\right)^2}\,\frac{\widehat{g}_{\left(3\right)}\,g_{\left(3\right)}}{R} 
	+ \frac{18}{\left(d_k\right)^2}\,\frac{\widehat{g}_{\left(5\right)}\,g_{\left(1\right)}}{R}
	+ 6\,\frac{\widehat{g}_{\left(5\right)}\,g_{\left(3\right)}}{R} 
	+ \frac{8}{\left(d_k\right)^4}\,\widehat{f}_{(3)}
	\nonumber\\ 
	&& - \frac{12}{\left(d_k\right)^2}\,\widehat{f}_{(5)} - \frac{945}{512}\,\frac{\widehat{g}_{(2)}}{\left(d_k\right)^6}
        + \frac{1733}{256}\,\frac{\widehat{g}_{(4)}}{\left(d_k\right)^4} - \frac{575}{64}\,\frac{\widehat{g}_{(6)}}{\left(d_k\right)^2}
        + \frac{21}{32}\,\widehat{g}_{(8)} - \frac{117}{4} \left(d_k\right)^2 \widehat{g}_{(10)}
	- \frac{945}{512}\,\frac{\widehat{h}_{(2)}}{\left(d_k\right)^7}\,, 
        \label{scalar_function_L_n_sigma_14}
        \\ 
        \widehat{L}_{\left(15\right)} &=& + \frac{40}{\left(d_k\right)^8} \left(1 + \widehat{g}_{\left(1\right)}\right) \frac{e_{\left(1\right)}}{R} 
	- \frac{20}{\left(d_k\right)^6} \left(1 + \widehat{g}_{\left(1\right)}\right) \frac{f_{\left(1\right)}}{R} 
	+ \frac{2}{\left(d_k\right)^4}\left(1 + \widehat{g}_{\left(1\right)}\right) \frac{f_{\left(3\right)}}{R} 
	- \frac{16}{\left(d_k\right)^8}\,\frac{\widehat{e}_{\left(1\right)}\,g_{\left(1\right)}}{R} 
	+ \frac{8}{\left(d_k\right)^6}\,\frac{\widehat{f}_{\left(1\right)}\,g_{\left(1\right)}}{R} 
	\nonumber\\ 
	&& + \frac{2}{\left(d_k\right)^4}\,\frac{\widehat{f}_{\left(3\right)}\,g_{\left(1\right)}}{R} 
	+ \frac{6}{\left(d_k\right)^2}\,\frac{\widehat{f}_{\left(5\right)}\,g_{\left(1\right)}}{R} 
	+ \frac{20}{\left(d_k\right)^6}\,\frac{\widehat{g}_{\left(3\right)}\,e_{\left(1\right)}}{R} 
	- \frac{10}{\left(d_k\right)^4}\,\frac{\widehat{g}_{\left(3\right)}\,f_{\left(1\right)}}{R} 
	+ \frac{1}{\left(d_k\right)^2}\,\frac{\widehat{g}_{\left(3\right)}\,f_{\left(3\right)}}{R}
	- \frac{6}{\left(d_k\right)^4}\,\frac{\widehat{g}_{\left(5\right)}\,e_{\left(1\right)}}{R}
	\nonumber\\ 
	&& + \frac{3}{\left(d_k\right)^2}\,\frac{\widehat{g}_{\left(5\right)}\,f_{\left(1\right)}}{R}
        + 6\,\frac{\widehat{g}_{\left(5\right)}\,f_{\left(3\right)}}{R} 
	- \frac{8}{\left(d_k\right)^6}\,\widehat{f}_{(2)}
        + \frac{16}{\left(d_k\right)^4}\,\widehat{f}_{(4)}
        - \frac{27}{2}\,\widehat{f}_{(8)}
        - \frac{27}{4} \left(d_k\right)^2 \widehat{f}_{(10)}
        - \frac{16}{\left(d_k\right)^8}\,\widehat{g}_{(1)}
        - \frac{16}{\left(d_k\right)^6}\,\widehat{g}_{(3)}
	\nonumber\\ 
	&& + \frac{6}{\left(d_k\right)^4}\,\widehat{g}_{(5)}\,,
        \label{scalar_function_L_n_sigma_15}
        \\
        \widehat{L}_{\left(16\right)} &=& - \frac{40}{\left(d_k\right)^8} \left(1 + \widehat{g}_{\left(1\right)}\right) \frac{e_{\left(1\right)}}{R} 
	+ \frac{20}{\left(d_k\right)^6} \left(1 + \widehat{g}_{\left(1\right)}\right) \frac{f_{\left(1\right)}}{R} 
	- \frac{4}{\left(d_k\right)^4} \left(1 + \widehat{g}_{\left(1\right)}\right) \frac{f_{\left(3\right)}}{R} 
        + \frac{16}{\left(d_k\right)^8}\,\frac{\widehat{e}_{\left(1\right)}\,g_{\left(1\right)}}{R} 
	+ \frac{24}{\left(d_k\right)^6}\,\frac{\widehat{e}_{\left(1\right)}\,g_{\left(3\right)}}{R}
	\nonumber\\ 
	&& - \frac{8}{\left(d_k\right)^6}\,\frac{\widehat{f}_{\left(1\right)}\,g_{\left(1\right)}}{R} 
	- \frac{12}{\left(d_k\right)^4}\,\frac{\widehat{f}_{\left(1\right)}\,g_{\left(3\right)}}{R}
	- \frac{12}{\left(d_k\right)^4}\,\frac{\widehat{f}_{\left(3\right)}\,g_{\left(1\right)}}{R} 
	- \frac{8}{\left(d_k\right)^2}\,\frac{\widehat{f}_{\left(3\right)}\,g_{\left(3\right)}}{R} 
        + \frac{12}{\left(d_k\right)^2}\,\frac{\widehat{f}_{\left(5\right)}\,g_{\left(1\right)}}{R} 
        + 12\,\frac{\widehat{f}_{\left(5\right)}\,g_{\left(1\right)}}{R}
	\nonumber\\ 
	&& - \frac{20}{\left(d_k\right)^6}\,\frac{\widehat{g}_{\left(3\right)}\,e_{\left(1\right)}}{R}
	+ \frac{10}{\left(d_k\right)^4}\,\frac{\widehat{g}_{\left(3\right)}\,f_{\left(1\right)}}{R}
	- \frac{2}{\left(d_k\right)^2}\,\frac{\widehat{g}_{\left(3\right)}\,f_{\left(3\right)}}{R}
	- \frac{24}{\left(d_k\right)^4}\,\frac{\widehat{g}_{\left(5\right)}\,e_{\left(1\right)}}{R}
	+ \frac{12}{\left(d_k\right)^2}\,\frac{\widehat{g}_{\left(5\right)}\,f_{\left(1\right)}}{R} 
	+ \frac{64}{\left(d_k\right)^8} \left(1 + \widehat{g}_{\left(1\right)}\right) 
	\nonumber\\ 
	&& - \frac{16}{\left(d_k\right)^6}\,\widehat{f}_{(2)} 
        + \frac{44}{\left(d_k\right)^4}\,\widehat{f}_{(4)} 
        - \frac{36}{\left(d_k\right)^2}\,\widehat{f}_{(6)} 
        - \frac{29}{2}\,\widehat{f}_{(8)} 
	- \frac{27}{2} \left(d_k\right)^2 \widehat{f}_{(10)}
        + \frac{16}{\left(d_k\right)^6}\,\widehat{g}_{(3)} 
	+ \frac{60}{\left(d_k\right)^4}\,\widehat{g}_{(5)}\,, 
        \label{scalar_function_L_n_sigma_16} 
      \end{eqnarray}
      \end{widetext}

        \begin{widetext}
        \begin{eqnarray}
	\widehat{L}_{\left(17\right)} &=& + \frac{20}{\left(d_k\right)^6}\,\frac{\widehat{f}_{\left(3\right)}\,e_{\left(1\right)}}{R} 
	- \frac{36}{\left(d_k\right)^4}\,\frac{\widehat{f}_{\left(5\right)}\,e_{\left(1\right)}}{R} 
	+ \frac{12}{\left(d_k\right)^6}\,\frac{\widehat{e}_{\left(1\right)}\,f_{\left(3\right)}}{R} 
	- \frac{6}{\left(d_k\right)^4}\,\frac{\widehat{f}_{\left(1\right)}\,f_{\left(3\right)}}{R}
	- \frac{10}{\left(d_k\right)^4}\,\frac{\widehat{f}_{\left(3\right)}\,f_{\left(1\right)}}{R}
	\nonumber\\ 
	&& - \frac{4}{\left(d_k\right)^2}\,\frac{\widehat{f}_{\left(3\right)}\,f_{\left(3\right)}}{R} 
	+ \frac{18}{\left(d_k\right)^2}\,\frac{\widehat{f}_{\left(5\right)}\,f_{\left(1\right)}}{R}
	+ 6\,\frac{\widehat{f}_{\left(5\right)}\,f_{\left(3\right)}}{R} 
	- \frac{8}{\left(d_k\right)^6} \left(1 + \widehat{g}_{\left(1\right)}\right) \frac{g_{\left(3\right)}}{R} 
	- \frac{4}{\left(d_k\right)^4}\,\frac{\widehat{g}_{\left(3\right)}\,g_{\left(3\right)}}{R} 
	\nonumber\\
	&& - \frac{6}{\left(d_k\right)^4}\,\frac{\widehat{g}_{\left(5\right)}\,g_{\left(1\right)}}{R} 
	- \frac{6}{\left(d_k\right)^2}\,\frac{\widehat{g}_{\left(5\right)}\,g_{\left(3\right)}}{R} 
	- \frac{32}{\left(d_k\right)^6}\,\widehat{f}_{(3)}
        + \frac{36}{\left(d_k\right)^4}\,\widehat{f}_{(5)}
        + \frac{2205}{512}\,\frac{\widehat{g}_{(2)}}{\left(d_k\right)^8}
        - \frac{7457}{256}\,\frac{\widehat{g}_{(4)}}{\left(d_k\right)^6}
	\nonumber\\
	&& + \frac{1427}{64}\,\frac{\widehat{g}_{(6)}}{\left(d_k\right)^4}
        + \frac{255}{32}\,\frac{\widehat{g}_{(8)}}{\left(d_k\right)^2}
        + \frac{63}{4}\,\widehat{g}_{(10)}
        + \frac{2205}{512}\,\frac{\widehat{h}_{(2)}}{\left(d_k\right)^9}\,,
        \label{scalar_function_L_n_sigma_17}
        \\
	\widehat{L}_{\left(18\right)} &=& - \frac{4}{\left(d_k\right)^6} \left(1 + \widehat{g}_{\left(1\right)}\right) \frac{f_{\left(3\right)}}{R} 
	- \frac{2}{\left(d_k\right)^4}\,\frac{\widehat{g}_{\left(3\right)}\,f_{\left(3\right)}}{R} 
	+ \frac{12}{\left(d_k\right)^6}\,\frac{\widehat{g}_{\left(5\right)}\,e_{\left(1\right)}}{R}
	- \frac{6}{\left(d_k\right)^4}\,\frac{\widehat{g}_{\left(5\right)}\,f_{\left(1\right)}}{R} 
	- \frac{3}{\left(d_k\right)^2}\,\frac{\widehat{g}_{\left(5\right)}\,f_{\left(3\right)}}{R} 
	\nonumber\\ 
	&& - \frac{36}{\left(d_k\right)^6}\,\widehat{f}_{\left(4\right)} 
	+ \frac{24}{\left(d_k\right)^4}\,\widehat{f}_{\left(6\right)} 
	+ \frac{15}{2}\,\frac{1}{\left(d_k\right)^2}\,\widehat{f}_{\left(8\right)} 
	+ \frac{9}{2}\,\widehat{f}_{\left(10\right)} 
	- \frac{28}{\left(d_k\right)^6}\,\widehat{g}_{\left(5\right)}\,, 
        \label{scalar_function_L_n_sigma_18}
        \\
        \widehat{L}_{\left(19\right)} &=& + \frac{5}{128}\,\frac{\widehat{g}_{(2)}}{\left(d_k\right)^6}
        + \frac{5}{192}\,\frac{\widehat{g}_{(4)}}{\left(d_k\right)^4}
        + \frac{1}{48}\,\frac{\widehat{g}_{(6)}}{\left(d_k\right)^2} 
        - \frac{5}{8}\,\widehat{g}_{(8)} 
        + \frac{5}{128}\,\frac{\widehat{h}_{(2)}}{\left(d_k\right)^7}\,,  
        \label{scalar_function_L_n_sigma_19}
        \\
        \widehat{L}_{\left(20\right)} &=& - \frac{8}{\left(d_k\right)^6} \left(1 + \widehat{g}_{\left(1\right)}\right) \frac{g_{\left(1\right)}}{R} 
	- \frac{4}{\left(d_k\right)^4}\,\frac{\widehat{g}_{\left(3\right)}\,g_{\left(1\right)}}{R} 
        + \frac{925}{256}\,\frac{\widehat{g}_{(2)}}{\left(d_k\right)^6} 
        + \frac{925}{384}\,\frac{\widehat{g}_{(4)}}{\left(d_k\right)^4} 
        + \frac{185}{96}\,\frac{\widehat{g}_{(6)}}{\left(d_k\right)^2}
        + \frac{95}{16}\,\widehat{g}_{(8)}
	\nonumber\\ 
	&& - \frac{15}{2} \left(d_k\right)^2 \widehat{g}_{(10)}  
        + \frac{925}{256}\,\frac{\widehat{h}_{(2)}}{\left(d_k\right)^7}\,,
        \label{scalar_function_L_n_sigma_20}
        \\
        \widehat{L}_{\left(21\right)} &=& + \frac{16}{\left(d_k\right)^8} \left(1 + \widehat{g}_{\left(1\right)}\right) \frac{e_{\left(1\right)}}{R}
	- \frac{8}{\left(d_k\right)^6} \left(1 + \widehat{g}_{\left(1\right)}\right) \frac{f_{\left(1\right)}}{R}
	+ \frac{32}{\left(d_k\right)^8}\,\frac{\widehat{e}_{\left(1\right)}\,g_{\left(1\right)}}{R} 
	- \frac{16}{\left(d_k\right)^6}\,\frac{\widehat{f}_{\left(1\right)}\,g_{\left(1\right)}}{R}
	- \frac{4}{\left(d_k\right)^4}\,\frac{\widehat{f}_{\left(3\right)}\,g_{\left(1\right)}}{R}
	\nonumber\\ 
	&& - \frac{12}{\left(d_k\right)^2}\,\frac{\widehat{f}_{\left(5\right)}\,g_{\left(1\right)}}{R} 
	- \frac{8}{\left(d_k\right)^6}\,\frac{\widehat{g}_{\left(3\right)}\,e_{\left(1\right)}}{R} 
	- \frac{4}{\left(d_k\right)^4}\,\frac{\widehat{g}_{\left(3\right)}\,f_{\left(1\right)}}{R} 
	- \frac{48}{\left(d_k\right)^8} \left(1 + \widehat{g}_{(1)}\right)
        + \frac{16}{\left(d_k\right)^6}\,\widehat{f}_{(2)}
        + \frac{10}{\left(d_k\right)^4}\,\widehat{f}_{(4)}
	\nonumber\\ 
	&& + \frac{4}{\left(d_k\right)^2}\,\widehat{f}_{(6)}
        + \frac{55}{4}\,\widehat{f}_{(8)}
	- 15 \left(d_k\right)^2 \widehat{f}_{(10)}
        - \frac{8}{\left(d_k\right)^6}\,\widehat{g}_{(3)}\,,
        \label{scalar_function_L_n_sigma_21}
        \\
	\widehat{L}_{\left(22\right)} &=& - \frac{64}{\left(d_k\right)^{10}}\,\frac{\widehat{e}_{\left(1\right)}\,e_{\left(1\right)}}{R} 
	+ \frac{32}{\left(d_k\right)^8}\,\frac{\widehat{e}_{\left(1\right)}\,f_{\left(1\right)}}{R} 
        + \frac{32}{\left(d_k\right)^8}\,\frac{\widehat{f}_{\left(1\right)}\,e_{\left(1\right)}}{R} 
	- \frac{16}{\left(d_k\right)^6}\,\frac{\widehat{f}_{\left(1\right)}\,f_{\left(1\right)}}{R} 
	+ \frac{8}{\left(d_k\right)^6}\,\frac{\widehat{f}_{\left(3\right)}\,e_{\left(1\right)}}{R}
	- \frac{4}{\left(d_k\right)^4}\,\frac{\widehat{f}_{\left(3\right)}\,f_{\left(1\right)}}{R}
	\nonumber\\ 
	&& - \frac{24}{\left(d_k\right)^4}\,\frac{\widehat{f}_{\left(5\right)}\,e_{\left(1\right)}}{R} 
	+ \frac{12}{\left(d_k\right)^2}\,\frac{\widehat{f}_{\left(5\right)}\,f_{\left(1\right)}}{R} 
	+ \frac{128}{\left(d_k\right)^{10}}\,\widehat{e}_{(1)}
        - \frac{64}{\left(d_k\right)^8}\,\widehat{f}_{(1)}
        - \frac{32}{\left(d_k\right)^6}\,\widehat{f}_{(3)}
        + \frac{24}{\left(d_k\right)^4}\,\widehat{f}_{(5)}
        - \frac{6895}{256}\,\frac{\widehat{g}_{(2)}}{\left(d_k\right)^8}
        \nonumber\\
	&& - \frac{13039}{384}\,\frac{\widehat{g}_{(4)}}{\left(d_k\right)^6}
        - \frac{227}{96}\,\frac{\widehat{g}_{(6)}}{\left(d_k\right)^4}
        - \frac{5}{16}\,\frac{\widehat{g}_{(8)}}{\left(d_k\right)^2}
        + \frac{15}{2}\,\widehat{g}_{(10)}
        - \frac{6895}{256}\,\frac{\widehat{h}_{(2)}}{\left(d_k\right)^9}\,,
        \label{scalar_function_L_n_sigma_22}
        \\
        \widehat{L}_{\left(23\right)} &=& - \frac{12}{\left(d_k\right)^8}\,\frac{\widehat{e}_{\left(1\right)}\,e_{\left(1\right)}}{R} 
	+ \frac{6}{\left(d_k\right)^6}\,\frac{\widehat{e}_{\left(1\right)}\,f_{\left(1\right)}}{R} 
	- \frac{6}{\left(d_k\right)^4}\,\frac{\widehat{e}_{\left(1\right)}\,f_{\left(3\right)}}{R} 
	+ \frac{8}{\left(d_k\right)^6} \left(1 + \widehat{g}_{\left(1\right)}\right) \frac{g_{\left(1\right)}}{R} 
	+ \frac{6}{\left(d_k\right)^6}\,\frac{\widehat{f}_{\left(1\right)}\,e_{\left(1\right)}}{R} 
	\nonumber\\ 
	&& - \frac{3}{\left(d_k\right)^4}\,\frac{\widehat{f}_{\left(1\right)}\,f_{\left(1\right)}}{R} 
	+ \frac{3}{\left(d_k\right)^2}\,\frac{\widehat{f}_{\left(1\right)}\,f_{\left(3\right)}}{R} 
	- \frac{2}{\left(d_k\right)^4}\,\frac{\widehat{f}_{\left(3\right)}\,e_{\left(1\right)}}{R} 
	+ \frac{1}{\left(d_k\right)^2}\,\frac{\widehat{f}_{\left(3\right)}\,f_{\left(1\right)}}{R} 
	- 1\,\frac{\widehat{f}_{\left(3\right)}\,f_{\left(3\right)}}{R} 
	+ \frac{4}{\left(d_k\right)^4}\,\frac{\widehat{g}_{\left(3\right)}\,g_{\left(1\right)}}{R}
	\nonumber\\ 
	&& + \frac{24}{\left(d_k\right)^8}\,\widehat{e}_{(1)} 
        - \frac{12}{\left(d_k\right)^6}\,\widehat{f}_{(1)}
        + \frac{14}{\left(d_k\right)^4}\,\widehat{f}_{(3)} 
        - \frac{6}{\left(d_k\right)^2}\,\widehat{f}_{(5)}
        - \frac{25875}{2048}\,\frac{\widehat{g}_{(2)}}{\left(d_k\right)^6}
        + \frac{8783}{1024}\,\frac{\widehat{g}_{(4)}}{\left(d_k\right)^4}
        - \frac{701}{256}\,\frac{\widehat{g}_{(6)}}{\left(d_k\right)^2}
        \nonumber\\ 
        && - \frac{2577}{128}\,\widehat{g}_{(8)}
        + \frac{399}{16} \left(d_k\right)^2 \widehat{g}_{(10)}
        - \frac{75}{8} \left(d_k\right)^4 \widehat{g}_{(12)}
        - \frac{25875}{2048}\,\frac{\widehat{h}_{(2)}}{\left(d_k\right)^7}\,, 
        \label{scalar_function_L_n_sigma_23}
        \\
	\widehat{L}_{\left(24\right)} &=& - \frac{24}{\left(d_k\right)^8}\,\frac{\widehat{e}_{\left(1\right)}\,g_{\left(1\right)}}{R}
	+ \frac{12}{\left(d_k\right)^6}\,\frac{\widehat{e}_{\left(1\right)}\,g_{\left(3\right)}}{R}
	+ \frac{8}{\left(d_k\right)^4} \left(1 + \widehat{g}_{\left(1\right)}\right) \frac{f_{\left(3\right)}}{R} 
        + \frac{12}{\left(d_k\right)^6}\,\frac{\widehat{f}_{\left(1\right)}\,g_{\left(1\right)}}{R}
	- \frac{6}{\left(d_k\right)^4}\,\frac{\widehat{f}_{\left(1\right)}\,g_{\left(3\right)}}{R}
	\nonumber\\ 
        && + \frac{10}{\left(d_k\right)^4}\,\frac{\widehat{f}_{\left(3\right)}\,g_{\left(1\right)}}{R}
	+ \frac{2}{\left(d_k\right)^2}\,\frac{\widehat{f}_{\left(3\right)}\,g_{\left(3\right)}}{R}
	- \frac{6}{\left(d_k\right)^2}\,\frac{\widehat{f}_{\left(5\right)}\,g_{\left(1\right)}}{R}
	+ \frac{4}{\left(d_k\right)^2}\,\frac{\widehat{g}_{\left(3\right)}\,f_{\left(3\right)}}{R}
	+ \frac{24}{\left(d_k\right)^8} \left(1 + \widehat{g}_{(1)}\right)
        - \frac{24}{\left(d_k\right)^6}\,\widehat{f}_{(2)}
	\nonumber\\ 
	&& + \frac{3}{\left(d_k\right)^4}\,\widehat{f}_{(4)}
        - \frac{273}{8}\,\widehat{f}_{(8)}
        + \frac{135}{2} \left(d_k\right)^2 \widehat{f}_{(10)}
        - \frac{75}{2} \left(d_k\right)^4 \widehat{f}_{(12)}
        - \frac{12}{\left(d_k\right)^6}\,\widehat{g}_{(3)}\,,
        \label{scalar_function_L_n_sigma_24}
        \end{eqnarray}
        \end{widetext} 

        \begin{widetext} 
        \begin{eqnarray}
	\widehat{L}_{\left(25\right)} &=& - \frac{64}{\left(d_k\right)^{10}}\,\frac{\widehat{e}_{\left(1\right)}\,e_{\left(1\right)}}{R}
        + \frac{32}{\left(d_k\right)^8}\,\frac{\widehat{e}_{\left(1\right)}\,f_{\left(1\right)}}{R}
        - \frac{18}{\left(d_k\right)^6}\,\frac{\widehat{e}_{\left(1\right)}\,f_{\left(3\right)}}{R}
        + \frac{32}{\left(d_k\right)^8}\,\frac{\widehat{f}_{\left(1\right)}\,e_{\left(1\right)}}{R}
        - \frac{16}{\left(d_k\right)^6}\,\frac{\widehat{f}_{\left(1\right)}\,f_{\left(1\right)}}{R}
        \nonumber\\
        && + \frac{9}{\left(d_k\right)^4}\,\frac{\widehat{f}_{\left(1\right)}\,f_{\left(3\right)}}{R}
        - \frac{6}{\left(d_k\right)^6}\,\frac{\widehat{f}_{\left(3\right)}\,e_{\left(1\right)}}{R}
        + \frac{3}{\left(d_k\right)^4}\,\frac{\widehat{f}_{\left(3\right)}\,f_{\left(1\right)}}{R}
        + \frac{4}{\left(d_k\right)^2}\,\frac{\widehat{f}_{\left(3\right)}\,f_{\left(3\right)}}{R}
        + \frac{12}{\left(d_k\right)^4}\,\frac{\widehat{f}_{\left(5\right)}\,e_{\left(1\right)}}{R}
        - \frac{6}{\left(d_k\right)^2}\,\frac{\widehat{f}_{\left(5\right)}\,f_{\left(1\right)}}{R}
        \nonumber\\
        && + \frac{128}{\left(d_k\right)^{10}}\,\widehat{e}_{(1)} 
        - \frac{64}{\left(d_k\right)^8}\,\widehat{f}_{(1)} 
        + \frac{24}{\left(d_k\right)^6}\,\widehat{f}_{(3)}
        - \frac{12}{\left(d_k\right)^4}\,\widehat{f}_{(5)}
        - \frac{19405}{1024}\,\frac{\widehat{g}_{(2)}}{\left(d_k\right)^8} 
        + \frac{42035}{1536}\,\frac{\widehat{g}_{(4)}}{\left(d_k\right)^6} 
        + \frac{2263}{384}\,\frac{\widehat{g}_{(6)}}{\left(d_k\right)^4}
        \nonumber\\ 
        && + \frac{49}{64}\,\frac{\widehat{g}_{(8)}}{\left(d_k\right)^2}
        - \frac{135}{8}\,\widehat{g}_{(10)} 
        + \frac{75}{4} \left(d_k\right)^2 \widehat{g}_{(12)} 
        - \frac{19405}{1024}\,\frac{\widehat{h}_{(2)}}{\left(d_k\right)^9}\,,
        \label{scalar_function_L_n_sigma_25}
	\\
	\widehat{L}_{\left(26\right)} &=& + \frac{64}{\left(d_k\right)^{10}}\,\frac{\widehat{e}_{\left(1\right)}\,e_{\left(1\right)}}{R} 
	- \frac{32}{\left(d_k\right)^8}\,\frac{\widehat{e}_{\left(1\right)}\,f_{\left(1\right)}}{R} 
	- \frac{16}{\left(d_k\right)^6} \left(1 + \widehat{g}_{\left(1\right)}\right) \frac{g_{\left(3\right)}}{R} 
	- \frac{32}{\left(d_k\right)^8}\,\frac{\widehat{f}_{\left(1\right)}\,e_{\left(1\right)}}{R} 
	+ \frac{16}{\left(d_k\right)^6}\,\frac{\widehat{f}_{\left(1\right)}\,f_{\left(1\right)}}{R}
	\nonumber\\
	&& - \frac{8}{\left(d_k\right)^6}\,\frac{\widehat{f}_{\left(3\right)}\,e_{\left(1\right)}}{R} 
	+ \frac{4}{\left(d_k\right)^4}\,\frac{\widehat{f}_{\left(3\right)}\,f_{\left(1\right)}}{R} 
	- \frac{8}{\left(d_k\right)^4}\,\frac{\widehat{g}_{\left(3\right)}\,g_{\left(3\right)}}{R}
	+ \frac{12}{\left(d_k\right)^4}\,\frac{\widehat{g}_{\left(5\right)}\,g_{\left(1\right)}}{R}
	- \frac{128}{\left(d_k\right)^{10}}\,\widehat{e}_{(1)} 
        + \frac{64}{\left(d_k\right)^8}\,\widehat{f}_{(1)}
	\nonumber\\
	&& + \frac{8}{\left(d_k\right)^6}\,\widehat{f}_{(3)} 
        + \frac{6395}{512}\,\frac{\widehat{g}_{(2)}}{\left(d_k\right)^8} 
        + \frac{251}{768}\,\frac{\widehat{g}_{(4)}}{\left(d_k\right)^6} 
        - \frac{1025}{192}\,\frac{\widehat{g}_{(6)}}{\left(d_k\right)^4} 
        + \frac{73}{32}\,\frac{\widehat{g}_{(8)}}{\left(d_k\right)^2}
        - \frac{135}{4}\,\widehat{g}_{(10)} 
        + \frac{75}{2} \left(d_k\right)^2 \widehat{g}_{(12)} 
	\nonumber\\
	&& + \frac{6395}{512}\,\frac{\widehat{h}_{(2)}}{\left(d_k\right)^9}\,,
        \label{scalar_function_L_n_sigma_26}
        \\
        \widehat{L}_{\left(27\right)} &=& + \frac{48}{\left(d_k\right)^8}\,\frac{\widehat{e}_{\left(1\right)}\,g_{\left(3\right)}}{R} 
	- \frac{8}{\left(d_k\right)^6} \left(1 + \widehat{g}_{\left(1\right)}\right) \frac{f_{\left(3\right)}}{R} 
	- \frac{24}{\left(d_k\right)^6}\,\frac{\widehat{f}_{\left(1\right)}\,g_{\left(3\right)}}{R}
	- \frac{6}{\left(d_k\right)^4}\,\frac{\widehat{f}_{\left(3\right)}\,g_{\left(3\right)}}{R}
	+ \frac{6}{\left(d_k\right)^4}\,\frac{\widehat{f}_{\left(5\right)}\,g_{\left(1\right)}}{R} 
	\nonumber\\ 
	&& - \frac{4}{\left(d_k\right)^6}\,\frac{\widehat{g}_{\left(3\right)}\,f_{\left(3\right)}}{R} 
	- \frac{24}{\left(d_k\right)^6}\,\frac{\widehat{g}_{\left(5\right)}\,e_{\left(1\right)}}{R}
	+ \frac{12}{\left(d_k\right)^4}\,\frac{\widehat{g}_{\left(5\right)}\,f_{\left(1\right)}}{R}
	- \frac{16}{\left(d_k\right)^{10}} \left(1 + \widehat{g}_{(1)}\right) 
        - \frac{48}{\left(d_k\right)^8}\,\widehat{f}_{(2)}
        + \frac{48}{\left(d_k\right)^6}\,\widehat{f}_{(4)} 
	\nonumber\\ 
	&& - \frac{6}{\left(d_k\right)^4}\,\widehat{f}_{(6)}
        - \frac{75}{2}\,\widehat{f}_{(10)} 
        + \frac{75}{2} \left(d_k\right)^2 \widehat{f}_{(12)}
        - \frac{56}{\left(d_k\right)^8}\,\widehat{g}_{(3)}
        + \frac{24}{\left(d_k\right)^6}\,\widehat{g}_{(5)}\,,
        \label{scalar_function_L_n_sigma_27}
        \\
        \widehat{L}_{\left(28\right)} &=& + \frac{24}{\left(d_k\right)^8}\,\frac{\widehat{e}_{\left(1\right)}\,f_{\left(3\right)}}{R} 
	- \frac{12}{\left(d_k\right)^6}\,\frac{\widehat{f}_{\left(1\right)}\,f_{\left(3\right)}}{R}
	- \frac{3}{\left(d_k\right)^4}\,\frac{\widehat{f}_{\left(3\right)}\,f_{\left(3\right)}}{R} 
	- \frac{12}{\left(d_k\right)^6}\,\frac{\widehat{f}_{\left(5\right)}\,e_{\left(1\right)}}{R}
	+ \frac{6}{\left(d_k\right)^4}\,\frac{\widehat{f}_{\left(5\right)}\,f_{\left(1\right)}}{R}
	- \frac{24}{\left(d_k\right)^8}\,\widehat{f}_{(3)} 
        + \frac{12}{\left(d_k\right)^6}\,\widehat{f}_{(5)}
	\nonumber\\ 
	&& + \frac{19845}{2048}\,\frac{\widehat{g}_{(2)}}{\left(d_k\right)^{10}}
	- \frac{17961}{1024}\,\frac{\widehat{g}_{(4)}}{\left(d_k\right)^8} 
        + \frac{1323}{256}\,\frac{\widehat{g}_{(6)}}{\left(d_k\right)^6} 
        + \frac{183}{128}\,\frac{\widehat{g}_{(8)}}{\left(d_k\right)^4}
        + \frac{15}{16}\,\frac{\widehat{g}_{(10)}}{\left(d_k\right)^2}
        - \frac{75}{8}\,\widehat{g}_{(12)} 
        + \frac{19845}{2048}\,\frac{\widehat{h}_{(2)}}{\left(d_k\right)^{11}}\,.  
        \label{scalar_function_L_n_sigma_28}
        \end{eqnarray}
        \end{widetext}

\section{The scalar functions in Eq.~(\ref{transformation_k_to_n})}\label{Appendix10}

The scalar functions in Eq.~(\ref{transformation_k_to_n}) are just given by the sum of 
the scalar functions of Eqs.~(\ref{transformation_k_to_sigma}) and (\ref{transformation_sigma_to_n}), 
\begin{eqnarray}
	\widehat{M}_{\left(n\right)} &=& \widehat{A}_{\left(n\right)} + \widehat{H}_{\left(n\right)}\,,
	\label{scalar_function_M_N}
	\\
	\widehat{N}_{\left(n\right)} &=& \widehat{B}_{\left(n\right)} + \widehat{I}_{\left(n\right)}\,,
        \label{scalar_function_N_N}
        \\
        \widehat{U}_{\left(n\right)} &=& \widehat{C}_{\left(n\right)} + \widehat{J}_{\left(n\right)}\,,
        \label{scalar_function_U_N}
        \\
        \widehat{V}_{\left(n\right)} &=& \widehat{D}_{\left(n\right)} + \widehat{K}_{\left(n\right)}\,,
        \label{scalar_function_V_N}
        \\
        \widehat{W}_{\left(n\right)} &=& \widehat{E}_{\left(n\right)} + \widehat{L}_{\left(n\right)}\,.
        \label{scalar_function_W_N}
\end{eqnarray}

\noindent
The explicit expressions of the scalar functions $\widehat{A}_{\left(n\right)}$, $\widehat{B}_{\left(n\right)}$, $\widehat{C}_{\left(n\right)}$, 
$\widehat{D}_{\left(n\right)}$, $\widehat{E}_{\left(n\right)}$ and $\widehat{H}_{\left(n\right)}$, $\widehat{I}_{\left(n\right)}$, $\widehat{J}_{\left(n\right)}$,
$\widehat{K}_{\left(n\right)}$, $\widehat{L}_{\left(n\right)}$ are given in the Appendixes~\ref{Appendix8} and \ref{Appendix9}, respectively.


\end{document}